\documentclass[referee]{jfm} 

\usepackage{graphicx}
\usepackage{natbib}
\usepackage{color}
\usepackage{url}
\usepackage{multirow}
\usepackage[table]{xcolor}
\usepackage{colortbl}

\ifCUPmtlplainloaded \else
  \checkfont{eurm10}
  \iffontfound
    \IfFileExists{upmath.sty}
      {\typeout{^^JFound AMS Euler Roman fonts on the system,
                   using the 'upmath' package.^^J}%
       \usepackage{upmath}}
      {\typeout{^^JFound AMS Euler Roman fonts on the system, but you
                   dont seem to have the}%
       \typeout{'upmath' package installed. JFM.cls can take advantage
                 of these fonts,^^Jif you use 'upmath' package.^^J}%
      }
  \else
  \fi
\fi

\ifCUPmtlplainloaded \else
  \checkfont{msam10}
  \iffontfound
    \IfFileExists{amssymb.sty}
      {\typeout{^^JFound AMS Symbol fonts on the system, using the
                'amssymb' package.^^J}%
       \usepackage{amssymb}%
       \let\le=\leqslant  
         
      }{}
  \fi
\fi

\ifCUPmtlplainloaded \else
  \IfFileExists{amsbsy.sty}
    {\typeout{^^JFound the 'amsbsy' package on the system, using it.^^J}%
     \usepackage{amsbsy}}
    {}
\fi

\newcommand\Rey{\mbox{\textit{Re}}}  
\newcommand\Ha{\mbox{\textit{Ha}}}  
\newcommand\N{\mbox{\textit{N}}} 

\newsavebox{\astrutbox}
\sbox{\astrutbox}{\rule[-5pt]{0pt}{20pt}}

\newcommand\eg{e.g.}

\newcommand{\bm}[1]{\mbox{\boldmath$#1$\unboldmath}}
\newcommand{\no}[1]{}
\def\drawline#1#2{\raise 2.5pt\vbox{\hrule width #1pt height #2pt}}

\title[Decay of turbulence in a duct with transverse magnetic field]{Decay of turbulence in a liquid metal duct flow with transverse magnetic field}

\author[Zikanov, Krasnov, Boeck and Sukoriansky]{Oleg Zikanov$^1$, Dmitry Krasnov$^2$, Thomas Boeck$^2$ and Semion Sukoriansky$^3$}
\affiliation{$^1$Department of Mechanical Engineering,  University of Michigan - Dearborn, MI 48128-1491, USA\\
$^2$Institute for Thermodynamics and Fluid Mechanics, Technische Universit\"at Ilmenau, P.O.Box 100565, D-98684 Ilmenau, Germany\\
$^3$Department of Mechanical Engineering, Ben-Gurion University of the Negev, Beer-Sheva, 84105, Israel}

\date{\today}

\begin{document}

\maketitle 

\begin{abstract}
Decay of honeycomb-generated turbulence in a duct with a static transverse magnetic field is studied via direct numerical simulations. The simulations follow the revealing experimental study of \citet{Sukoriansky:1986}, in particular the paradoxical observation of high-amplitude velocity fluctuations, which exist in the downstream portion of the flow when the strong transverse magnetic field is imposed in the entire duct including the honeycomb exit, but not in other configurations. It is shown that the fluctuations are caused by the large-scale quasi-two-dimensional structures forming in the flow at the initial stages of the decay and surviving the magnetic suppression. Statistical turbulence properties, such as the energy decay curves, two-point correlations and typical length scales are computed. The study demonstrates that turbulence decay in the presence of a magnetic field is a complex phenomenon critically depending on the state of the flow at the moment the field is introduced.
\end{abstract}


\section{Introduction} \label{sec:intro}
This paper addresses decay of turbulence in an electrically conducting fluid in the presence of an imposed static magnetic field. The parameters typical for technological and laboratory flows of liquid metals are considered, so the quasi-static (also called non-inductive) approximation, according to which the magnetohydrodynamic flow-field interaction is reduced to the effect of the imposed field on a flow, is adopted \citep[see, \eg,][for the derivation and a discussion of validity of the approximation]{Davidson:2016}. 

In any three-dimensional flow of an electrically conducting fluid, an imposed magnetic field suppresses turbulent fluctuations via the Joule dissipation of induced electric currents. Unlike its viscous counterpart, the Joule dissipation is  active irrespective of the length scale, and anisotropic in the sense that  its rate is proportional to the square of the gradient of velocity along the magnetic field lines. As described by \citet{Moffatt:1967}, this transforms an initially isotropic flow into a flow with reduced or even zero velocity gradients along the magnetic field lines. In flows with walls, the picture is more complex due to the effect of walls on the velocity and electric currents. In particular, in the case of an MHD duct, the mean flow is changed by the Lorentz force, and special boundary layers appear \citep[see, \eg,][]{Branover:1978,Mueller:2001}. The principal features of the transformation of turbulence still remain \emph{(1)} suppression of fluctuations, so the MHD flows are found in a laminar or transitional state at much higher Reynolds numbers than their hydrodynamic counterparts \citep[see, \eg,][for a review]{ZikanovASME:2014}, and \emph{(2)} dimensional anisotropy with weaker velocity gradients along the field lines than across them \citep[see, \eg,][]{Moffatt:1967,Davidson:1997,Zikanov:1998,Vorobev:2005,Krasnov:2008a,Reddy:2014,Verma:2017}. The anisotropy may reach the asymptotic state of flow's quasi- (i.e. to the degree allowed by the boundary conditions) two-dimensionality if the magnetic field is sufficiently strong to suppress three-dimensional instabilities inherently present in such a flow \citep{Thess:2007}.

The term anisotropy is used in this paper with the meaning commonly employed in the research of MHD turbulence \citep[see, \eg,][]{Zikanov:1998,Vorobev:2005,Knaepen:Moin:2004,Knaepen:2004}--as the persistent inequality of the typical length scales of the flow structures in the directions along and across the magnetic field. The anisotropy of the Reynolds stress tensor (the inequality of velocity components) is, as discussed, e.g., by  \citet{Burattini:2010,Favier:2010,Favier:2011,Verma:2015} not caused directly by the magnetic field and strongly affected by the presence of walls and other features of a particular flow, as well as the typical length scale at which the velocity is considered.

It must be stressed that while the picture outlined above is generally correct for any transformation of conventional three-dimensional turbulence, 
MHD flows exhibit complex and often counterintuitive behavior.
Good examples are the flow regimes with spatially localized or intermittent turbulence reported by \citet{ZikanovASME:2014,Boeck:2008,KrasnovPRL:2013,Brethouwer:2012,Krasnov:2012,ZikanovTCFD:2013}, and the experimental demonstration by \citet{Potherat:2014,Potherat:2017} that under certain circumstances the magnetic field can, in fact, enhance turbulence.

A starting point of the modern understanding of the decay of quasi-static MHD turbulence in a uniform 
 field is the theoretical analysis of \citet{Moffatt:1967}. A linearized model based on the assumption of a very strong magnetic field acting on an initially isotropic flow was used and 
the concept of the magnetically induced anisotropy was established, which  largely formed the basis of the future work. The other results of \citet{Moffatt:1967}, such as the power law of the energy decay $\sim t^{-1/2}$ and the asymptotically reached energy partition such that the energy of the field-parallel velocity component becomes two times larger than in the transverse components, have later been 
 found to be non-universal  \citep[see, \eg, discussion in][]{Burattini:2010}. \cite{Verma:2017} also critically reviews some of Moffatt's ideas.

A theoretical model of decaying homogeneous turbulence was developed by \citet{Davidson:1997} \citep[see also][]{Davidson:2016}. Estimates of the rates of viscous and Joule dissipation in terms of the integral length scales along and across the magnetic field have led to a simple model of the decay. It shows that  power-law scaling of energy with time is only possible when one dissipation mechanism is much stronger than the other. 
In general, the decay rate varies with the flow's anisotropy in the course of the process. 

Numerical simulations of homogeneous decaying MHD turbulence in the framework of the periodic box model were performed by \citet{Schumann:1976,Knaepen:Moin:2004,Burattini:2010,Favier:2011}. The results of simulations of \citet{Schumann:1976} and, to a lesser degree, of \citet{Favier:2011} were limited to the behavior at small Reynolds numbers due to the DNS accuracy requirements and the rapid magnetic suppression of turbulence. The limitations were avoided by \citet{Knaepen:Moin:2004} and \citet{Burattini:2010} via the use of the dynamic Smagorinsky LES model, which had been demonstrated to be reliably accurate for the MHD quasi-static turbulence by \citet{Knaepen:Moin:2004,Vorobev:2005,vorobev:2007b}. It was confirmed by \citet{Burattini:2010} that the linear model of \citet{Moffatt:1967} is only valid for very strong magnetic suppression and during short (less than one turnover time) transformation of the flow. Otherwise, the evolution is complex and strongly influenced by the large-scale anisotropic structures forming in the flow during the initial decay period. This implies inevitable influence of the boundaries and, in general, lack of universality of the decay behavior. 

Experimental reproduction of the decay of homogeneous MHD turbulence was attempted by \citet{Alemany:1979}. Turbulence was generated by a grid falling through a cylindrical vessel filled with mercury and positioned within a uniform axially oriented magnetic field. At moderate distances $x$ from the grid, the fluctuation energy of the field-parallel velocity component was found to fall as $\sim x^{-1}$. Farther from the grid, the decay accelerated to approximately $\sim x^{-1.7}$. This change of behaviour was attributed by \citet{Alemany:1979} to the increase of the effective local magnetic interaction parameter $N$ (we define the parameter in section 2.1). An interesting result was found for the energy power spectra, whose slope gradually approached $\sim k^{-3}$ indicating strong anisotropy or even approximate two-dimensionality. It is pertinent to mention in view of our following discussion that in the experiment of \citet{Alemany:1979} turbulence was  generated entirely within the zone of the applied magnetic field. Furthermore, we note that the energy spectrum is difficult to ascertain in strongly suppressed flows at high $\N$. Dependencies other than $\sim k^{-3}$, for example,  exponential decay $\sim \exp(-bk)$ with a decay length $b$, are also found to be  consistent with the experimental and computational data for the energy power spectrum \citep[]{Verma:2017}.

A series of similarly configured experiments with GaInSn as a working fluid was reported by \citet{Voronchikhin:1985}. Several parameters of these experiments make them potentially more interesting for our study than those of \citet{Alemany:1979}. In particular, the use of stationary velocity probes allowed the authors to record longer decay histories. Similarly to our study, two types of decay were considered. In one, as in \citet{Alemany:1979},  turbulence was generated within the magnetic field. In the other, the magnetic field was imposed after full passage of the grid through the cylinder, i. e. on an already developed turbulent flow.

Only a limited portion of the data obtained in the course of the experiments was reported by \citet{Voronchikhin:1985}. This prevents an in-depth comparison between their results and the computational results reported in this paper.  One important conclusion directly relevant to our study was, however, made. The effect of accelerated decay of turbulence caused by the magnetic field was found to be much stronger when the field was imposed on the developed turbulent flow than when  turbulence formed within the field.

Extensive experimental studies of the mercury flows in ducts with imposed transverse magnetic fields were carried out from the late 1960s to  1980s in Riga \citep[see \eg][]{Branover:1970,Kolesnikov:1974,Votsish:1976a,Votsish:1976b,Kljukin:1989}. A major motivation of the experiments was 
to explain the so-called residual fluctuations of velocity found in the flows with strong magnetic fields when the measurements of pressure drop indicated full laminarization. It was hypothesized that the fluctuations were manifestations of nearly two-dimensional flow structures forming in the flow. It was argued that the decay rate of turbulence would be reduced by the presence of such structures in two ways. Their quasi-two-dimensionality would mean that they are only weakly suppressed by the magnetic field. Furthermore, the strong anisotropy would imply reduction of the energy cascade to small length scales or inversion of the cascade, thus leading to reduction of the viscous dissipation rate. 

The existence of quasi-two-dimensional structures was confirmed in the experiments. The flow 
was also found to be strongly influenced by the mechanism of turbulence generation. A particularly interesting example was the experiment of \citet{Kljukin:1989}. Turbulence in a duct was generated by a grid combining two sets of cylindrical bars, one parallel and one perpendicular to the magnetic field. Two experiments were performed: with the bars parallel to the magnetic field located on the downstream or the upstream side of the grid. No significant difference between the two flows was found at weak magnetic fields. In the strong field case, however, the flow with the field-parallel bars on the downstream side of the grid demonstrated residual fluctuations with intensity decreasing very slowly along the duct. No such behavior was found in the flow with the field-parallel bars located on the upstream side of the grid. The effect was attributed by \citet{Kljukin:1989} to formation of strong quasi-two-dimensional vortical structures in the former case.

Recent numerical simulations of MHD duct flows by \citet{ZikanovASME:2014,KrasnovPRL:2013,Krasnov:2012,ZikanovTCFD:2013} have shown that the presence of velocity perturbations at apparently laminar pressure drop along the duct can also be caused by turbulence in the sidewall (parallel to the magnetic field) boundary layers, which survives at much stronger magnetic fields than the turbulence in the core of the duct and in the Hartmann boundary layers normal to the field. Such turbulence could not be registered
in the experiments 
of \citet{Branover:1970,Kolesnikov:1974,Votsish:1976a,Votsish:1976b}, where the measured 
 pressure drop was dominated by the friction in the thin Hartmann layers. At the same time, the alternative explanation proposed by the Riga researchers certainly had substantial experimental support. 

\begin{figure}
\begin{center}

\parbox{0.2\linewidth}{$(a)$\vskip-17mm}\parbox{0.8\linewidth}{$~~~$}\\
\hskip9mm
\includegraphics[width=0.80\textwidth,clip=]{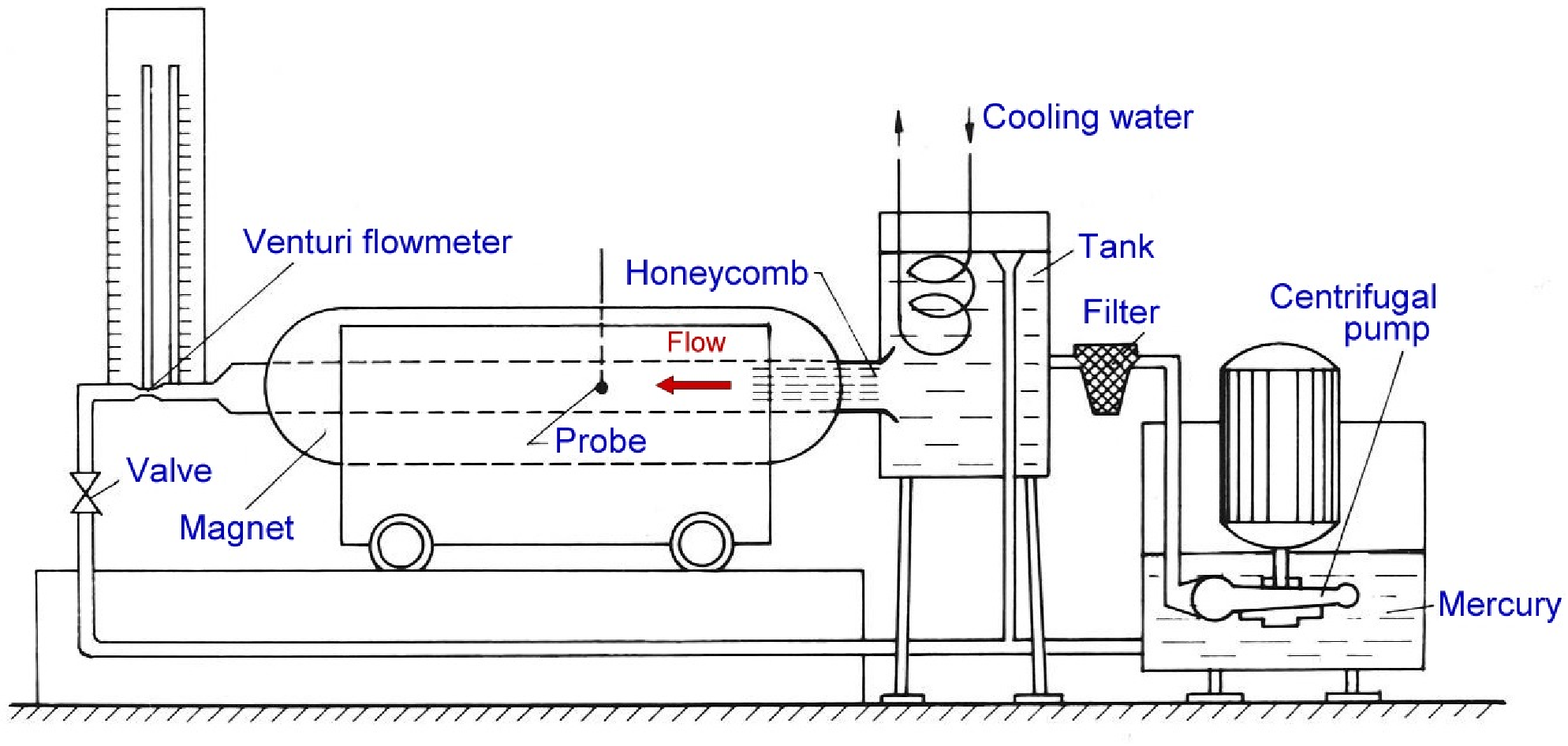}

\parbox{0.2\linewidth}{$(b)$\vskip-10mm}\parbox{0.8\linewidth}{$~~~$}\\
\includegraphics[width=0.80\textwidth,clip=]{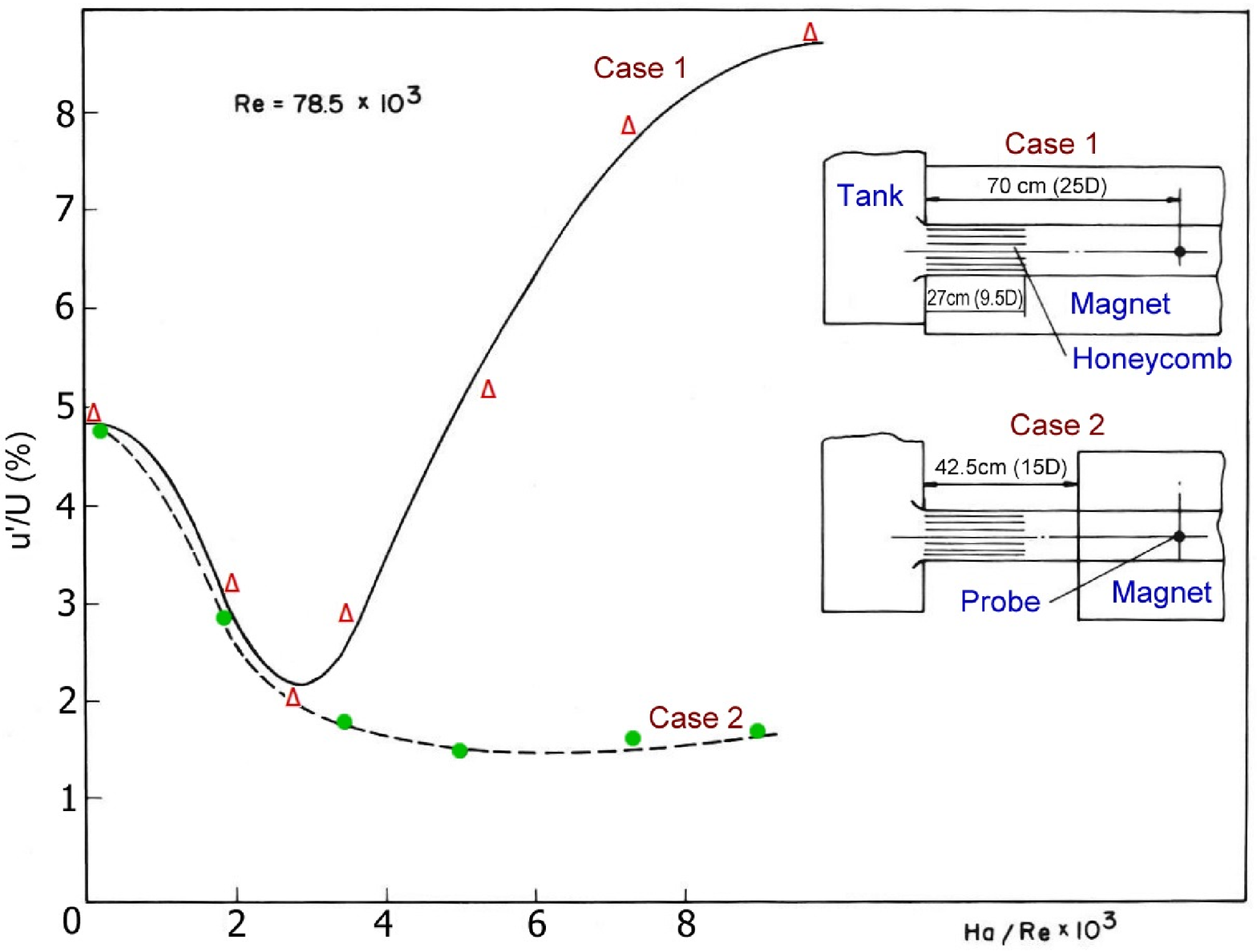}

\end{center}
\caption{\emph{(a)}, Schematic diagram of the experimental facility of \citet{Sukoriansky:1986}.
\emph{(b)}, Experimental results. Turbulence intensities on the duct axis as functions of $Ha/Re$
at different magnet positions (reprinted with the permission of Springer).}
\label{fig:experiment}
\end{figure}

The present work follows  closely the experiments of \citet{Sukoriansky:1986}, in which the phenomenon  of turbulent fluctuations persisting along the duct in the presence of a strong magnetic field was revisited on a higher level of accuracy and technical sophistication. Flows of mercury in a duct of $2\times 4.8$ cm cross-section were studied. Magnetic field of strength up to 1.1 T with the main component transverse to the flow's direction and parallel to the shorter side of the duct was imposed in the test section by a long (pole length about $ 90$ cm) electromagnet. 
The inlet into the test section was equipped with a honeycomb consisting of densely packed round tubes of diameter 2.4 mm with electrically insulating 0.5 mm thick walls (common drinking straws). The purpose of the honeycomb was two-fold. It generated approximately isotropic and uniform field of velocity fluctuations and reduced or even prevented the M-shaped mean velocity profile normally forming at the entrance into the magnetic field \citep[see \eg][]{Branover:1978}. The Reynolds and Hartmann numbers were
\begin{equation}
\label{rehar}
\Rey_D\equiv\frac{DU}{\nu}=7.85\times 10^4, \: \Ha_D\equiv BD\left(\frac{\sigma}{\rho \nu}\right)^{1/2}=0,\ldots,780,
\end{equation}
where $D$ was the duct's hydraulic diameter, $U$ was the mean velocity, and $\nu$, $\sigma$, and $\rho$ were the kinematic viscosity, electric conductivity, and density of the fluid.  The experimental setup and the key results are shown in figures \ref{fig:experiment}a and b, respectively.

The striking and, at first glance, paradoxical results were obtained in the hot-film measurements of velocity fluctuations 43 cm downstream  of the honeycomb's exit. The measurements showed completely different signals for the two distributions of the magnetic field illustrated in figure \ref{fig:experiment}a. In the situation identified in \citet{Sukoriansky:1986} and this paper as Case 1, the entire length (27 cm) of the honeycomb was located between the magnet poles (see the upper schematic illustration in figure \ref{fig:experiment}b), and turbulence was generated and decayed entirely within the practically uniform transverse magnetic field. In the situation identified as Case 2, the magnet poles were shifted downstream so that the axial distance between the honeycomb's exit and the nearest corner of the pole was 15.5 cm (see the lower schematic illustration in figure \ref{fig:experiment}b). In this case, turbulence was generated at negligible magnetic field and traveled about 5.5 convective times $D/U$ before entering the space between the poles and thus experiencing the full magnetic suppression effect. 

The key results are shown 
 in figure \ref{fig:experiment}b reproduced from figure 5 of  \citet{Sukoriansky:1986}. The curves show the turbulence intensity $u^{\prime} /U$ based on the streamwise velocity fluctuations measured on the duct axis 43 cm downstream of the honeycomb, i.e. well in the zone of the uniform magnetic field. The signals measured in the two cases are about the same for weak magnetic fields, approximately at $\Ha_D/\Rey_D<3\times 10^{-3}$. 
 Owing to the turbulence suppression by the magnetic field, the intensities decrease with growing $\Ha_D$ reaching $\sim 0.02$ at  $\Ha_D/\Rey_D=3\times 10^{-3}$. For stronger magnetic fields, however, the signals show entirely different trends. In the case 2, the intensity continues to decrease to about $0.015$ at high $\Ha_D$. In the case 1, the intensity grows rapidly with growing $\Ha_D$ and reaches 0.09 (almost twice the intensity in the flow without magnetic field) at $\Ha_D/\Rey_D=10^{-2}$. 

The appearance of high-amplitude fluctuations for strong magnetic fields in the case 1 configuration was explained in \citet{Sukoriansky:1986} by the effect described above, i.e by development of quasi-two-dimensional flow structures with  weak gradients along the magnetic field lines. Such structures would experience weak magnetic suppression and a reduced energy cascade to small length scales thus preserving the strength of the associated velocity fluctuations  as the fluid moved downstream. The explanation is consistent with 
other experiments, e.g. of \citet{Kljukin:1989}. No direct  evidence of this scenario has, however, been obtained. The type of the flow structures and the degree of their anisotropy could also not be determined in the experiments and has not been a subject of numerical analysis.

In this paper, we present high-resolution numerical simulations designed to explore validity of the hypothesized scenario leading to the residual velocity fluctuations and to produce a detailed description of the flow. The numerical model reproduces the geometry and parameters of the experiment of \citet{Sukoriansky:1986} with one adjustment. For the purpose of understanding the effect of walls on decaying turbulence, two orientations of the transverse magnetic field, along the shorter (as in \citet{Sukoriansky:1986}) and longer sides of the duct are considered. The role of the anisotropy  introduced by the honeycomb is also addressed. The problem formulation, parameters and numerical procedure
are described in section \ref{sec:model}. The structure and statistical properties of the computed flows are presented in section  \ref{sec:results} . The concluding remarks are provided in section \ref{sec:concl}.

\section{Problem formulation, method and parameters}
\label{sec:model}

\subsection{Problem formulation}

\begin{figure}
\begin{center}

\includegraphics[width=1.0\textwidth,clip=]{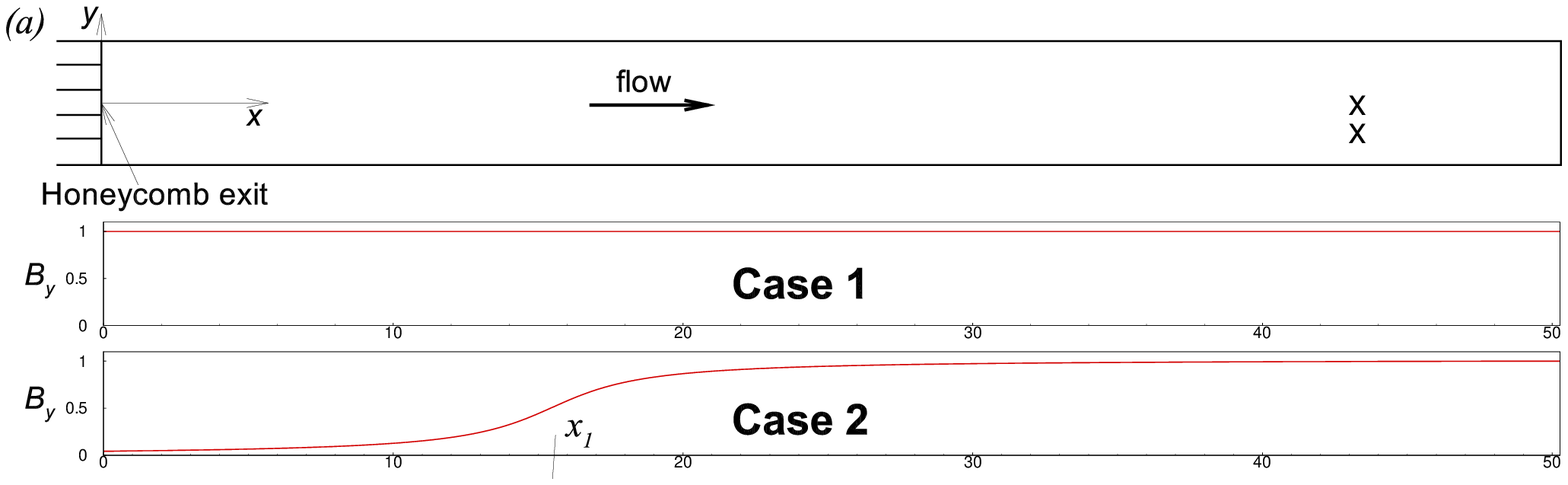}\\

\parbox{0.5\linewidth}{\hskip-10mm $(b)$ Type A}\parbox{0.3\linewidth}{\hskip-10mm Type B}
\includegraphics[width=0.48\textwidth,bb=20 20 592 278,clip=]{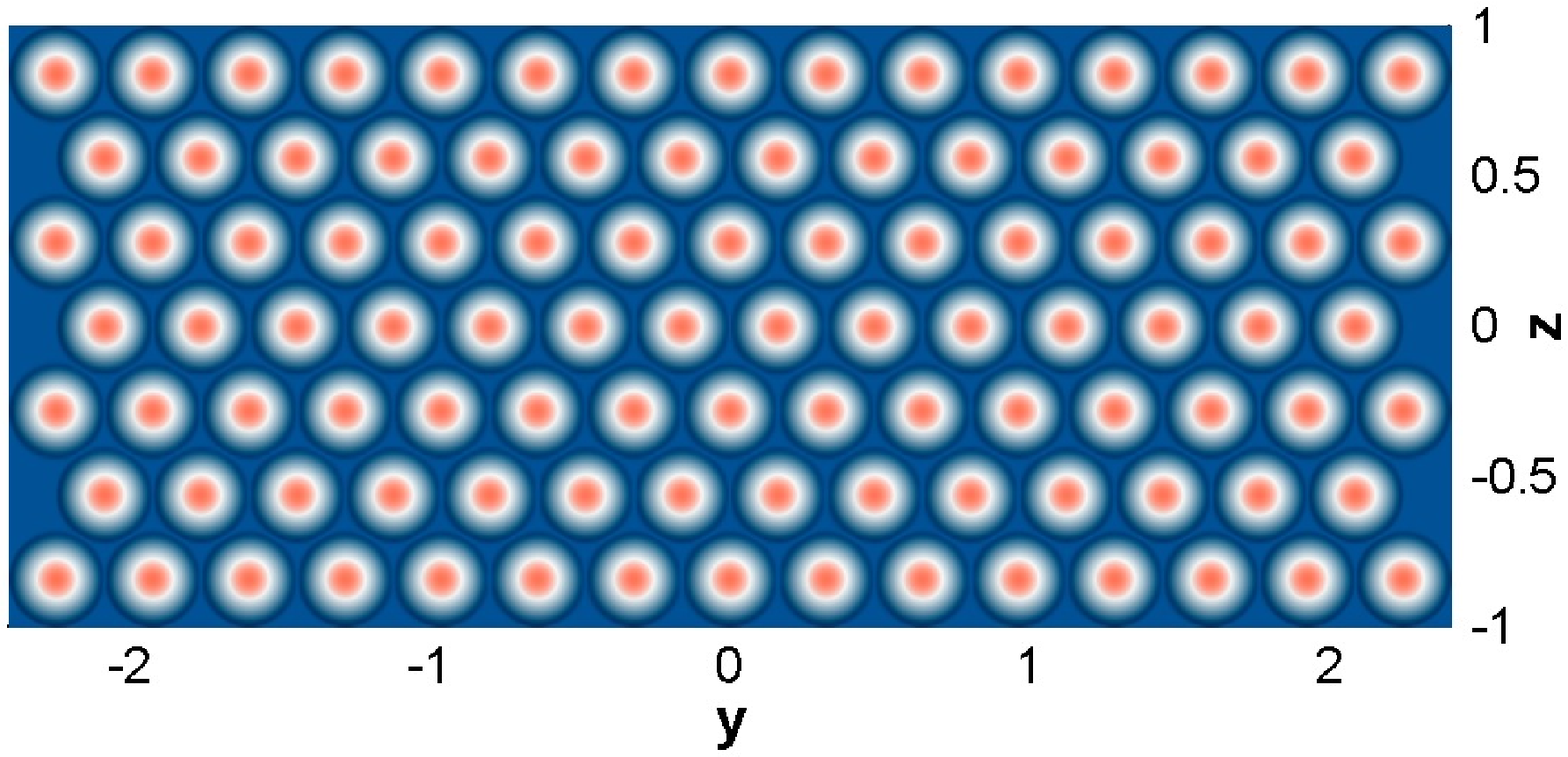}
\includegraphics[width=0.48\textwidth,bb=20 20 592 278,clip=]{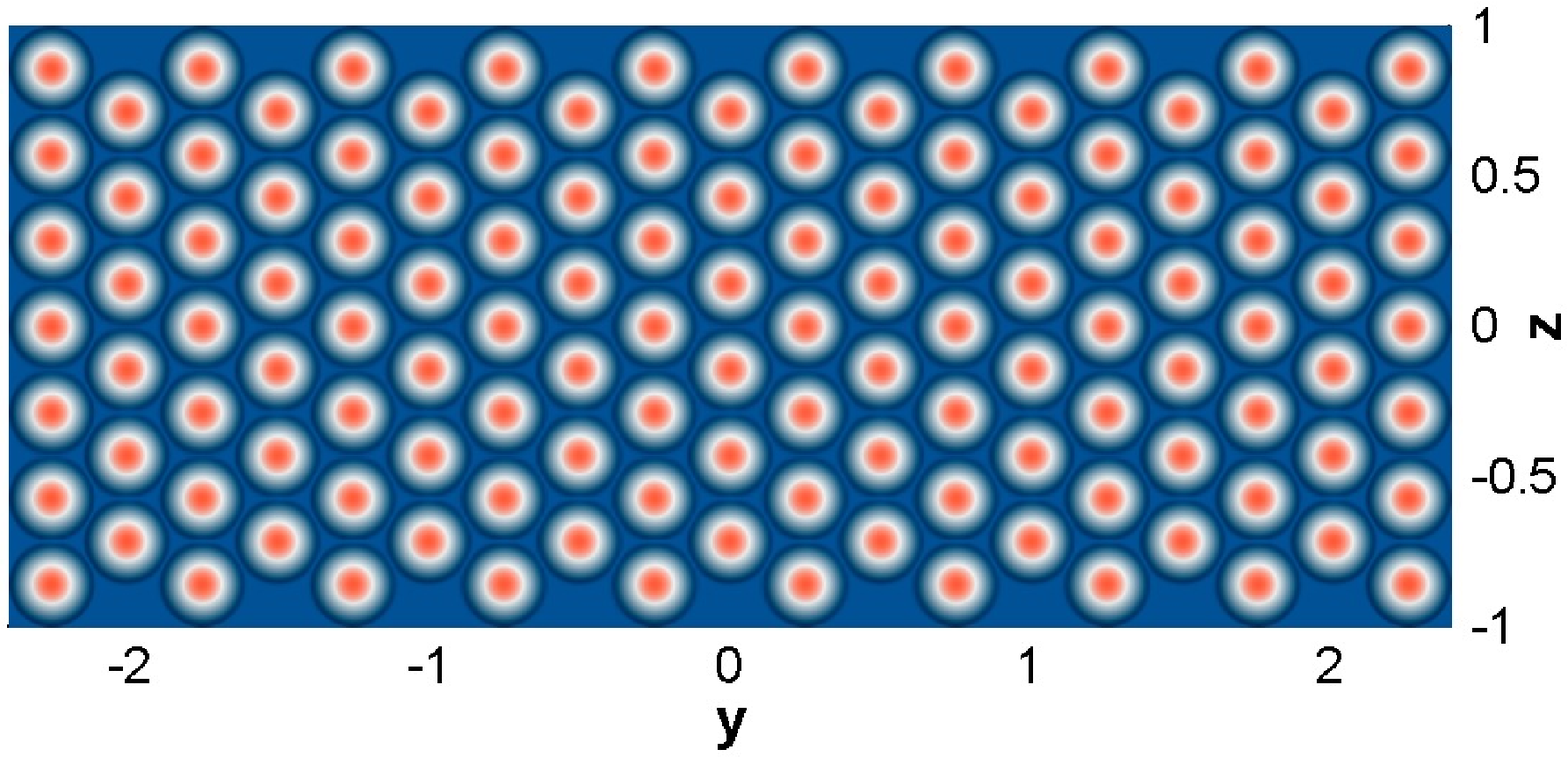}

\includegraphics[width=0.45\textwidth,bb=50 20 580 80,clip=]{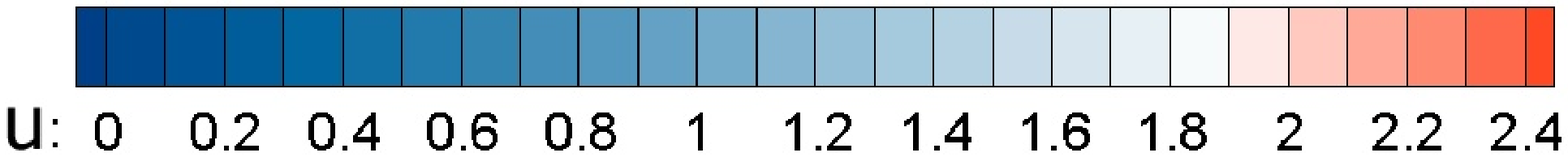}
\vskip-4mm

\end{center}
\caption{Setting of the problem. \emph{(a)} Scheme of the computational domain shown in the $x-y$ cross-section.
The $x$- and $y$-axes of the coordinate system used in the simulations are shown. The non-dimensional width of
the domain in the $z$-direction is 2.0. The profiles of the main component $B$ of the magnetic field computed
according to the model of \citet{Votyakov:2009} are shown (see text). $x_1$ is the location of the upstream corner
of the magnet pole-pieces in the case 2. The two crosses in the downstream part of the flow domain indicate the locations
where the velocity fluctuation signals are recorded in the experiment of \citet{Sukoriansky:1986} and in the simulations.
\emph{(b)} Distribution of the streamwise velocity $u_{inlet}$ imposed at the inlet to imitate the flow exiting
the honeycombs of type A and type B (see text).
}
\label{fig:geom}
\end{figure}

An isothermal flow of an incompressible electrically conducting Newtonian fluid in a duct of rectangular cross-section is considered. A transverse magnetic field, the exact configuration of which is specified below, is imposed. Assuming the asymptotic limit of low magnetic Reynolds and Prandtl numbers, the quasi-static (non-inductive) approximation of the magnetohydrodynamic interactions \citep[see \eg][]{Davidson:2016} is used. The non-dimensional governing equations are  
\begin{eqnarray}
\label{eq:gov_Navier}
\frac{\partial \bm{v}}{\partial t} + (\bm{v} {\cdot \nabla})\bm{v} =
- {\nabla p} + \frac{1}{Re} {\nabla}^2 \bm{v}
+ \frac{\Ha^2}{\Rey} \left(-\nabla \phi \times \bm{B} + (\bm{v} \times \bm{B}) \times \bm{B} \right),\\
\label{eq:gov_div}
\nabla \cdot \bm{v} = 0,\\
\label{eq:gov_phi}
\nabla^2{\phi} = \nabla \cdot ({ \bm{v} \times  \bm{B}}),
\end{eqnarray}
where $\bm{v}$, $p$, and $\phi$ are the fields of velocity, pressure and electric potential and $\bm{B}$ is the non-dimensionalized magnetic field. The typical scales used to derive (\ref{eq:gov_Navier})--(\ref{eq:gov_phi}) are the mean streamwise velocity $U$ for velocity, shorter half-width $H$ of the duct for length, $H/U$ for time, $\rho U^2$ for pressure, the maximum strength $B_0$ of the transverse component  for the magnetic field, and $U B_0 H$ for electric potential. The non-dimensional parameters are the Reynolds number
\begin{equation}
\label{Rey}
 	\Rey \equiv \frac{U H}{\nu}
\end{equation}
and the Hartmann number 
\begin{equation}
\label{Hart}
\Ha \equiv B_0 H\sqrt{\sigma/\rho\nu}
\end{equation}
related to the parameters (\ref{rehar}) based on the hydraulic diameter as $\Rey=0.3542\Rey_D$ and $\Ha=0.3542\Ha_D$. 

We will also use the magnetic interaction parameter
\begin{equation}
\label{Stuart}
\N \equiv \frac{\Ha^2}{\Rey} = \frac{B_0^2H\sigma}{\rho U}.
\end{equation}

Further settings of the problem are illustrated in figure \ref{fig:geom}. The computational domain reproduces the test section of the experiment of \citet{Sukoriansky:1986}. It is a duct segment of length $0 \le x \le L_x$ and cross-section $-L_y/2\le y\le L_y/2$, $-L_z/2\le z\le L_z/2$ with $L_x=16\pi$, $L_y=4.8$ and $L_z=2.0$. 

The sidewalls are of zero slip and perfect electric insulation:
\begin{equation}
\label{eq:bc_walls}
\bm{v}=0, \: \frac{\partial \phi}{\partial n}=0 \mbox{ at sidewalls}.
\end{equation}

At the inlet $x=0$, we require that 
\begin{equation}
\label{eq:bc_phiinlet}
\frac{\partial \phi }{\partial x}=0.
\end{equation}
A velocity distribution imitating the flow exiting the  honeycomb is applied. In the experiment, the tubes of the honeycomb are densely packed and have the inner diameter $d\approx 2.4$ mm and wall thickness about 0.5 mm. The  parameters for the flow in a single tube are $\Rey_d=6600$ and $\Ha_d=45$ and the non-dimensional pipe length is $L_d/d=112.5$. At such parameters, the flow is expected to be weakly turbulent in the case 2. In the case 1, the magnetic field suppresses turbulence  and slightly deforms the streamwise velocity profile \citep[see][]{ZikanovASME:2014,Mueller:2001,Li:2013}. The numerical model ignores the differences and uses the same velocity distribution in  the two cases (see figure \ref{fig:geom}b). The flow in the space between the tubes present in the experiment is also ignored. 

To compute the velocity distribution, the inlet plane is covered by hexagons, into which circles of inner diameters and wall thickness corresponding to those of the honeycomb tubes are fitted. The axisymmetric parabolic profile of streamwise velocity is imposed within each tube. At each time step, random three-dimensional velocity perturbations of relative amplitude $10^{-4}$ are added, after which the entire distribution is rescaled so that the mean streamwise velocity is equal to 1.0. 

As illustrated in figure \ref{fig:geom}b, the tubes of the honeycomb can be packed so that they form straight rows along the longer (the honeycomb type A in the following discussion) or shorter (type B) walls of the duct.  The results of the simulations presented in section 3.2 demonstrate that the two arrangements produce noticeably different flows.

Soft  boundary conditions 
\begin{equation}
\label{eq:bc_exit}
\frac{\partial \bm{v}}{\partial x}=\frac{\partial \phi}{\partial x}=0
\end{equation}
are applied at the exit $x=L_x$ of the computational domain.

Two orientations of the main component of the magnetic field, parallel to the longer ($B_y$) or shorter ($B_z$) walls of the duct are used. In each case, 
the distribution of the magnetic field is approximated in the simulations using the model suggested by \citet{Votyakov:2009}. The model provides simple formulas for divergence-free, two-dimensional, two-component field created by a magnet with two infinitely wide rectangular pole-pieces. The accuracy of the model was verified in comparison with measurements in \citet{ZikanovJFM:2013}. The input parameters of the model are the coordinates of the corners of the pole-pieces, for which we take $y=\pm 2.6$ (for $B_y$) or $z=\pm 1.2$ (for $B_z$), and $x_1=-27$, $x_2=63$ in the case 1 and $x_1=15.5$, $x_2=105.5$ in the case 2. The resulting magnetic field has the main component illustrated in figure \ref{fig:geom}a and the component $B_x$, which is much weaker and only significant within the flow domain around the entrance into the magnetic field in the case 2.

The problem is solved numerically using the finite-difference scheme first described as the scheme B in \citet{Krasnov:FD:2011} and extended to spatially evolving flows in a duct {\eg }  in \citet{ZikanovTCFD:2013}. 
The solver has been successfully applied in numerous simulations of turbulent and transitional MHD flows at high $\Rey$ and $\Ha$ \citep[see \eg][]{ZikanovASME:2014,KrasnovPRL:2013,Krasnov:2012,ZikanovTCFD:2013,Li:2013}. 
The scheme is explicit and of the second order in time and space. The discretization is on the structured collocated grid built along the lines of the Cartesian coordinate system. The exact conservation of mass, momentum, and electric charge, as well as near-conservation of kinetic energy are achieved  by using the velocity and current fluxes obtained by interpolation to staggered grid points. The standard projection technique is applied to compute pressure and enforce incompressibility. The numerical algorithm is parallelized using the hybrid MPI-OpenMP approach. 

The  modification of the algorithm in comparison with the original version of \citet{Krasnov:FD:2011} concerns the solution of the Poisson equations for pressure and electric potential. The fast cosine decomposition is used in the streamwise direction, for which the right-hand side of the equation is modified to achieve homogeneous Neumann boundary conditions at $x=0$ and $x=L_x$. The direct cyclic reduction solver implemented in the subroutines of the library FishPack \citep{fishpack} is used in the $y-z$-plane.

The computational results reported below are obtained on the grid consisting of
$N_x\times N_y \times N_z=3072\times 512 \times 192$ points. The points are clustered
towards the duct's walls using the coordinate transformation
\begin{equation}
\label{eq:coordtrans}
y = \frac{L_y}{2}\left[0.9\sin\left(\frac{\pi}{2}\eta\right)+0.1\eta\right], \: z = \frac{L_z}{2}\left[0.9\sin\left(\frac{\pi}{2}\zeta\right)+0.1\zeta\right],
\end{equation}
where $-1 \le \eta \le 1$ and $-1 \le \zeta \le 1$ are the transformed coordinates,
in which the grid is uniform.

A grid sensitivity study was performed to determine that the model sufficiently accurately reproduced
the essential features of the flow, such as mixing and instabilities of the honeycomb jets, generation
of turbulence, and its decay in the presence of the magnetic field. Additional simulations for the case 1
and case 2 configurations with the magnetic field parallel to the longer sides of the duct on the smaller grid with $N_x\times N_y \times N_z=2048\times 384 \times 128$
and the same clustering scheme were carried out. The results were qualitatively the same as on the larger grid
with minor quantitative differences. In particular, the time-averaged  wall friction coefficients computed for the entire flow domain
changed by less than 1\%. The effect of the numerical resolution on the results is further discussed in section \ref{sec:concl}.

Several additional tests were performed at $\Ha=0$ to analyze the effect of the grid size, grid clustering,
and the amplitude of the noise added at the inlet on the instability and mixing of jets in the portion
of the duct just downstream of the inlet. It has been found that at the grid clustering associated with
(\ref{eq:coordtrans}) further increase of the grid size and further decrease of the noise amplitude
do not result in visible changes in the formation of turbulence.

\section{Results}
\label{sec:results}
The parameters of the simulations are listed in Table \ref{table:param}.
For convenience of the readers, the runs are numbered such that odd indices 1,3,5,7 correspond to
case 1 with homogeneous field, whereas even indices 2,4,6,8 -- to case 2 with non-homogeneous field (see Fig. \ref{fig:geom}a).
Each simulation is initialized  with a laminar state and continued for $100$ non-dimensional time units, 
whereby a fully developed flow is established. Subsequently, the simulation is continued for a ``production phase''
of $100$ (in the runs 1-4) or $50$ (in the runs 5-8) time units. The turbulence statistics in this paper are based on
the, respectively, $1000$ or $500$ flow samples collected during this phase with the time interval $0.1$.

The simulations $1$ and $2$ are for $\Ha/\Rey=\Ha_D/\Rey_D=2.0\times 10^{-3}$, i.e., for the parameters in
the range of moderate magnetic fields where a strong (two-fold) reduction of turbulence intensity was detected
in the experiment of \citet{Sukoriansky:1986} for both the field configurations (see {figure \ref{fig:experiment}b).
The simulations $3$-$8$ are for $\Ha/\Rey=\Ha_D/\Rey_D=7.0\times 10^{-3}$. For this strong magnetic field,
the experiment shows anomalous behavior with the turbulence intensity in the case 2 configuration remaining low,
but the intensity in the case 1 configuration growing to a level about 50\% higher than without the magnetic field.

In the following discussion, the properties of the computed flows  are analyzed using the fields of turbulent fluctuations defined as
\begin{equation}\label{turbfluctdef}
\bm{v}'=\bm{v}-\langle \bm{v}\rangle,
\end{equation}
where $\langle \bm{v}\rangle(\bm{x})$ is the mean velocity obtained by time-averaging over the entire production phase of the run.

\definecolor{Lgray}{gray}{0.9}
\definecolor{Lcyan}{rgb}{0.8, 1.0, 1.0}
\definecolor{Lpink}{rgb}{1.0, 0.8, 1.0}
\definecolor{Maroon}{cmyk}{0,0.87,0.68,0.32}

\begin{table}
\begin{center}
  \begin{tabular}{@{}c|c|c|c|c|c|c|c@{}}
           &     Field    &     Field     & Honeycomb &        &       & & \\[-0.8em]
    Run \# &  Orientation & Configuration &   Type    & $\Rey$ & $\Ha$ & $\N=\frac{\Ha^2}{\Rey}$ & $R=\frac{\Rey}{\Ha}$ \\
     \hline
    1   & $B_y$ & Case 1 & A &  27800 &  55 & 0.1088 & 505.5 \\
    2   & $B_y$ & Case 2 & A &  27800 &  55 & 0.1088 & 505.5 \\
    3   & $B_y$ & Case 1 & A &  27800 & 195 & 1.368  & 142.6 \\
    4  & $B_y$ & Case 2 & A &  27800 & 195 & 1.368  & 142.6 \\
    5 & $B_z$ & Case 1 & A &  27800 & 195 & 1.368  & 142.6 \\
    6 & $B_z$ & Case 2 & A &  27800 & 195 & 1.368  & 142.6 \\
    7 & $B_z$ & Case 1 & B &  27800 & 195 & 1.368  & 142.6 \\
    8 & $B_z$ & Case 2 & B &  27800 & 195 & 1.368  & 142.6 \\
  \end{tabular}
\end{center}
\vskip2mm
\caption{Simulation parameters.}
\vskip-2mm
\label{table:param}
\end{table}

We start the discussion with the main results summarized in Table \ref{table:signals}. The time-averaged root-mean-square amplitudes of the velocity fluctuations computed at $x=43$, $z=0$ and two values of $y$ are shown.  The values for $u'$ correspond to the experimental measurements of \citet{Sukoriansky:1986} (see figure \ref{fig:experiment}b) and show that the seemingly paradoxical dependence of the fluctuation amplitude on the strength of the magnetic field and magnet's location is reproduced by the simulations.  Weak fluctuations of all the velocity components are found in the runs 1 and 2 
 performed at $\Ha=55$. Equally weak fluctuations are  found in the runs 4, 6, and 8 
  performed at $\Ha=195$ when the poles of the magnet shifted downstream (the case 2 configuration in figure \ref{fig:geom}). Anomalously high fluctuation amplitudes are found in the runs 3, 5, and 7, i.e.
  in the flows with $\Ha=195$ and the honeycomb exit located within the zone of uniform magnetic field (the  case 1 configuration in figure \ref{fig:geom}). The amplitudes of two velocity components are increased: the streamwise component $u'$ and the component orthogonal to the magnetic field ($w'$ in the run 3 
  and $v'$ in the runs 5 and 7).
  The increase in comparison to the other cases is about four-fold in the runs 3 
  and 7 
  and  two-fold in the run 5.

Table \ref{table:signals} shows that the flow's behaviour is affected by the magnetic field strength, magnet location, orientation of the magnetic field with respect to the duct walls, and the honeycomb arrangement. The following discussion is separated into two parts. The mechanism of the generation of high-amplitude fluctuations is explained and illustrated in section \ref{sec:results_a} on the basis of the results obtained in the runs 1-4.
Further investigation of the fluctuations is presented in section \ref{sec:results_b}, where 
the influence of the magnetic field orientation and honeycomb arrangement is analyzed using the data from the runs 5-8.

A comment is in order concerning the comparison between the simulations and the experiments of \citet{Sukoriansky:1986}. 
As we have already mentioned and discuss in detail below, the qualitative agreement is quite satisfactory. 
The quantitative agreement is, however, poor.  
From table \ref{table:signals} and 
{figure \ref{fig:experiment}b} and figure 6 of \citet{Sukoriansky:1986} we see that in all the simulations the computed rms fluctuations are about five times lower than in the experiment. Possible reasons for this are discussed in section \ref{sec:concl}.

\begin{table}
\begin{center}
  
{\small
\begin{tabular}{@{}r@{}@{}c@{}c@{}c@{}@{}c@{}c@{}c@{}}

\multirow{2}{*}{Run \#~~} & \multicolumn{3}{c}{center ($y=0$)} & \multicolumn{3}{c}{off center ($y=-1.4$)}\\  
{} & $u'$ & $v'$ & $w'$  & $u'$ & $v'$ & $w'$  \\\hline
1 \hskip20mm & ~$4.35\times 10^{-3}$~ & ~$3.90\times 10^{-3}$~ & ~$3.47\times 10^{-3}$~ \hskip20mm & ~$4.40\times 10^{-3}$~ & ~$4.41\times 10^{-3}$~ & ~$3.79\times 10^{-3}$~ \\
 2 \hskip20mm & ~$4.67\times 10^{-3}$~ & ~$4.55\times 10^{-3}$~ & ~$3.55\times 10^{-3}$~ \hskip20mm & ~$4.82\times 10^{-3}$~ & ~$5.23\times 10^{-3}$~ & ~$4.12\times 10^{-3}$~ \\
\rowcolor{Lgray!80}
3 \hskip20mm & ~$1.35\times 10^{-2}$~ & ~$4.45\times 10^{-3}$~ & ~$1.13\times 10^{-2}$~ \hskip20mm & ~$1.35\times 10^{-2}$~ & ~$4.30\times 10^{-3}$~ & ~$1.21\times 10^{-2}$~ \\
4 \hskip20mm & ~$3.35\times 10^{-3}$~ & ~$2.65\times 10^{-3}$~ & ~$2.38\times 10^{-3}$~ \hskip20mm & ~$4.11\times 10^{-3}$~ & ~$2.56\times 10^{-3}$~ & ~$2.78\times 10^{-3}$~ \\
\rowcolor{Lgray!80}
5 \hskip20mm & ~$5.30\times 10^{-3}$~ & ~$5.40\times 10^{-3}$~ & ~$1.61\times 10^{-3}$~ \hskip20mm & ~$5.62\times 10^{-3}$~ & ~$5.75\times 10^{-3}$~ & ~$1.61\times 10^{-3}$~ \\
6 \hskip20mm & ~$3.76\times 10^{-3}$~ & ~$3.87\times 10^{-3}$~ & ~$1.87\times 10^{-3}$~ \hskip20mm & ~$3.84\times 10^{-3}$~ & ~$3.27\times 10^{-3}$~ & ~$2.78\times 10^{-3}$~ \\
\rowcolor{Lgray!80}
7 \hskip20mm & ~$1.23\times 10^{-2}$~ & ~$1.05\times 10^{-2}$~ & ~$3.46\times 10^{-3}$~ \hskip20mm & ~$1.18\times 10^{-2}$~ & ~$1.25\times 10^{-2}$~ & ~$3.59\times 10^{-3}$~ \\
8 \hskip20mm & ~$5.03\times 10^{-3}$~ & ~$4.66\times 10^{-3}$~ & ~$1.48\times 10^{-3}$~ \hskip20mm & ~$5.17\times 10^{-3}$~ & ~$4.69\times 10^{-3}$~ & ~$2.81\times 10^{-3}$~
\end{tabular}
}

\end{center}
\vskip2mm
\caption{RMS amplitudes of fluctuations of velocity components at the points $x=43$, $z=0$, $y=0$ (center) and $x=43$, $z=0$, $y=-1.4$ (off center) computed using the entire signals of fully developed flow. Since the time-averaged streamwise velocity at these points is about 1.0 in our units, the values approximately correspond to the respective turbulence intensities. The data for flows with anomalously high fluctuation amplitudes are marked by gray colour.
}
\vskip-2mm
\label{table:signals}
\end{table}

\subsection{Effect of magnetic field on turbulence decay}
\label{sec:results_a}
The following discussion is primarily based on 
simulations 1-4.

\subsubsection{Velocity fluctuations}
\label{sec:signals}

Figure \ref{fig:signals} shows the time signals of the velocity components computed at the point
$x=43$, $y=z=0$ corresponding to the point of velocity measurements in the experiment of \citet{Sukoriansky:1986} (see figure \ref{fig:experiment}b). 
The rms amplitudes listed in table \ref{table:signals} are calculated using these signals and similar signals recorded at $x=43$, $y=-1.4$, $z=0$. 
We see that the behaviour indicated by the rms data is not subject to significant variations at long time scales. 
Consistent anomalously high fluctuation amplitudes of streamwise ($u$) and field-normal transverse ($w$) velocity components are found in the run 3 
when the magnetic field is strong and has the case 1 configuration.

\begin{figure}
\vskip3mm

\parbox{0.5\linewidth}{\hskip5mm $\Ha=55$}\parbox{0.3\linewidth}{\hskip5mm $\Ha=195$}
\centerline{
\includegraphics[width=0.49\textwidth,clip=]{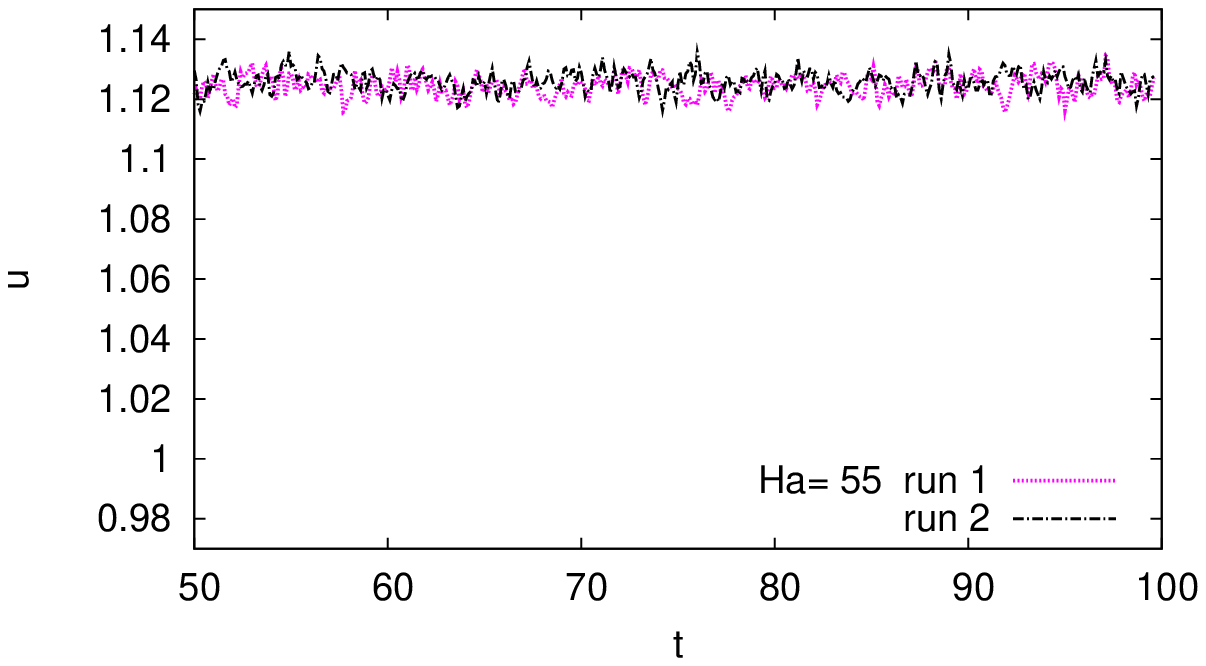}
\includegraphics[width=0.49\textwidth,clip=]{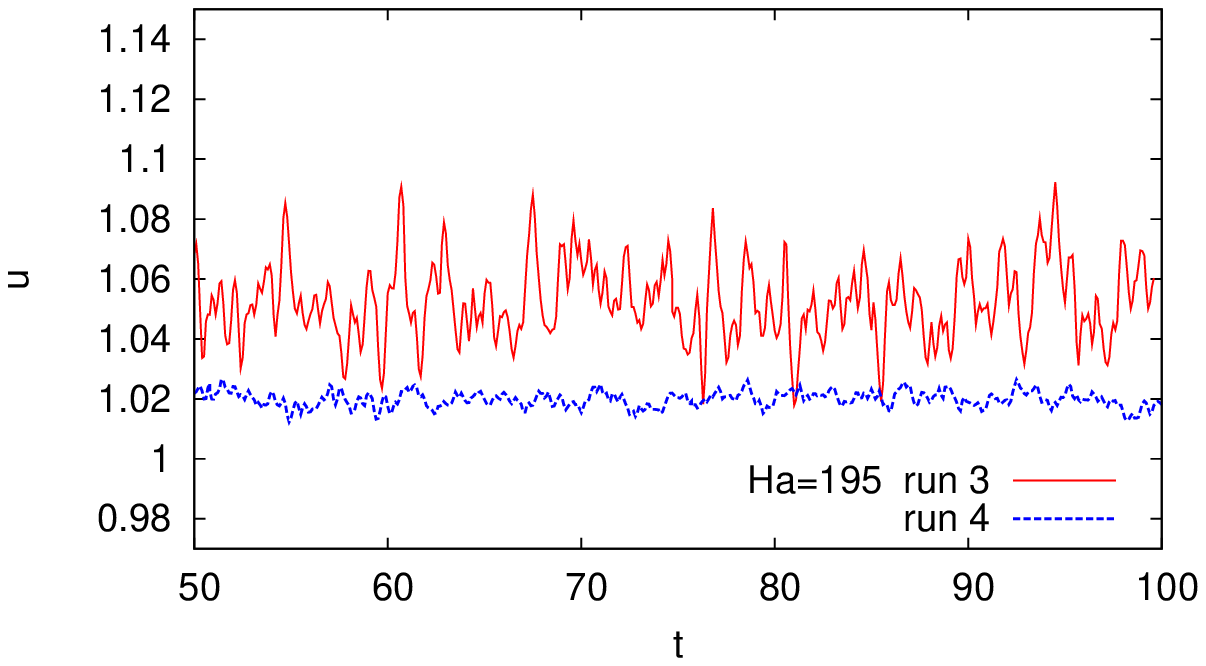}
}
\centerline{
\includegraphics[width=0.49\textwidth,clip=]{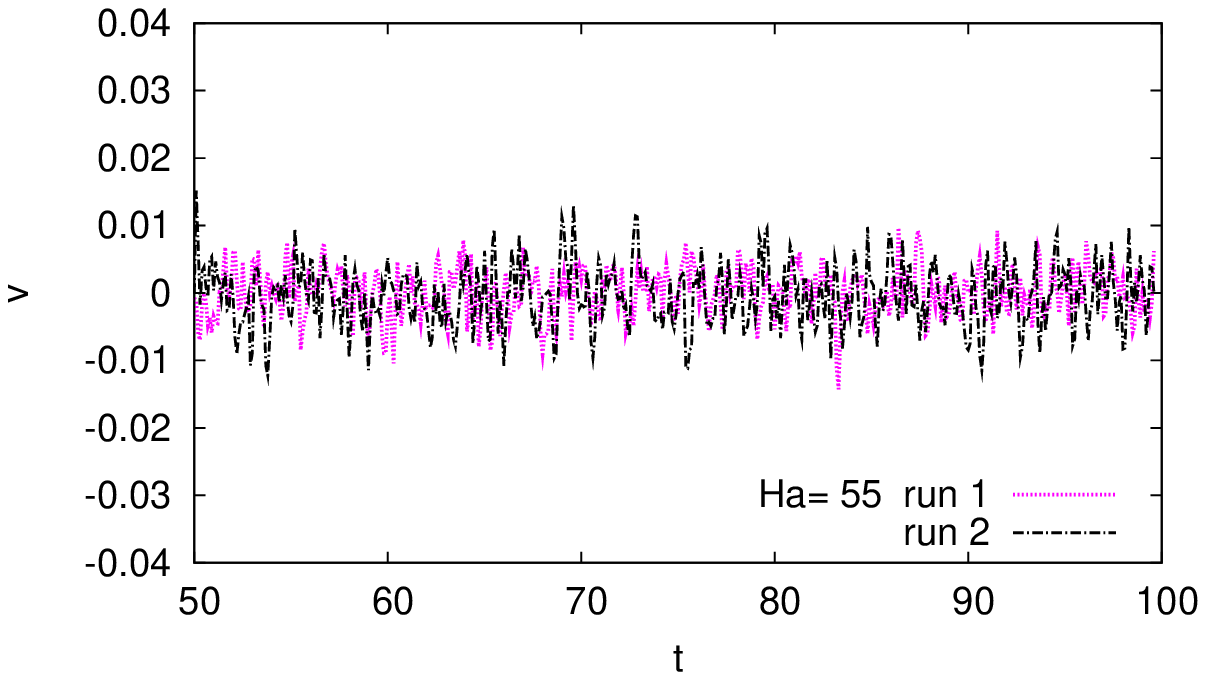}
\includegraphics[width=0.49\textwidth,clip=]{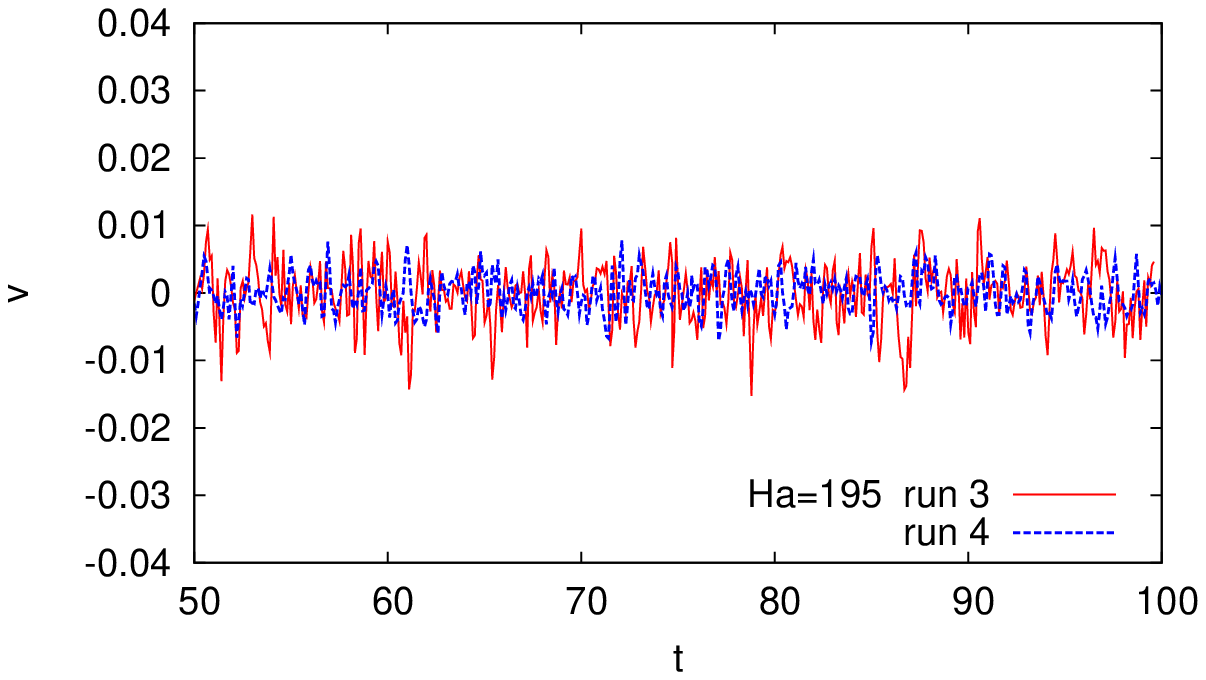}
}
\centerline{
\includegraphics[width=0.49\textwidth,clip=]{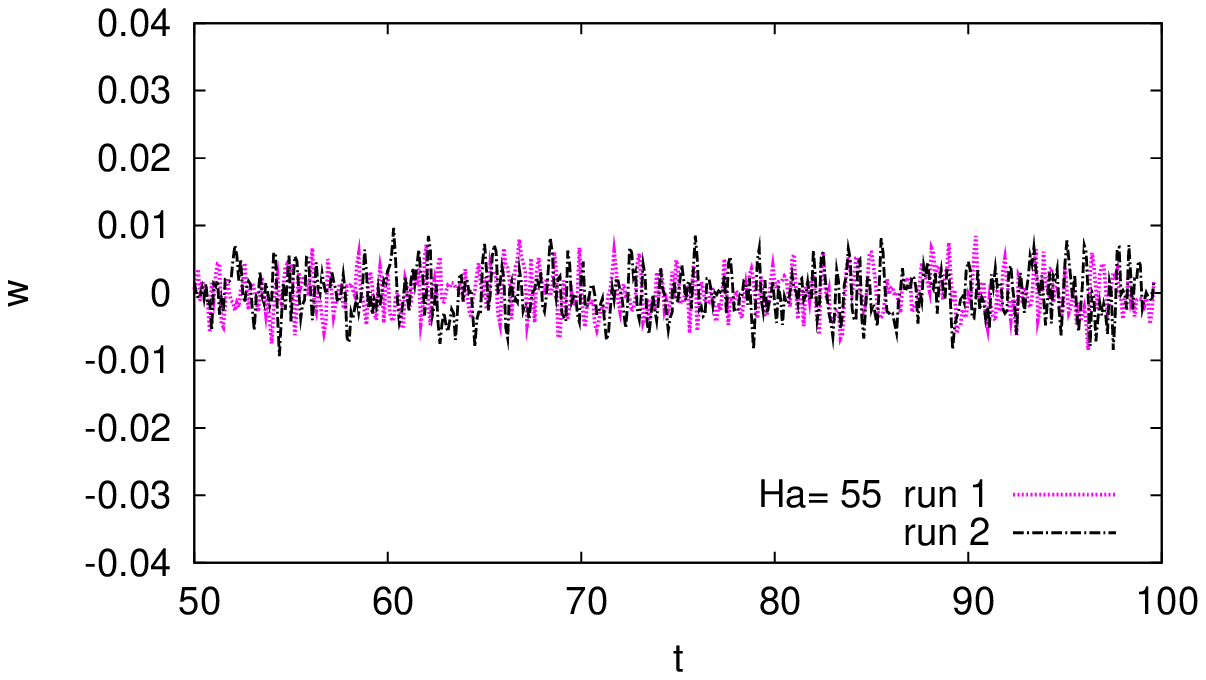}
\includegraphics[width=0.49\textwidth,clip=]{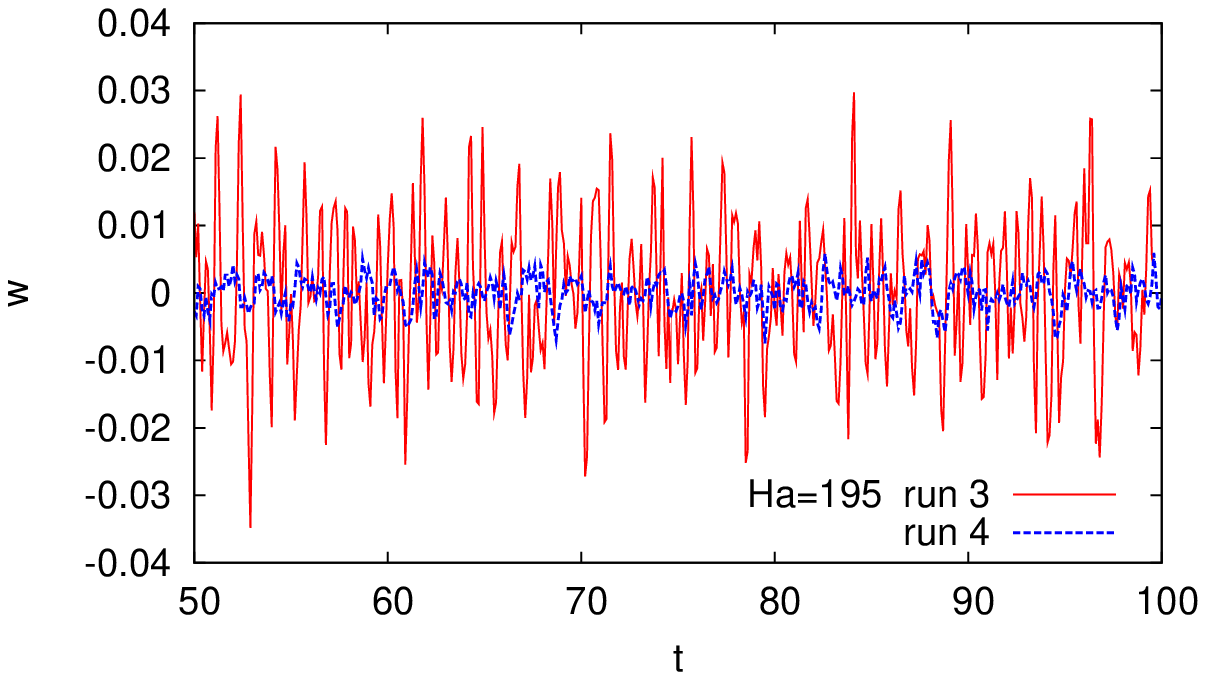}
}
\vskip-4mm
\caption{Time signals of velocity components computed at $x=43$, $z=0$ and $y=0$
shown for the second half of the fully developed flow stages of the simulations
1-4. 
Runs 1,2 at $\Ha=55$ 
and  runs 3,4 at $\Ha=195$ 
 are shown in, respectively, left and right columns.
From top to bottom: streamwise $u$, spanwise $v$ (transverse and parallel to
the main component of the magnetic field) and vertical $w$ (transverse and
perpendicular to the main component of the magnetic field) velocity components.
}
\label{fig:signals}
\end{figure}

\subsubsection{Flow structure}
\label{sec:structures}

\begin{figure}
\vskip3mm

\parbox{1.1\linewidth}{\hskip70mm Run 1}\parbox{0.0\linewidth}{\hskip-40mm Run 2}
\centerline{
\includegraphics[width=0.95\textwidth,clip=]{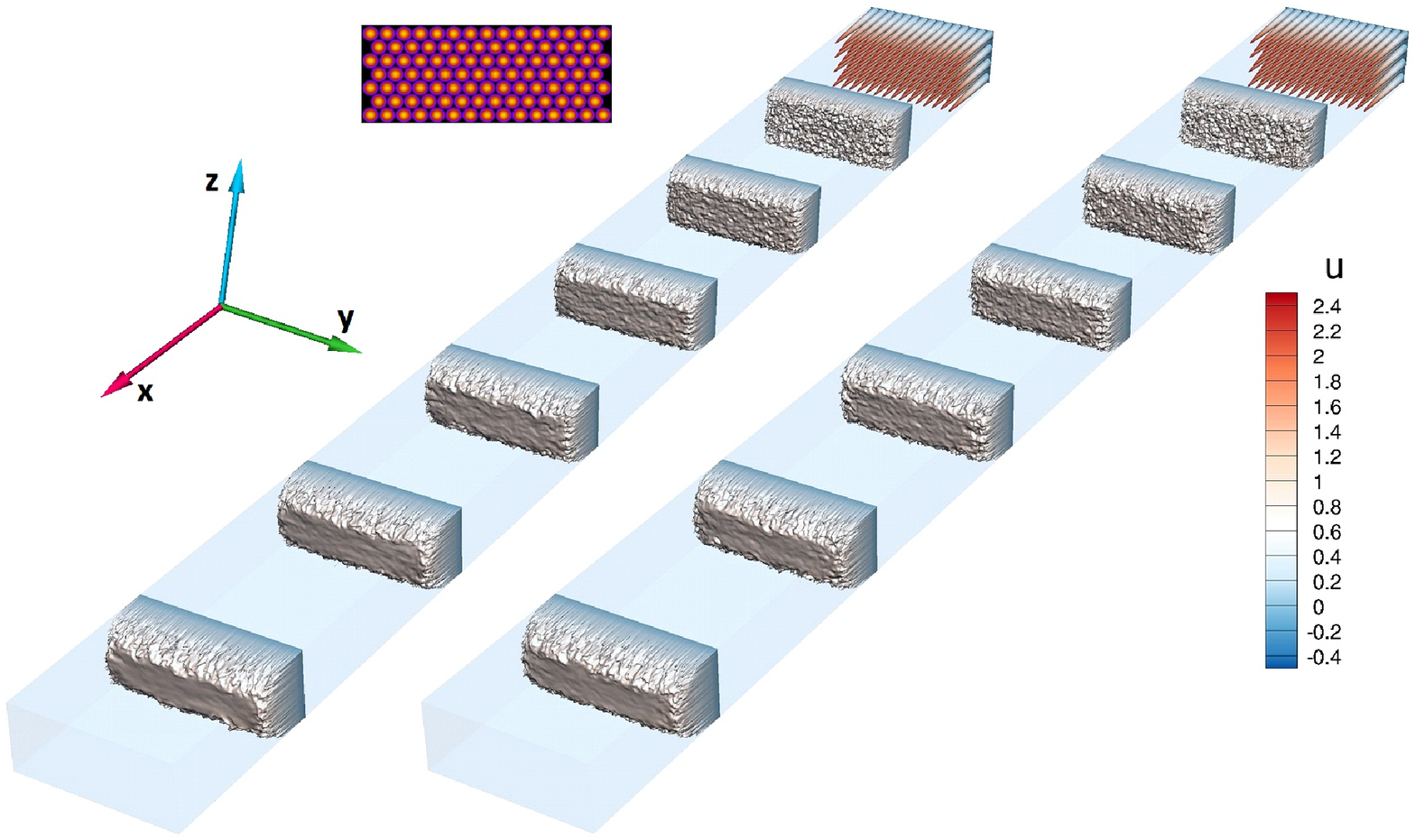}
}

\parbox{1.1\linewidth}{\hskip70mm Run 3}\parbox{0.0\linewidth}{\hskip-40mm Run 4}
\centerline{
\includegraphics[width=0.95\textwidth,clip=]{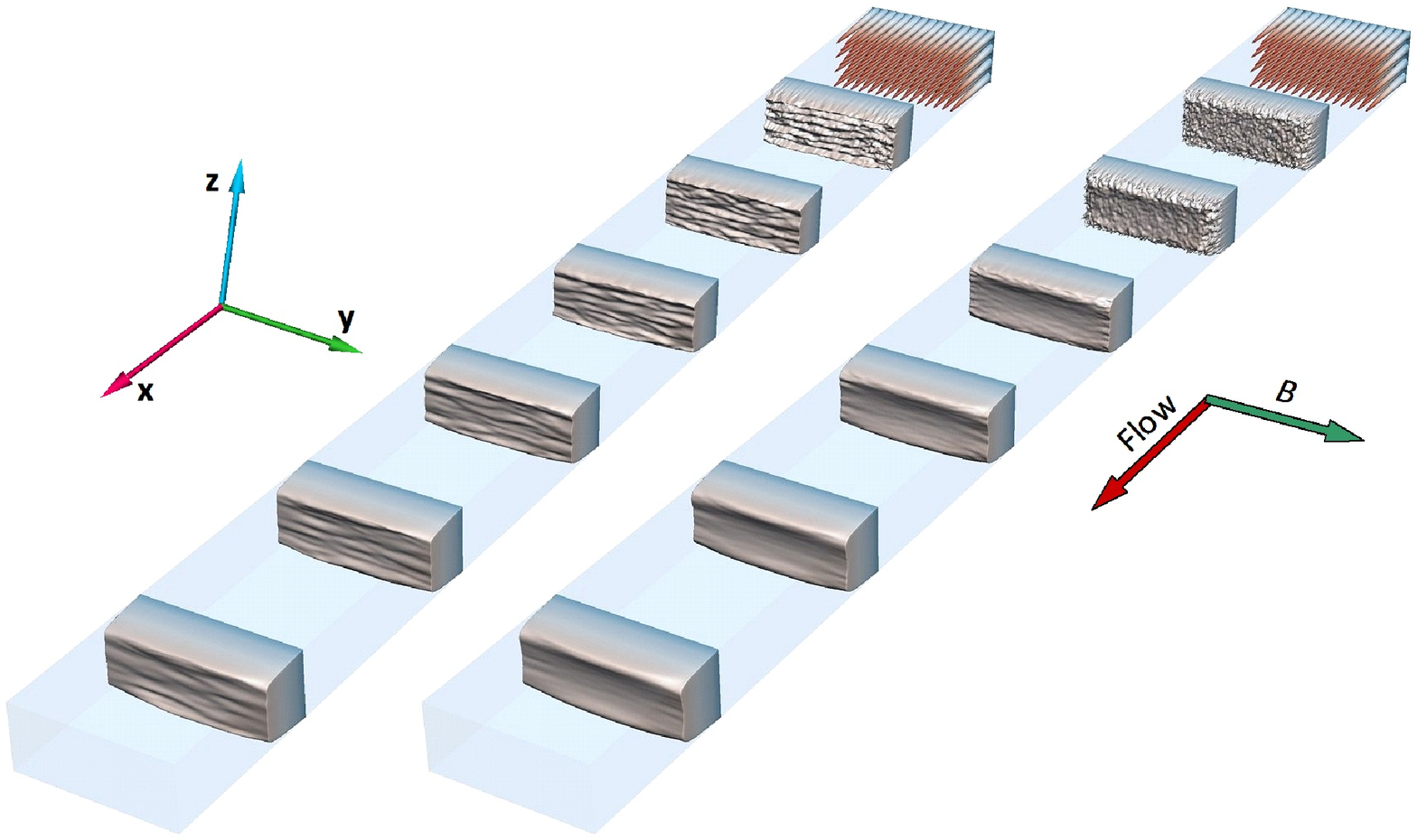}
}

\vskip-1mm
\caption{Instantaneous distributions of the streamwise velocity $u$ at several
locations along the duct shown for the fully developed flows in the simulations 1-4
(see table \ref{table:param} for the flow parameters).
}
\label{fig:prof}
\end{figure}

\begin{figure}
\parbox{1.1\linewidth}{\hskip70mm Run 3}\parbox{0.0\linewidth}{\hskip-31mm Run 4}\vskip1mm
\centerline{
\includegraphics[width=0.99\textwidth,bb=20 45 590 270,clip=]{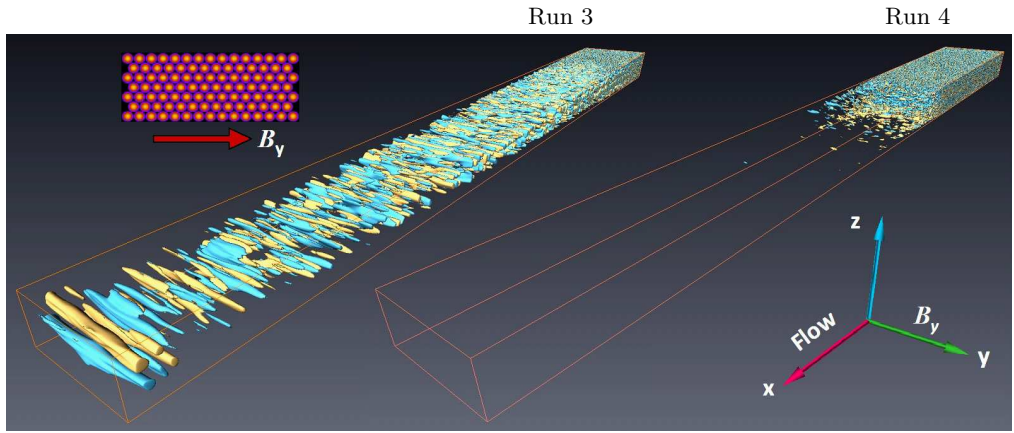}
}
\vskip-3mm
\caption{Isosurfaces of the vertical velocity component $w$ (transverse and perpendicular to
the main component of the magnetic field $B_{y}$) for the  runs 3 and 4.
Two iso-levels of the same magnitude and opposite signs (yellow -- positive, blue -- negative)
are visualized. The insert on the left shows the honeycomb pattern and the main component
of the magnetic field $B_{y}$.}
\label{fig:vz3d}
\end{figure}

\begin{figure}
\centering

\includegraphics[width=0.90\textwidth,bb=20 20 590 200,clip=]{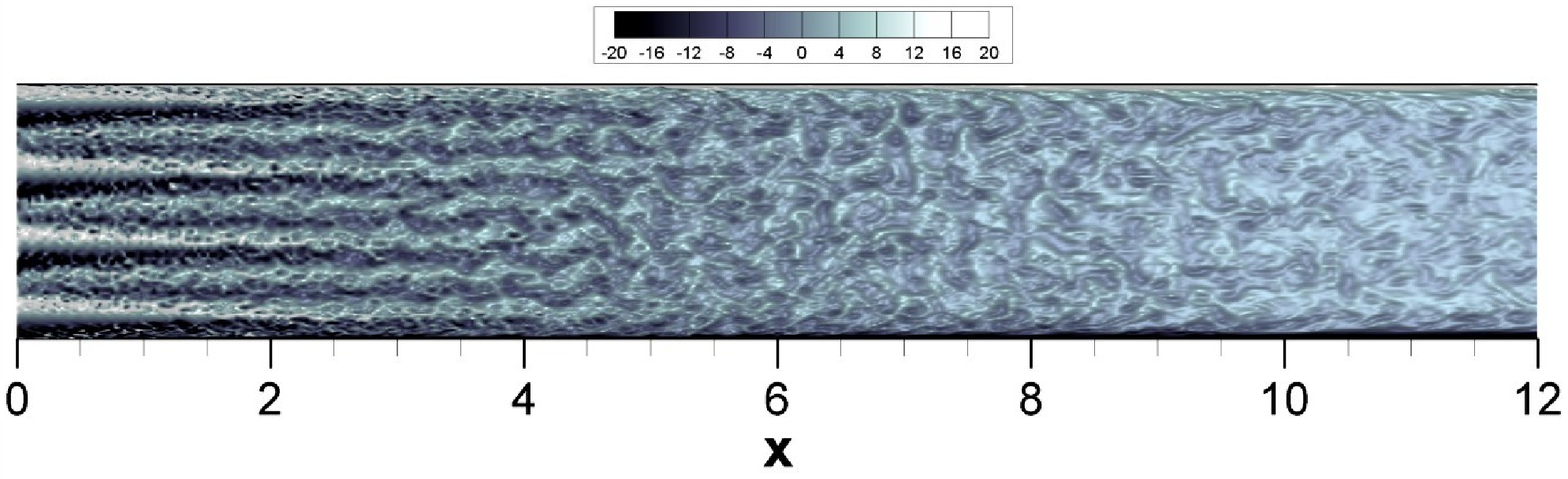}

\vskip-5mm
\caption{Instantaneous distribution of the vorticity component $\omega_{y}$ parallel to
the magnetic field in the $(x,z)$ cross-section through the duct's axis. 
The transformation of jets into vortices is shown for the run 3 by a close-up of 
the inlet region at $0 \le x \le 12$.}
\label{fig:omegaY-z}
\end{figure}

The spatial structures of the fully developed flows in the simulation runs 1-4 are illustrated in
figures \ref{fig:prof} and \ref{fig:vz3d}. We see that at $\Ha=55$ (runs 1 and 2 in figure \ref{fig:prof})
the flows remain turbulent, although the velocity fields are significantly modified by the magnetic fields.
The modifications include development of the mean flow profile with a nearly flat core and characteristic
Hartmann and sidewall boundary layers (see figure \ref{fig:prof}) and reduction of turbulence intensity.
Since the Reynolds number based on the Hartmann thickness $R\equiv \Rey/\Ha=505$, this result is in
agreement with the earlier studies of the flows in long ducts with uniform transverse magnetic field.
As discussed, for example, in the review by \citet{ZikanovASME:2014}, fully laminar and fully turbulent
flows are typically found at, respectively, $R<200$ and $R>400$, with the transitional range at $200<R<400$.
We also note that at $\Ha=55$ no substantial differences are observed between the case 1 and case 2
configurations except that the flow modification happens farther downstream in the run 2.

In the simulations 3 and 4 performed at $\Ha=195$, we have $R=143$, which is below the laminar-turbulent
transition range. Turbulence is, therefore, suppressed (albeit not completely, as we will see in the following
analysis) as the fluid moves through the magnetic field (see figure \ref{fig:prof}). The flows obtained for
the two configurations of the magnetic field are, however, clearly different.

In the case 2 configuration, there is a distance between the honeycomb and the beginning of the zone
of full-amplitude magnetic field. The plots for the run 4 in figures \ref{fig:prof} and \ref{fig:vz3d}
clearly show that the distance is sufficient for the instability and mixing of the jets generated by
the honeycomb. Three-dimensional turbulence develops. Upon entering the magnetic field, the turbulent
fluctuations are quickly suppressed, which is reflected by the strong reduction of the rms velocity
fluctuations at $x=43$ shown in table \ref{table:signals}.

In the case 1 configuration, the formation of turbulence near the honeycomb exit occurs in the presence
of a full-amplitude magnetic field. As shown in figures \ref{fig:prof} and \ref{fig:vz3d}, the velocity
field in the run 3 quickly becomes strongly anisotropic. The instability of the honeycomb jets does not
lead to a three-dimensional turbulent state, but to a quasi-two-dimensional flow dominated by structures
aligned with the magnetic field.

The illustrations in figures \ref{fig:prof} and \ref{fig:vz3d}, the distribution of the vorticity component
$\omega_y$ parallel to the magnetic field in the $(x,z)$ cross-section of the duct shown in figure \ref{fig:omegaY-z},
and the additional visualizations analyzed in the course of our work (not shown) suggest the following scenario
of the evolution of the spatial structure of the flow. In the inlet portion of the duct, approximately at $x < 3$,
the dominant feature of the evolution is the transformation of the round jets exiting the honeycomb into
quasi-two-dimensional planar (nearly parallel to the $(x,y)$ plane) jets. Already in the course of this
transformation, the jets experience the Kelvin-Helmholtz instability that leads to noticeable waviness
at $x$ between $3$ and $4$ and to roll-up into quasi-two-dimensional vortices at around $x = 5$. The following
evolution is characterized by quasi-two-dimensional vortices superimposed on the plug-like profile of
the streamwise velocity. It is indicated by figures \ref{fig:prof}-\ref{fig:omegaY-z} and confirmed by
the quantitative analysis presented later in this paper that the dynamics of the vortices is that of
quasi-two-dimensional turbulence.

The last preceding paragraph summarizes our key observation. It provides the basis for the explanation suggested
earlier for the anomalously strong velocity fluctuations observed in the experiments of \citet{Sukoriansky:1986}
and, likely, other experiments such as those of \cite{Kljukin:1989}. Due to their weak gradients along
the magnetic field lines, the quasi-two-dimensional vortices do not generate strong Joule dissipation.
Furthermore, the quasi-two-dimensionality reduces the energy flux from large to small length scales,
which implies weaker viscous dissipation. The flow structures are still suppressed by the Joule and
viscous dissipation in the boundary layers, but the effect is not strong. The quasi-two-dimensional
vortices are visible till the end of the flow domain (see figures \ref{fig:prof} and \ref{fig:vz3d}),
and are responsible for the generation of high-amplitude velocity fluctuations at far downstream locations.

\subsubsection{Turbulence decay along the duct}\label{sec:decay}
The distributions of the turbulent kinetic energy in each velocity component $\langle {u^{\prime}}^2\rangle$, $\langle {v}^2\rangle$, $\langle {w}^2\rangle$  are computed as functions of $x$ along the lines $y=z=0$ and $y=-1.4$, $z=0$. 

The turbulence decay curves obtained at $y=z=0$ are shown in figures \ref{fig:decay} and \ref{fig:decay2}. The intervals $0\le x< 0.1$ in figure \ref{fig:decay} and  $0\le x < 1$  in figure \ref{fig:decay2} are excluded to highlight the decay stage of the flow evolution and to eliminate the initial stage of jet instability and mixing, at which the data are strongly influenced by the position of the point $y=z=0$ with respect to the honeycomb pattern. The slope lines are plotted to illustrate the decay rate rather than to suggest a specific scaling.

\begin{figure}
\centerline{
\includegraphics[width=0.7\textwidth,clip=]{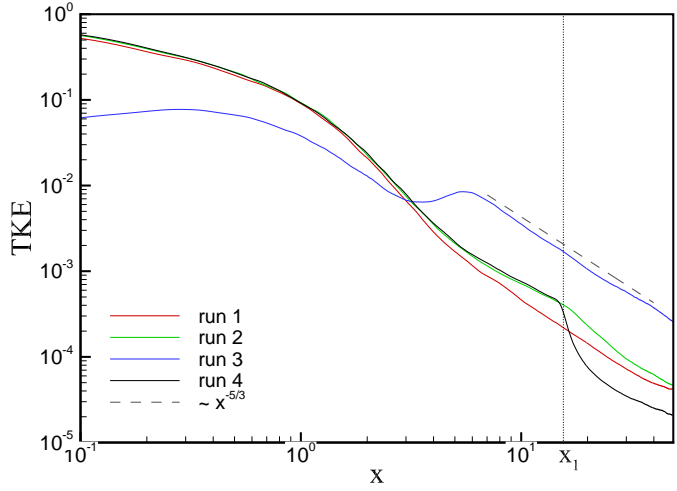}
}
\caption{Time-averaged turbulent kinetic energy $\langle {u^{\prime}}^2+{v^{\prime}}^2+{w^{\prime}}^2\rangle$
as a function of $x$ along the centerline of the duct $y=z=0$. The vertical dotted line indicates the location
of the corners of the magnet pole-pieces in the runs 2 and 4. The slope line $\sim x^{-5/3}$ is shown for comparison.}
\label{fig:decay}
\end{figure}

\begin{figure}
\vskip3mm
\parbox{0.5\linewidth}{(a)}\parbox{0.3\linewidth}{(b)}
\centerline{
\includegraphics[width=0.52\textwidth,clip=]{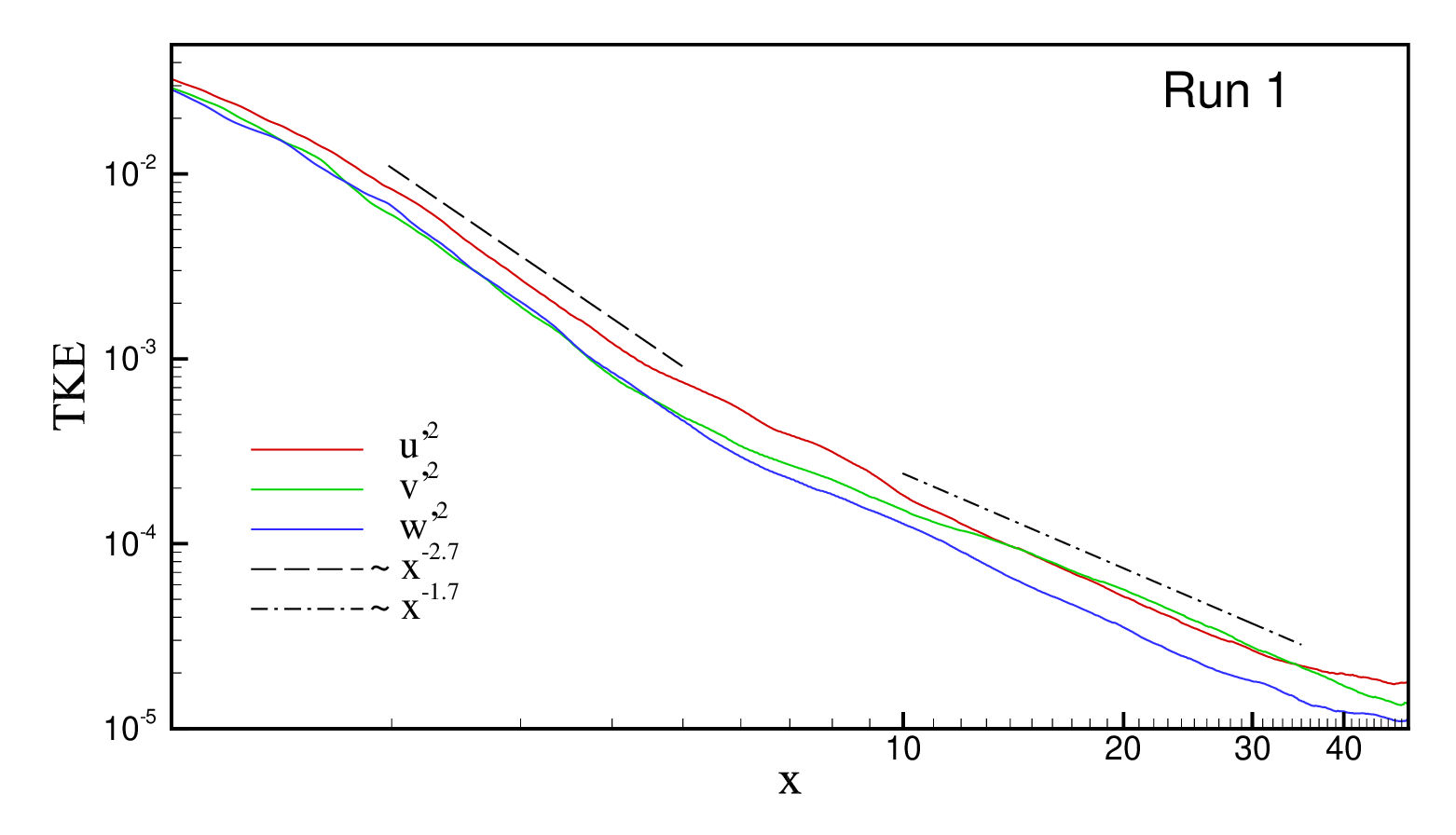}
\includegraphics[width=0.52\textwidth,clip=]{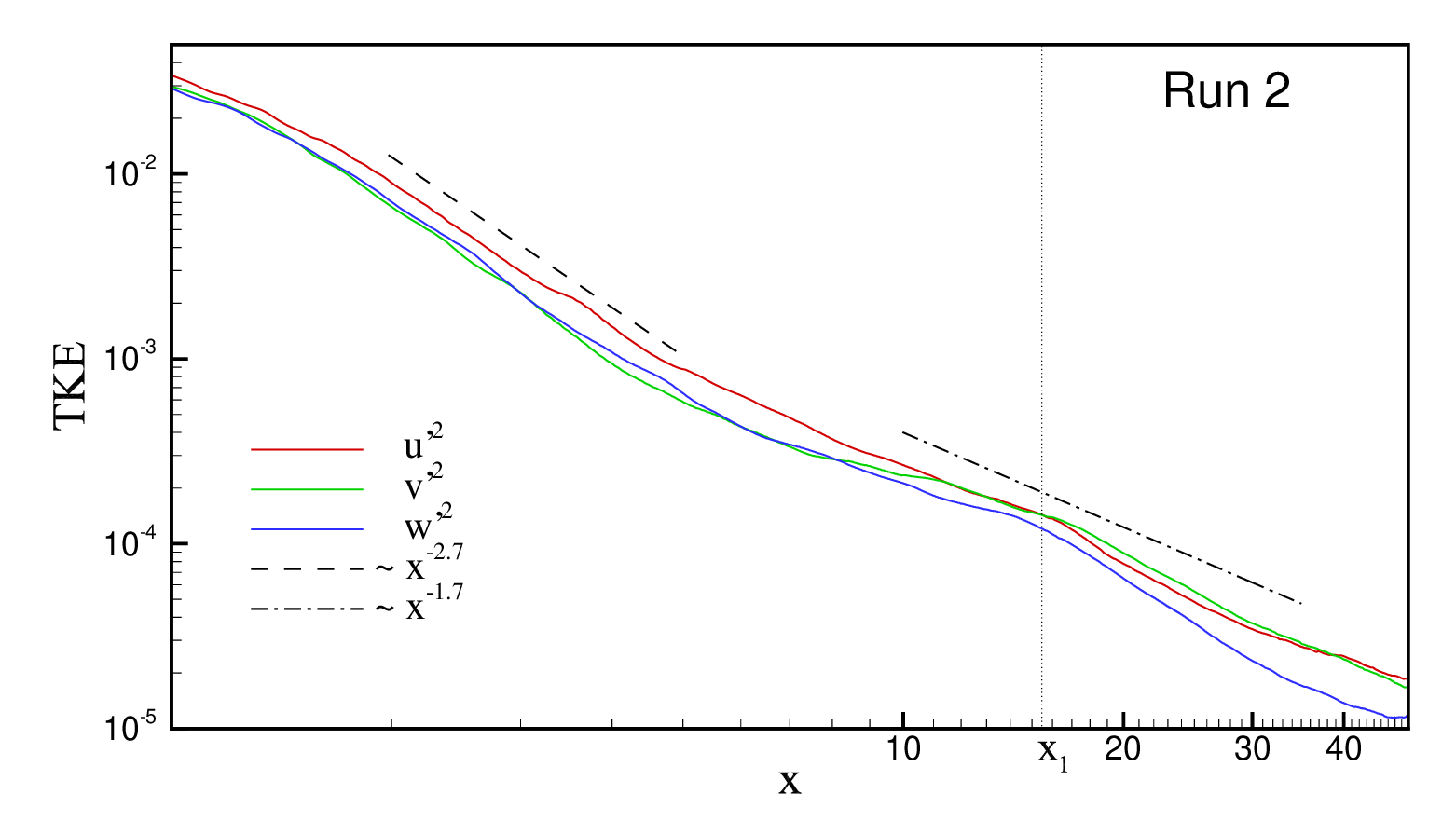}
}

\vskip-4mm
\parbox{0.5\linewidth}{(c)}\parbox{0.3\linewidth}{(d)}
\centerline{
\includegraphics[width=0.52\textwidth,clip=]{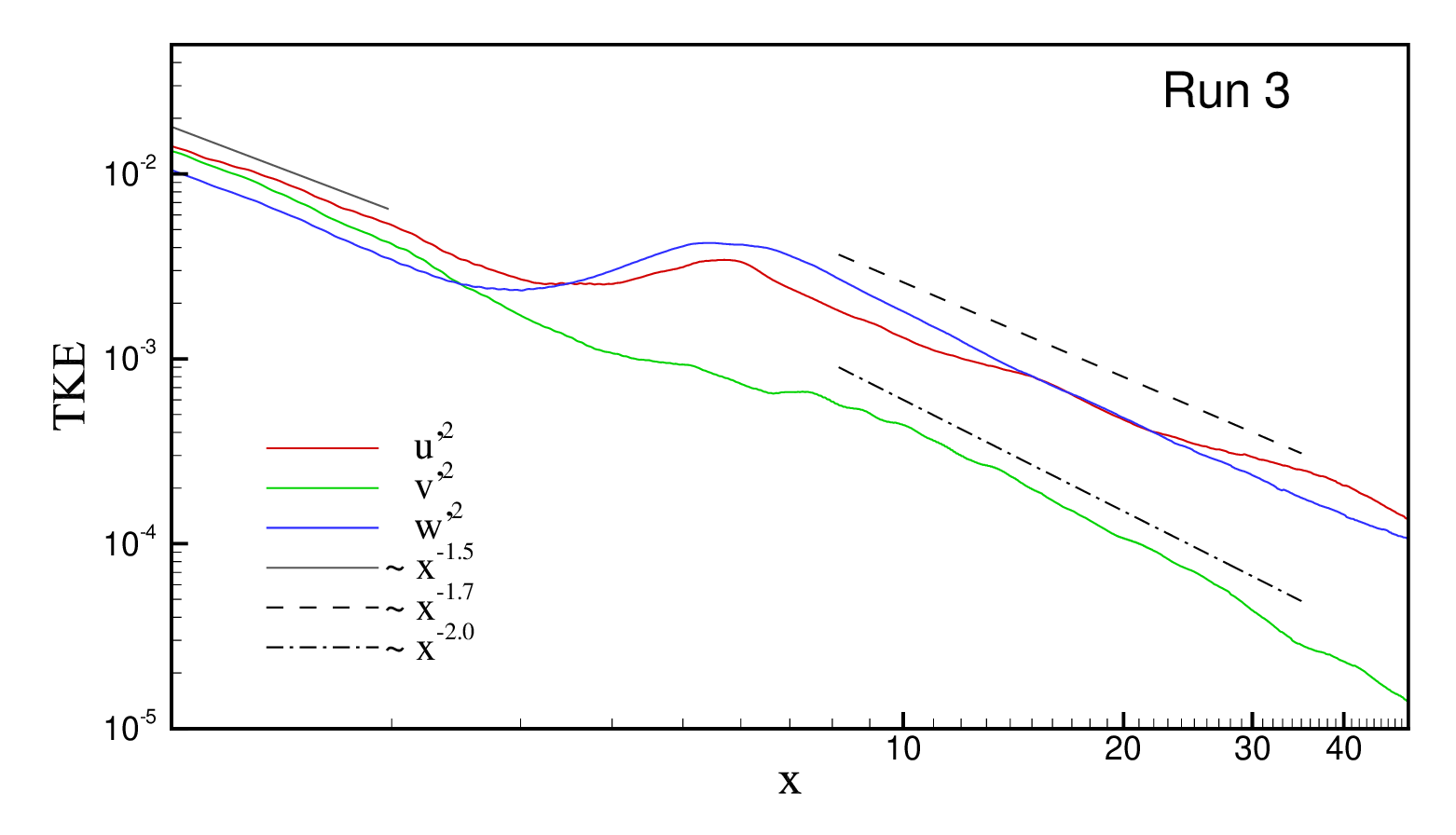}
\includegraphics[width=0.52\textwidth,clip=]{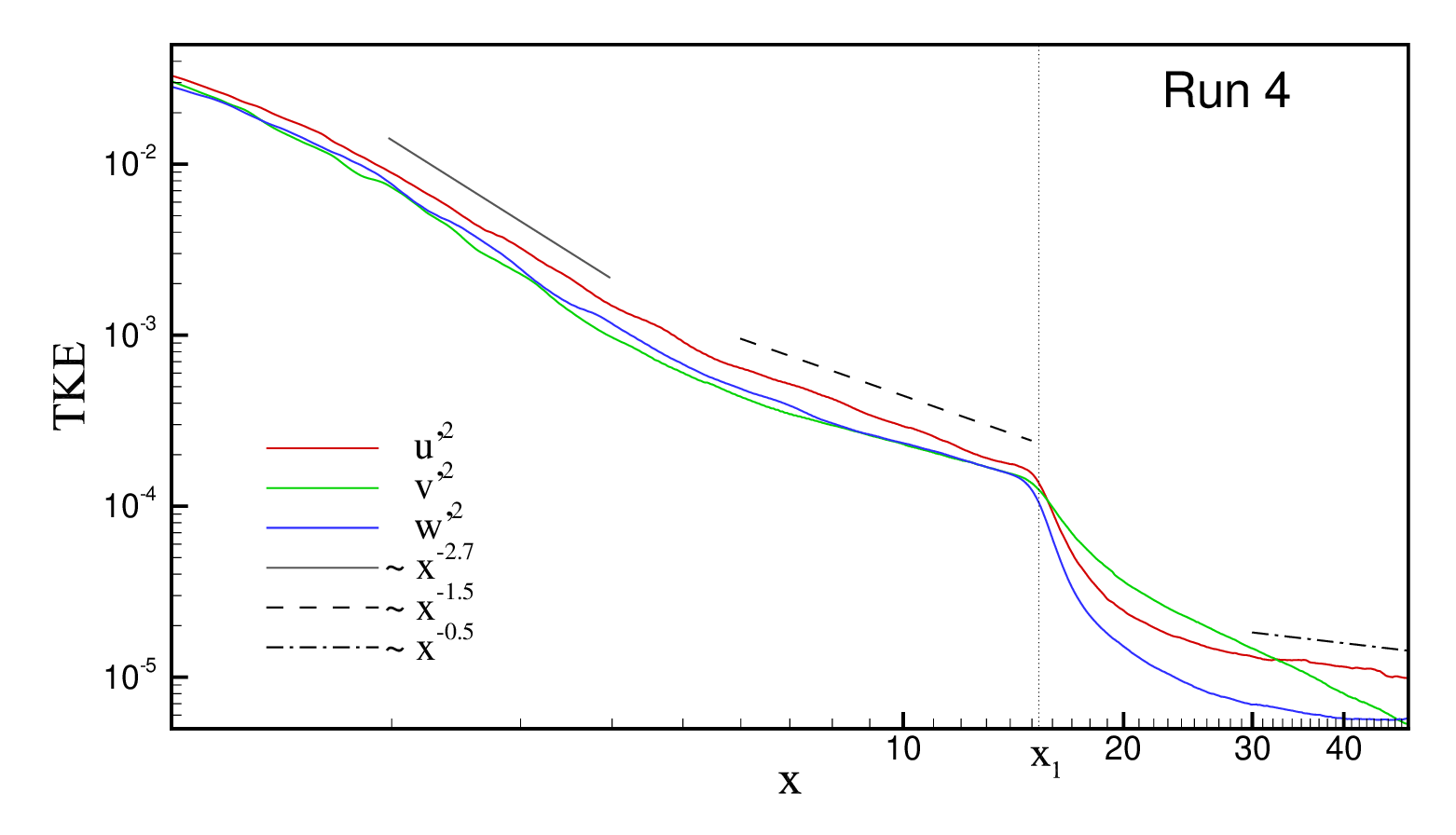}
}
\vskip-3mm
\caption{Time-averaged turbulent kinetic energies in separated velocity components
$\langle {u^{\prime}}^2 \rangle$, $\langle {v^{\prime}}^2 \rangle$, $\langle {w^{\prime}}^2 \rangle$
as functions of $x$ along the centerline of the duct $y=z=0$. The inlet section
of the duct $0\le x\le 1$ is excluded. Slope lines are shown for comparison.
The vertical dotted line indicates the location of the corners of the magnet
pole-pieces in the runs 2 and 4. }
\label{fig:decay2}
\end{figure}

For the runs 1 and 2, the energy decay curves obtained at two locations of the magnet are not
very different from each other. This suggests weak influence of the magnetic field in agreement
with the low magnetic interaction parameter $\N=0.1088$. For small $x$, the magnetic damping causes
somewhat more rapid decay in the run 1 than in the run 2. At larger $x$, approximately at $x > x_1$
where the strength of the magnetic field is about the same in the two flows, turbulence decays faster
for the run 2. We attribute that to the stronger Joule dissipation caused by the stronger velocity
gradients in the field direction retained by the flow. At the end of the duct, the turbulent kinetic
energy in the two flows decreases to approximately the same level.

The curves in figure \ref{fig:decay2}a,b show significant level of fluctuations in all three
velocity components. This is in agreement with the three-dimensional fully turbulent nature
of the flow visualized in figure \ref{fig:prof}. At the same time, the Reynolds stress tensor
is not isotropic. At small $x$, $\langle {u^{\prime}}^2\rangle > \langle {v^{\prime}}^2\rangle\sim\langle {w^{\prime}}^2\rangle$.
At larger $x$, approximately at $x>12$ in the flow 1 and $x>20$ in the flow 2, we see significant
anisotropy with $\langle {u^{\prime}}^2\rangle \sim \langle {v^{\prime}}^2\rangle > \langle {w^{\prime}}^2\rangle$.

The effect of the magnetic field is much more pronounced in the flows 3 and 4. For the run 4,
the energy decay curves is practically indistinguishable from those for the run 2 curve for
$x < x_1$ (see figure \ref{fig:decay}). For $x > x_1$, the strong imposed magnetic field 
results in rapid decay and the lowest value of the turbulent kinetic energy at the duct exit
among all the simulations 1-4. Interestingly, during the initial stages of this decay,
in the interval $15 < x < 30$, the fluctuations of the velocity component $v$ parallel to
the magnetic field remain stronger than the fluctuations of the other two components
(see figure \ref{fig:decay2}d). We do not have data that would allow us to precisely identify
the specific flow structures responsible for this effect. We note that the behaviour is consistent
with the evolution of homogeneous, initially isotropic turbulence after sudden application of
a strong magnetic field. As predicted by \citet{Moffatt:1967} and confirmed by \citet{Burattini:2010}
and \citet{Favier:2010}, the initial stages of the decay are characterized by the energy of field-parallel
velocity fluctuation component substantially larger (two times larger in the asymptotic limit $\N\gg 1$)
than the energy of the field-perpendicular components. Far downstream, approximately for $x > 30$,
the remaining fluctuations $u^{\prime}$ and $w^{\prime}$ decay very slowly, with the rate approaching
$\langle {u^{\prime}}^2\rangle \sim \langle {w^{\prime}}^2\rangle \sim x^{-0.5}$.

For the most interesting simulation 3, figures \ref{fig:decay} and \ref{fig:decay2} show a very strong
effect of the magnetic field. In the entrance portion of the duct, the generation of turbulence is
inhibited and the turbulent kinetic energy is an order of magnitude smaller than in the other three cases.
The energy grows slightly for $x < 0.3$ and then decays, but much slower than in the other cases.
The energy becomes larger than in the other flows at $x\approx 3$.

Interesting behaviour is observed in the interval $3 < x < 6$. While the fluctuation energy
$\langle {v^{\prime}}^2\rangle$ of the field-parallel velocity component continues to decay along the duct,
the fluctuation energies of the other two components grow. This behaviour manifests substantial energy
transfer from the mean flow to the fluctuations. The visualizations of the flow structure in figures
\ref{fig:prof}-\ref{fig:omegaY-z} allow us to attribute it to the Kelvin-Helmholtz instability of
the quasi-two-dimensional planar jets, which develop quite rapidly at already $x \approx 3$,
and the resulting formation of quasi-two-dimensional vortices.

The turbulence decay at $x > 6$ is characterized by
$\langle {u^{\prime}}^2\rangle \sim \langle {w^{\prime}}^2\rangle \gg \langle {v^{\prime}}^2\rangle$
(see figure  \ref{fig:decay2}c), which is expected for quasi-two-dimensional vortical structures extending
wall-to-wall in the field direction. The energy remains much larger than in the other three flows.
For $x > 8$, the decay is well approximated by the power law $\sim x^{-5/3}$ (see figure \ref{fig:decay}).
It should be stressed that we do not have theoretical arguments supporting this decay rate.
The same is true for the decay rates indicated by the slope lines in figure \ref{fig:decay2}.
The lines are shown purely for comparison, as illustrations of the decay trends obtained in the simulations.

\subsubsection{Turbulence statistics}
\label{sec:stat}
The velocity fields computed in the runs 1-4 for fully developed flows at $100 < t < 200$
are used to accumulate the turbulence statistics discussed in this section. 
Energy power spectra are calculated from the velocity
fluctuation signals at $x=43$, $y=z=0$ (see figure \ref{fig:signals}).
To comply with the periodicity condition, we have used a window function $w(\tau)$, based on
a superposition of two hyperbolic tangents $w(\tau) = \tanh(a \tau^3) + \tanh(a (T_m - \tau)^3) - 1$
with $a = 0.03$. Here it is assumed that the argument $\tau$ varies from $0$ to the maximum $T_m = 100$.
This function provides smooth transition from zero to unity at both ends and retains more than 90\%
of the unmodified sequence.

A possible alternative to this approach would be to compute the spatial wavenumber spectra
in the cross-section $x=const$. For that, we would have to use the data recorded in the core
(excluding the boundary layers) portion of the cross-section. The data would have to be interpolated
to a uniform grid and time-averaged. We see our approach as preferable for the following several reasons.
It is free from the errors associated with the interpolation and the variation of flow properties
in the cross-section. The spectra based on the time signal directly correspond to the measurements
made in the experiment. Finally, one-dimensional spectra are more informative in the case of strongly
anisotropic turbulence than three-dimensional or two-dimensional ones.

The spectra are shown in figure \ref{fig:spectra}.
We see that even at $\Ha=195$ the spectra are continuously populated in a wide range of frequencies $\omega$,
so the flows can be classified as turbulent. The inertial ranges cannot be reliably determined due to
their shortness typical for turbulence decay in the presence of MHD suppression. Still, one sees portions
of the spectra with the slope close to $\sim \omega^{-5/3}$ at $\Ha=55$ and $\sim \omega^{-3}$ at $\Ha=195$.
The latter can be viewed as an indication of the quasi-two-dimensional character of the turbulence,
although, as argued by \citet{Alemany:1979} and \citet{Sommeria:Moreau:1982}, the same spectrum may
appear as a result of the equilibrium between the local angular energy transfer and the Joule dissipation
in the core flow or the Hartmann boundary layers.

\begin{figure}

\vskip3mm
\parbox{0.5\linewidth}{(a)}\parbox{0.3\linewidth}{(b)}\vskip-1mm
\centerline{
\includegraphics[width=0.49\textwidth,clip=]{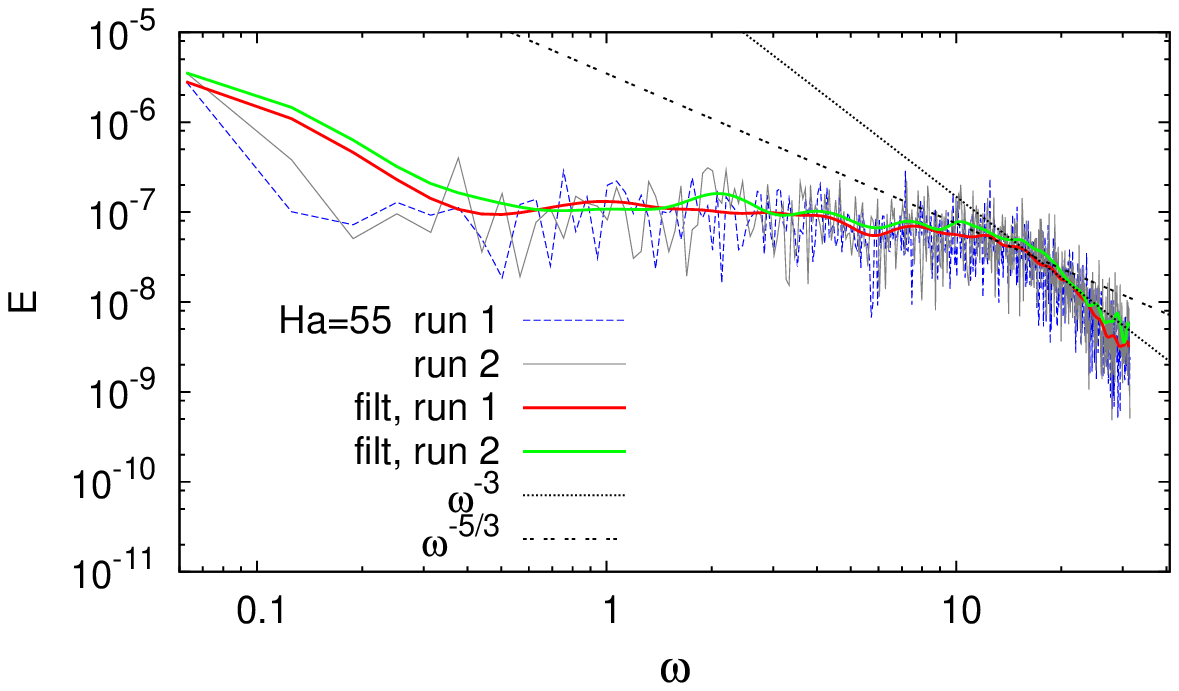}
\includegraphics[width=0.49\textwidth,clip=]{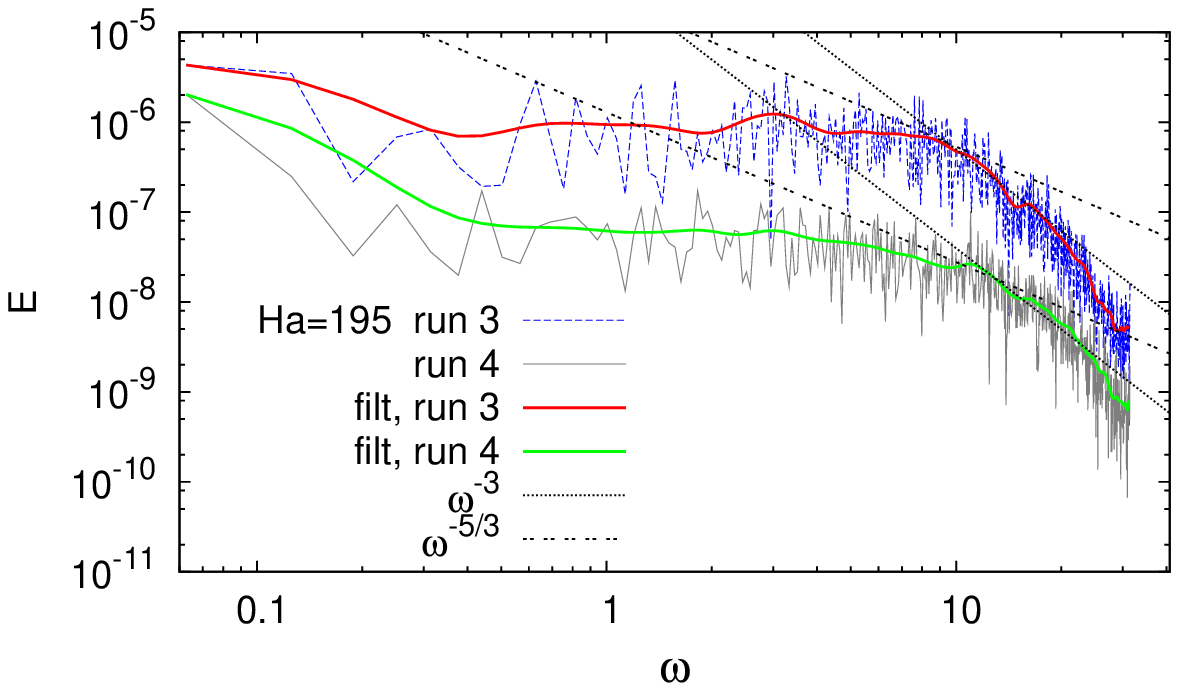}
}

\vskip-4mm
\parbox{0.5\linewidth}{(c)}\parbox{0.3\linewidth}{(d)}\vskip-1mm
\centerline{
\includegraphics[width=0.49\textwidth,clip=]{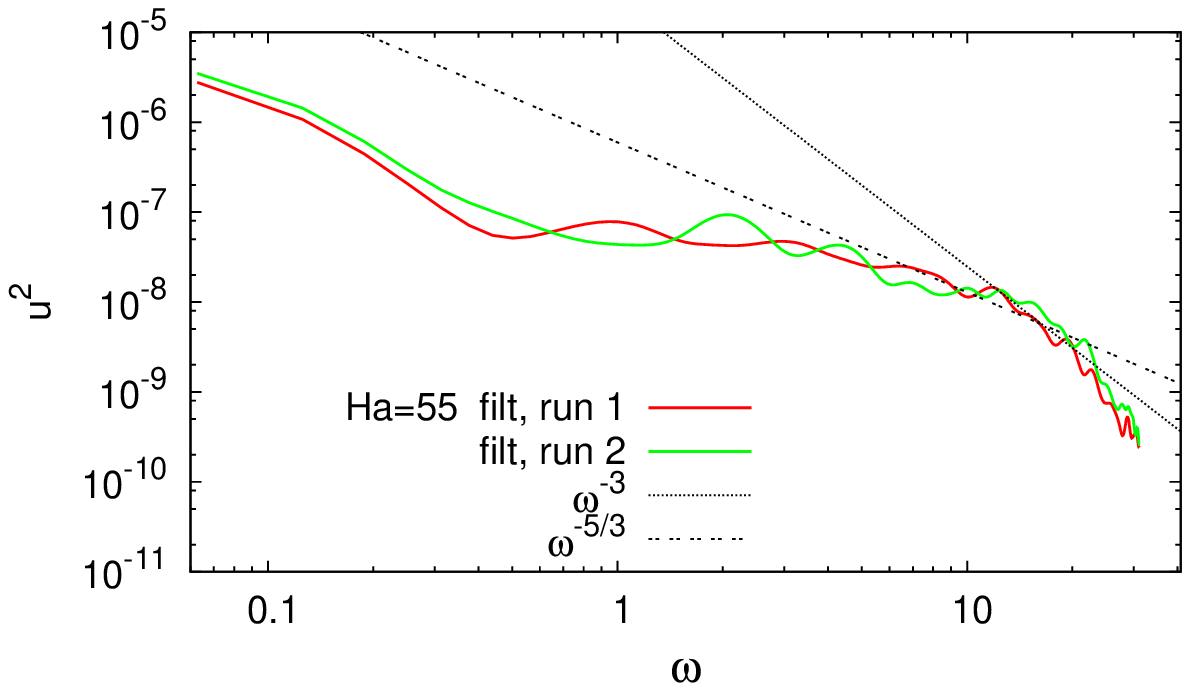}
\includegraphics[width=0.49\textwidth,clip=]{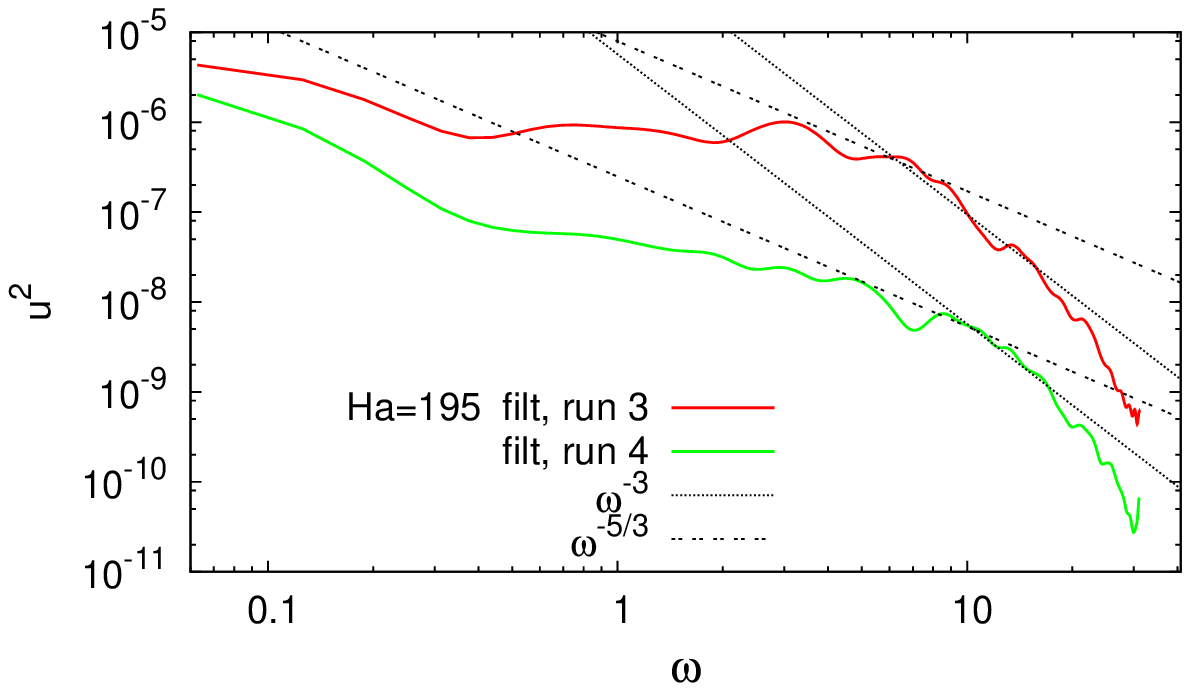}
}

\vskip-4mm
\parbox{0.5\linewidth}{(e)}\parbox{0.3\linewidth}{(f)}\vskip-1mm
\centerline{
\includegraphics[width=0.49\textwidth,clip=]{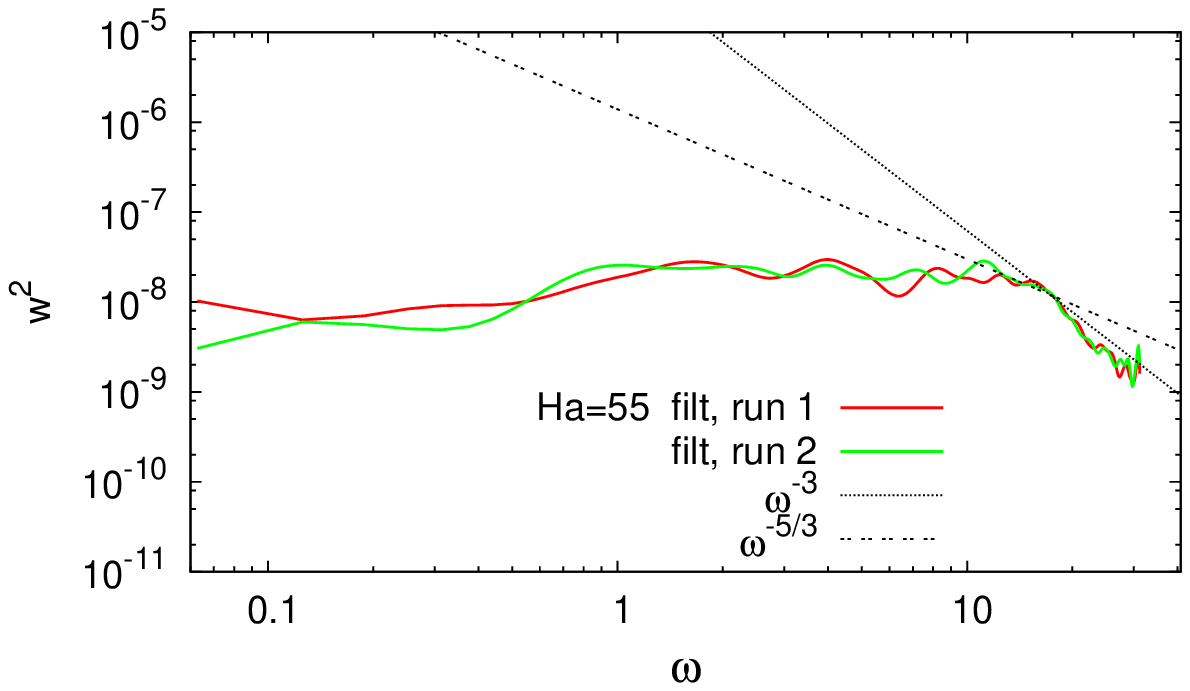}
\includegraphics[width=0.49\textwidth,clip=]{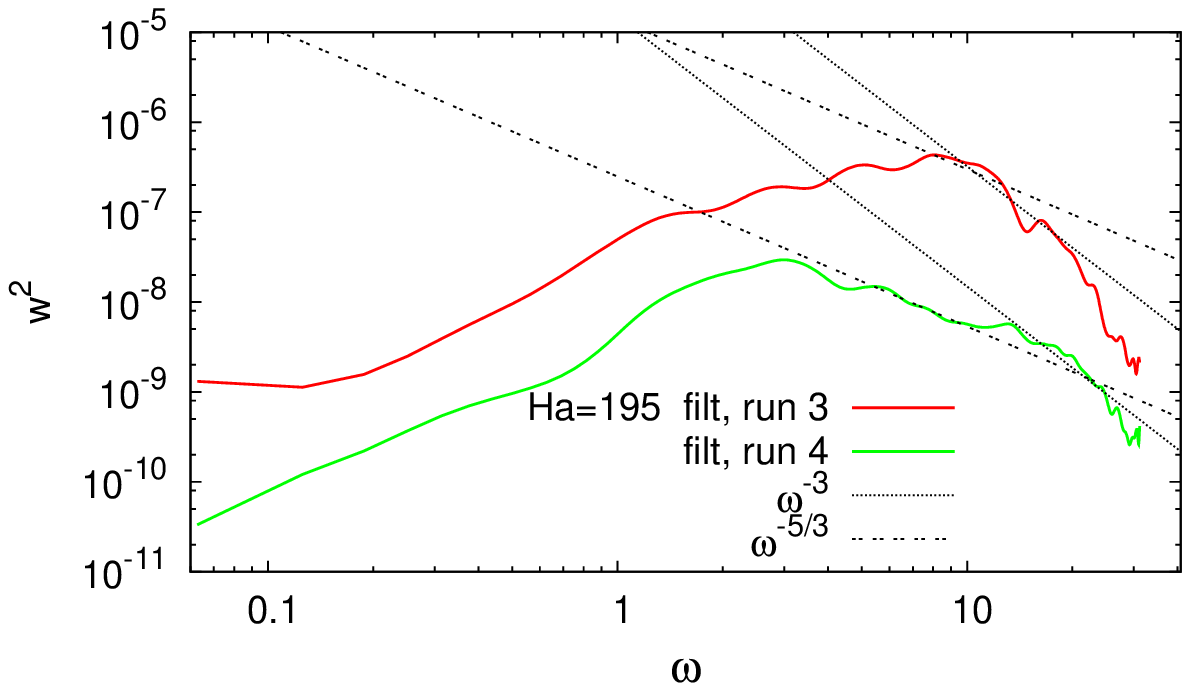}
}
\vskip-4mm
\caption{Power spectra of the kinetic energy based on the velocity signals computed
at $x=43$, $y=z=0$ in the fully developed flows at $\Ha=55$ (runs 1,2 shown in the left column)
and $\Ha=195$ (runs 3 and 4 shown in the right column). The spectra of the total kinetic energy
$E = u^2 + v^2 + w^2$ and the energy in two velocity components $u$ and $w$ are shown.
For the sake of clarity, the filtered spectra (using Bezier spline) are shown, the original
raw data are only demonstrated on plots (a,b). Also shown, for the sake of comparison,
are the power laws $\sim \omega^{-3}$ and $\sim\omega^{-5/3}$. The spectra of the energy
in the velocity component $v$ (not shown) demonstrate practically no difference between
the four flows.}
\label{fig:spectra}
\end{figure}

The spectrum of $w^2$ is particularly convenient for characterization of the anomalous
high-amplitude turbulent fluctuations observed in the flow 3 (see figure \ref{fig:spectra}f).
The energy peak at $\omega\approx 10$ is evidently associated with the characteristic
streamwise size of the vortices (see figure \ref{fig:vz3d}).

\begin{figure}

\parbox{0.5\linewidth}{(a)}\parbox{0.3\linewidth}{(b)}
\centerline{
\includegraphics[width=0.45\textwidth,clip=]{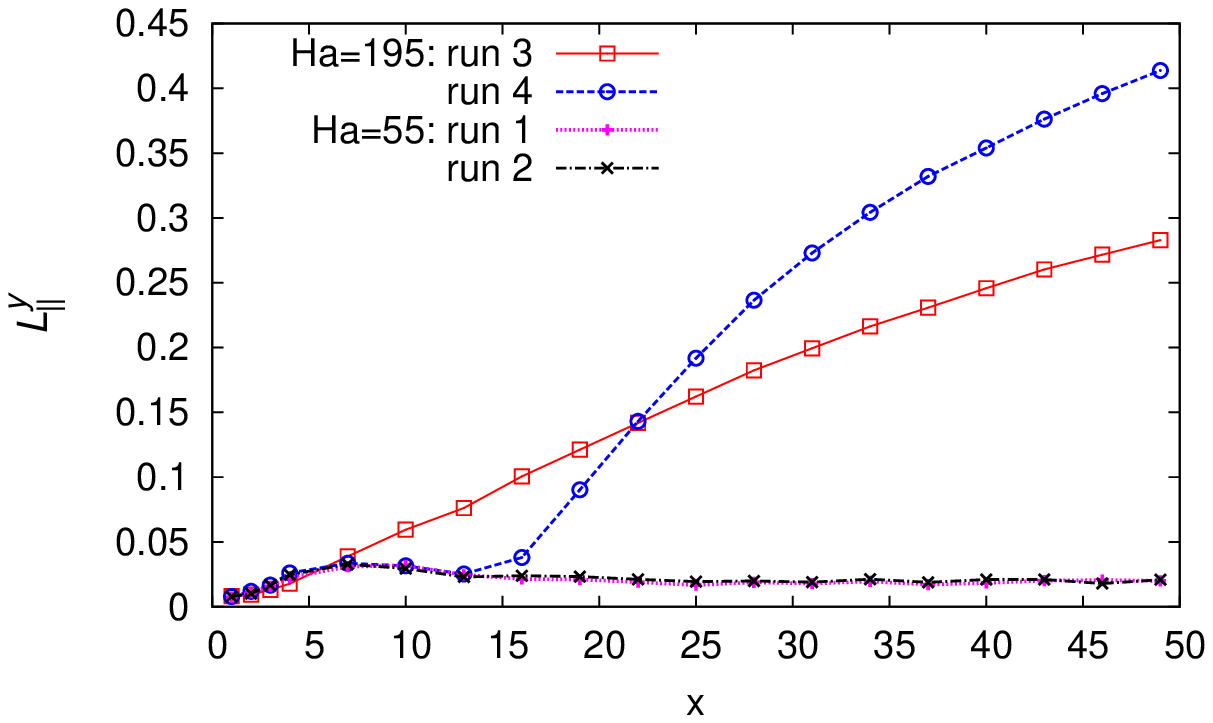}
\includegraphics[width=0.45\textwidth,clip=]{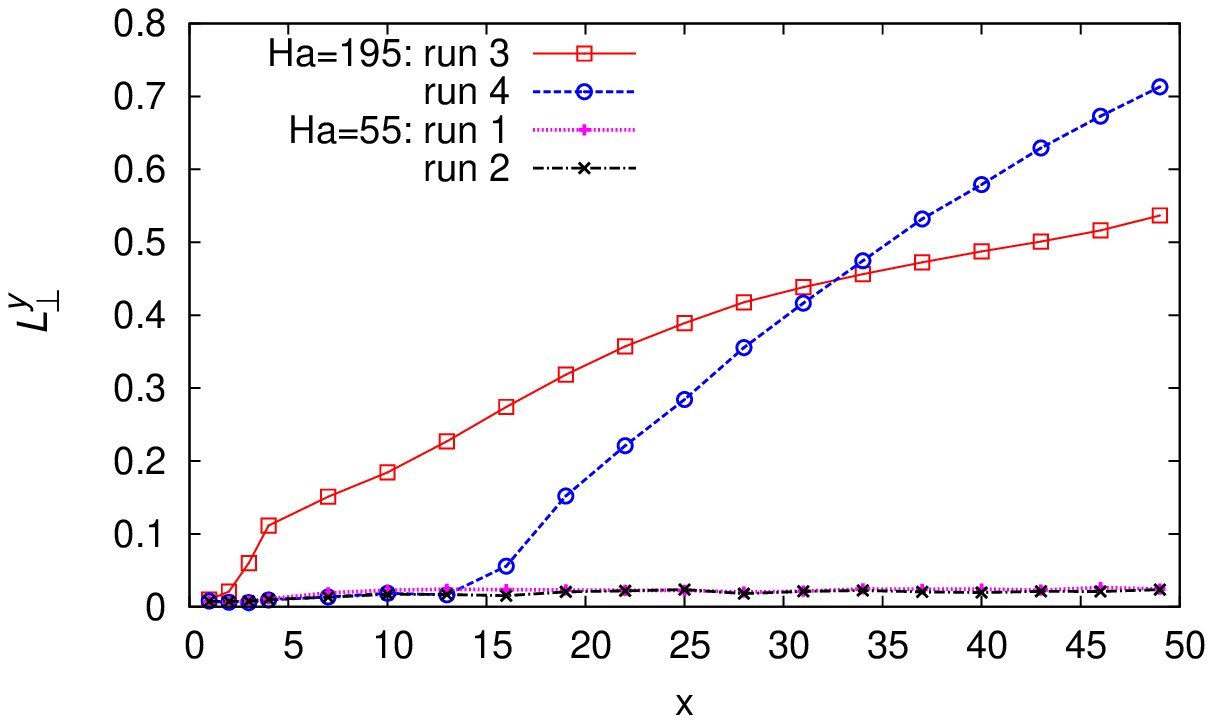}
}

\parbox{0.5\linewidth}{(c)}\parbox{0.3\linewidth}{(d)}
\centerline{
\includegraphics[width=0.45\textwidth,clip=]{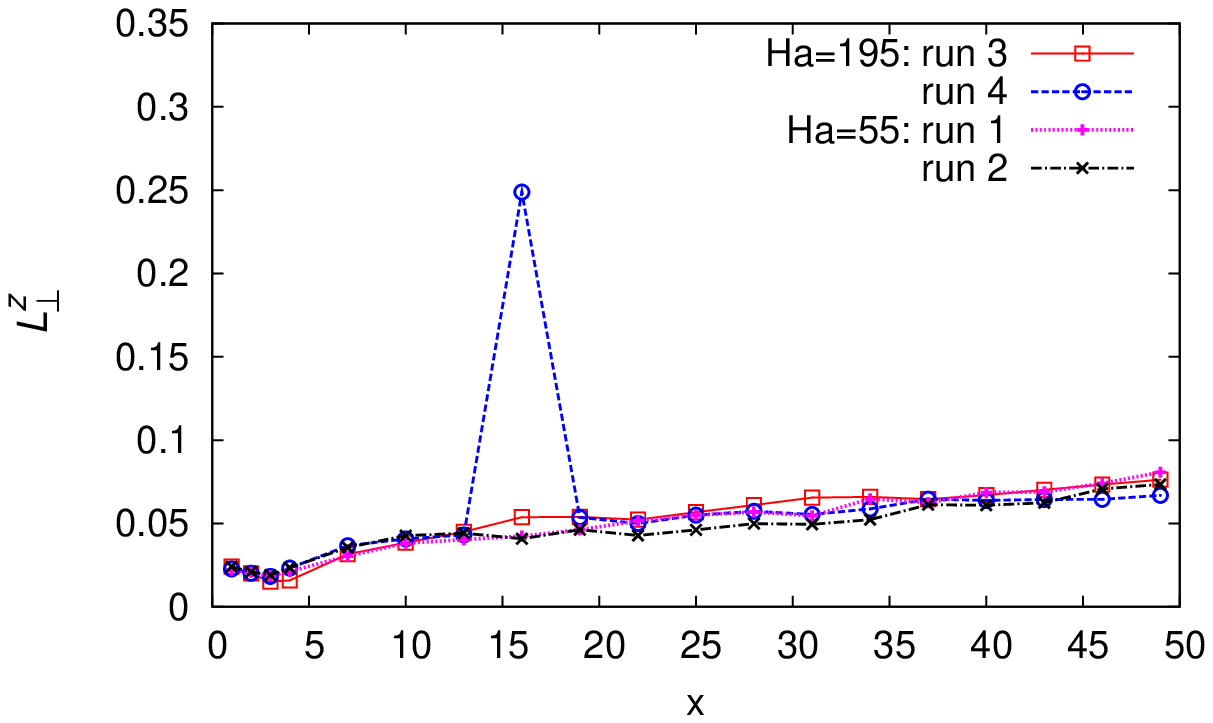}
\includegraphics[width=0.45\textwidth,clip=]{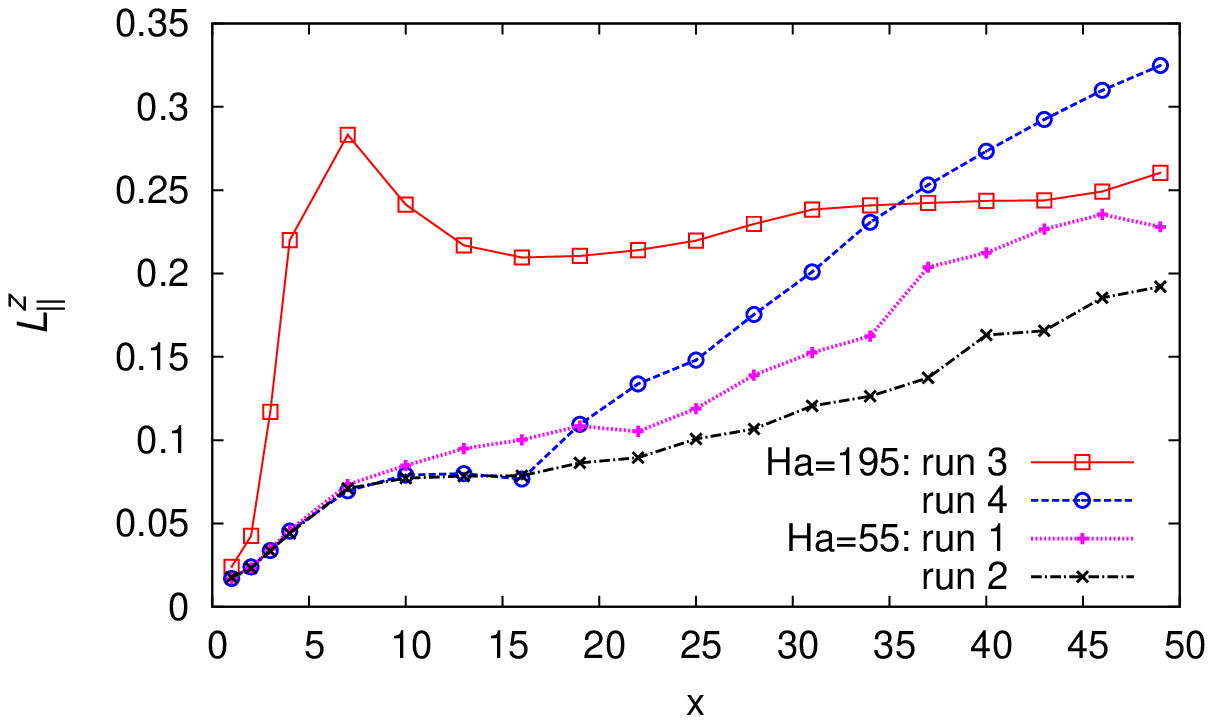}
}

\vskip-4mm
\caption{Integral length scales based on the correlation data obtained in the runs 1-4:
$(a,b)$ parallel to the magnetic field, $(c,d)$ perpendicular to the field.
The scales $L_{\parallel}^y$, $L_{\perp}^y$ and $L_{\parallel}^z$, $L_{\perp}^z$
are shown as functions of the streamwise coordinate $x$. The nature of the peak at $x=16$ in figure $(c)$ is explained in appendix \ref{appA}.}
\label{fig:len_scale}
\end{figure}

We have also evaluated  two-point velocity correlation functions  along the direction parallel ($y$) and perpendicular ($z$) to the magnetic field.
The coefficients are defined as (here for the velocity component $w$)
\begin{eqnarray}
\label{corr1}
R^w(\ell_y) & = & \frac{\int_{-L_z+\delta_z}^{L_z-\delta_z} w(x^*,0,z)w(x^*,\ell_y,z)dz+\int_{-L_z+\delta_z}^{L_z-\delta_z} w(x^*,0,z)w(x^*,-\ell_y,z)dz}{2\int_{-L_z+\delta_z}^{L_z-\delta_z} w^2(x^*,0,z)dz}\\
\label{corr2}
R^w(\ell_z) & = & \frac{\int_{-L_y+\delta_y}^{L_y-\delta_y} w(x^*,y,0)w(x^*,y,\ell_z)dy+\int_{-L_y+\delta_y}^{L_y-\delta_y} w(x^*,y,0)w(x^*,y,-\ell_z)dy}{2\int_{-L_y+\delta_y}^{L_y-\delta_y} w^2(x^*,y,0)dy}.
\end{eqnarray}
The magnetohydrodynamic boundary layers of thicknesses $\delta_y=L_y/\Ha$ and $\delta_z=L_z/\Ha^{1/2}$ are excluded from the integration, 
so the estimation of the correlations is limited
to the zone of approximately homogeneous turbulence in the core flow. The integrals are calculated
at the time moments separated by $0.1$ and time-averaged over the period of fully developed flow.
The calculations are performed for several duct's cross-sections $x=x^*$, namely at $x^*=1,2,3,4$
and at $7\le x^*\le 49$ with a step of $3$.

The computed correlation curves provide detailed information on the development of the dimensional anisotropy along the duct.
The results are presented in appendix \ref{appA}. 
Here, we discuss the longitudinal ($L_{\|}$) and transverse ($L_{\bot}$) length scales along ($y$) and across ($z$) the magnetic field derived as:
\begin{eqnarray}
\label{int1} L_{\|}^y& =& \int_0^{1-\delta_y}R^v(\ell_y) d\ell_y,\\
\label{int2} L_{\bot}^y& =& \int_0^{1-\delta_y}R^w(\ell_y) d\ell_y,\\
\label{int3} L_{\|}^z& =& \int_0^{1-\delta_z}R^w(\ell_z) d\ell_z,\\
\label{int4} L_{\bot}^z& =& \int_0^{1-\delta_z}R^v(\ell_z) d\ell_z.
\end{eqnarray}
 
In isotropic turbulence, we would find $L_{\|}^y \approx L_{\|}^z\approx 2L_{\bot}^y \approx 2 L_{\bot}^z$.
 These relationships are, quite expectedly, not satisfied by the flows 3 and 4
 with strong magnetic field. For the flows 1 and 2
  with weak magnetic field, the relationships hold for $L_{\|}^z$ and $L_{\bot}^z$ at large distances from the inlet, 
  where the honeycomb-created jets are properly mixed (see figures \ref{fig:len_scale}c and d), but not for $L_{\|}^y$ and $L_{\bot}^y$ (not clearly visible  in figures \ref{fig:len_scale}a and b, but verified in  our analysis).
We also see that at weak magnetic field the scales $L_{\|}^y$ and $L_{\bot}^y$ remain practically constant,
while $L_{\|}^z$ and $L_{\bot}^z$ grow downstream. The outlying point in figure \ref{fig:len_scale}c corresponds
to the effect of the local flow transformation in the run 4 discussed in appendix  \ref{appA}.

In the runs 3 and 4,
 the strong magnetic field causes rapid growth of $L_{\|}^y$, $L_{\bot}^y$,
and $L_{\|}^z$, but not $L_{\bot}^z$. The most interesting for us are
the length scales  $L_{\bot}^y$ and $L_{\|}^z$ computed on the basis of the fluctuations of the velocity
component $w$.
We see that the length scale $L_{\bot}^y$ along the magnetic field grows monotonically downstream after
the full-strength magnetic field is introduced (at $x=0$ in the run 3 
and at $x=x_1$ in the run 4)
as an indication of flow's transition into strongly anisotropic form. Interestingly, the 
large vortices developing in the flow 3 result in slower growth, so at the end of the domain,
$L_{\bot}^y$ is smaller 
than in the flow 4.
 The length scale $L_{\|}^z$ in the direction
perpendicular to the magnetic field grows very rapidly at small $x$ in the flow 3
and stabilizes at about
$0.25$ at $x$ above approximately $20$. This value as associated with the typical 
transverse size of the quasi-two-dimensional vortices.
On the contrary, in the flow 4,
where the vortices do not form, $L_{\|}^z$ grows continuously downstream.

\subsection{Effect of walls and anisotropy of inlet conditions}
\label{sec:results_b}
The discussion of section \ref{sec:results_a} as well as previous works by various authors 
\citep[see \eg][]{Moffatt:1967,Sukoriansky:1986,Kljukin:1989,Burattini:2010} 
suggest that the development and persistence of quasi-two-dimensional  structures aligned 
with the strong imposed magnetic field is a general physical phenomenon to be observed, in some form, 
in all decaying MHD turbulent flows. At the same time,  features of the flow's configuration may 
strongly affect the realization of the phenomenon in a specific case. For our system, the most
important such features are: \emph{(i)} the location of the duct's walls non-parallel to the magnetic field, which
limit the longitudinal size of the quasi-two-dimensional flow structures and \emph{(ii)} the design of the honeycomb, 
which may introduce anisotropy into the initial state of the flow. 

The importance of these features is due to the presence of the strong transverse magnetic field.
Without the field,  approximately homogeneous and isotropic turbulence insensitive to such details of the system's geometry is
expected to form in the core of the duct  downstream of the honeycomb's exit.

The two effects are explored in our study in the simulation runs 5-8
(see table \ref{table:param} for parameters).
The strong magnetic field corresponding to $\Ha=195$ is applied in all the simulations, so we expect the 
behaviour similar to that observed earlier in the simulations 3 and 4.
 The main
component of the magnetic field is oriented along the shorter side of the duct ($B_z$) and not along the longer side as before.
 The case 1 and case 2 
distributions of the magnetic field along the duct are considered. In addition to allowing us to see the effect of the
distance between the field-crossing walls, the new simulations  provide 
a direct 
comparison with the experiment of \citet{Sukoriansky:1986}, in which the magnetic field is 
in the $z$-direction. 

Two arrangements of the honeycomb tubes are considered. As illustrated in figure \ref{fig:geom}b, 
the tubes are arranged into straight rows along the longer (Type A) or shorter (Type B) sides of the duct.
This implies different anisotropies of the flows exiting the honeycomb. The type A (runs 5 and 6)
produces structures with weaker average gradients in the $y$-direction, i.e. perpendicularly to the magnetic field. The type B 
(runs 7 and 8)
 results in the flow structures with weaker gradient in the $z$-direction, i.e. 
the direction of the magnetic field.

The rms velocity fluctuations in fully developed flows are presented in 
table \ref{table:signals}.  We see that the situation is generally similar to that
observed earlier in the simulations 3 and 4.
The anomalously strong velocity fluctuations
appear when the magnetic field has the configuration of case 1 (runs 5 and 7)
but not of case 2 (runs 6 and 8).
Also as before, the strong fluctuations develop in the streamwise velocity 
component $u$ and the transverse component perpendicular to the magnetic field $v$. 

The effect of the anisotropy introduced by the honeycomb is clearly visible. The fluctuation amplitude 
in the run 7 
is about the same as in the run 3,
while it is about two times smaller in the run 5.

\begin{figure}
\vskip3mm

\parbox{1.1\linewidth}{\hskip70mm Run 5}\parbox{0.0\linewidth}{\hskip-40mm Run 6}
\centerline{
\includegraphics[width=0.95\textwidth,clip=]{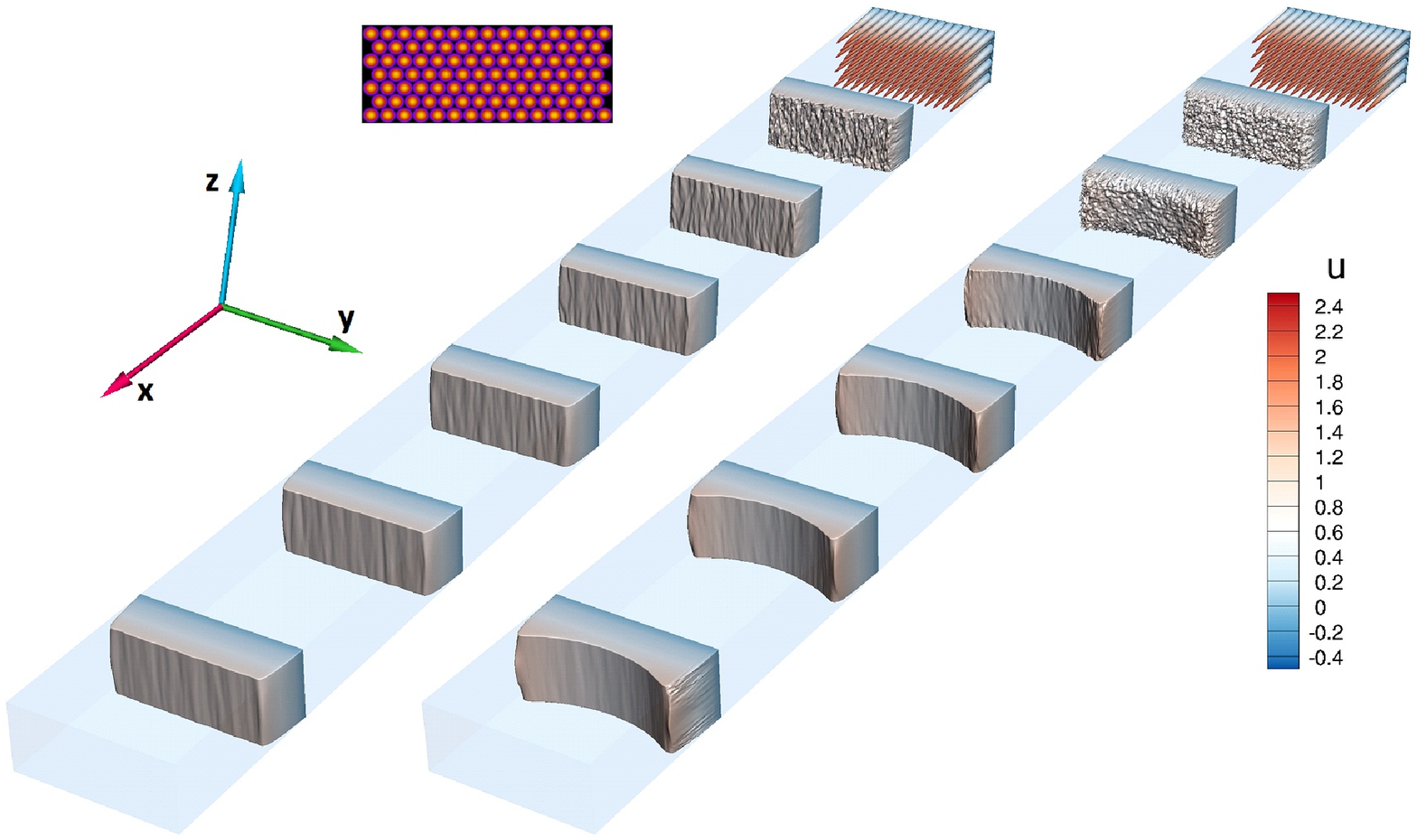}
}

\parbox{1.1\linewidth}{\hskip70mm Run 7}\parbox{0.0\linewidth}{\hskip-40mm Run 8}
\centerline{
\includegraphics[width=0.95\textwidth,clip=]{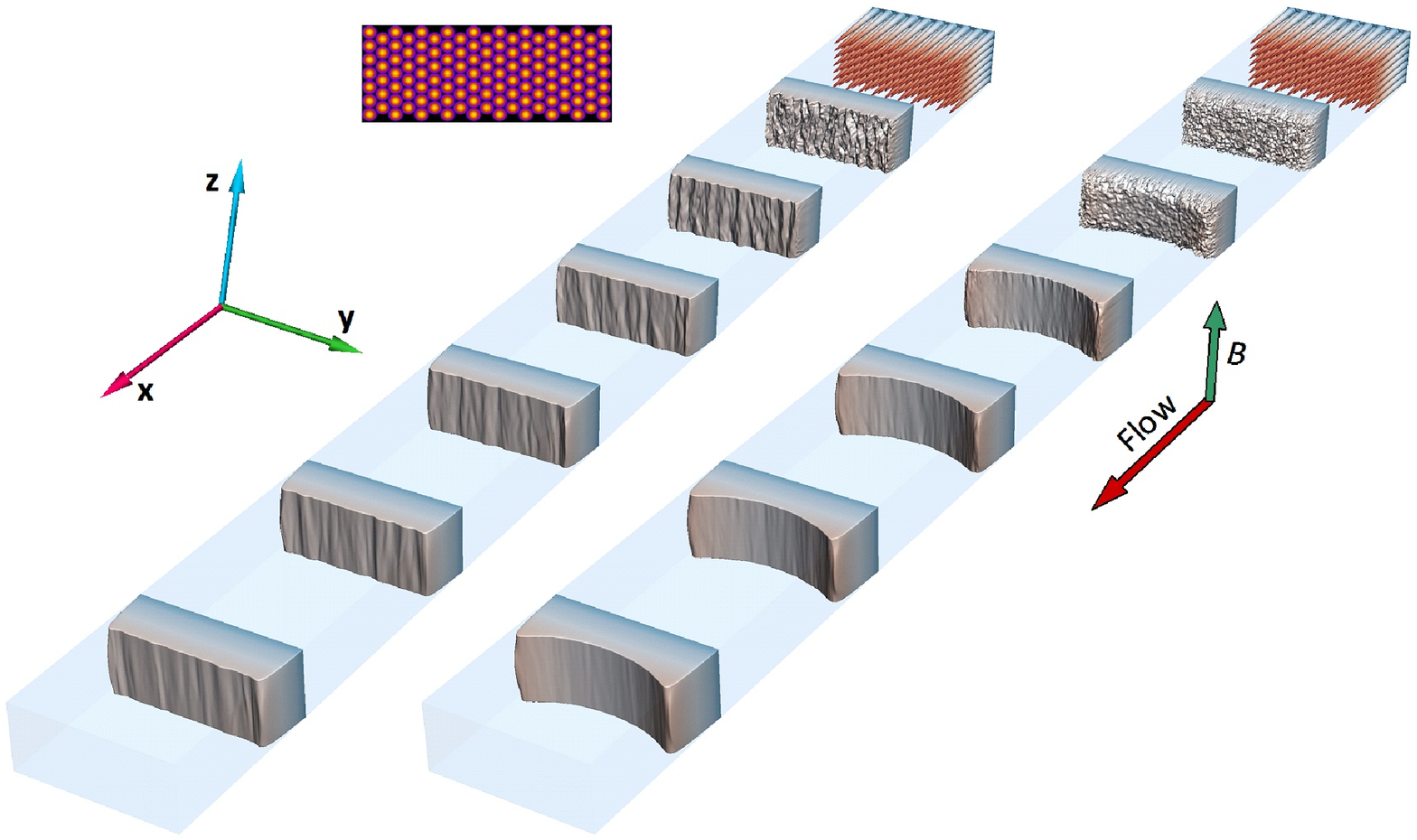}
}

\vskip-1mm
\caption{Instantaneous distributions of the streamwise velocity $u$ at several locations
along the duct shown for the fully developed flows in simulations 5-8 
(see table \ref{table:param} for flow parameters).
}
\label{fig:prof_Bz}
\end{figure}

\begin{figure}
\parbox{1.1\linewidth}{\hskip70mm Run 5}\parbox{0.0\linewidth}{\hskip-31mm Run 6}\vskip1mm
\centerline{
\includegraphics[width=0.99\textwidth,bb=20 45 590 270,clip=]{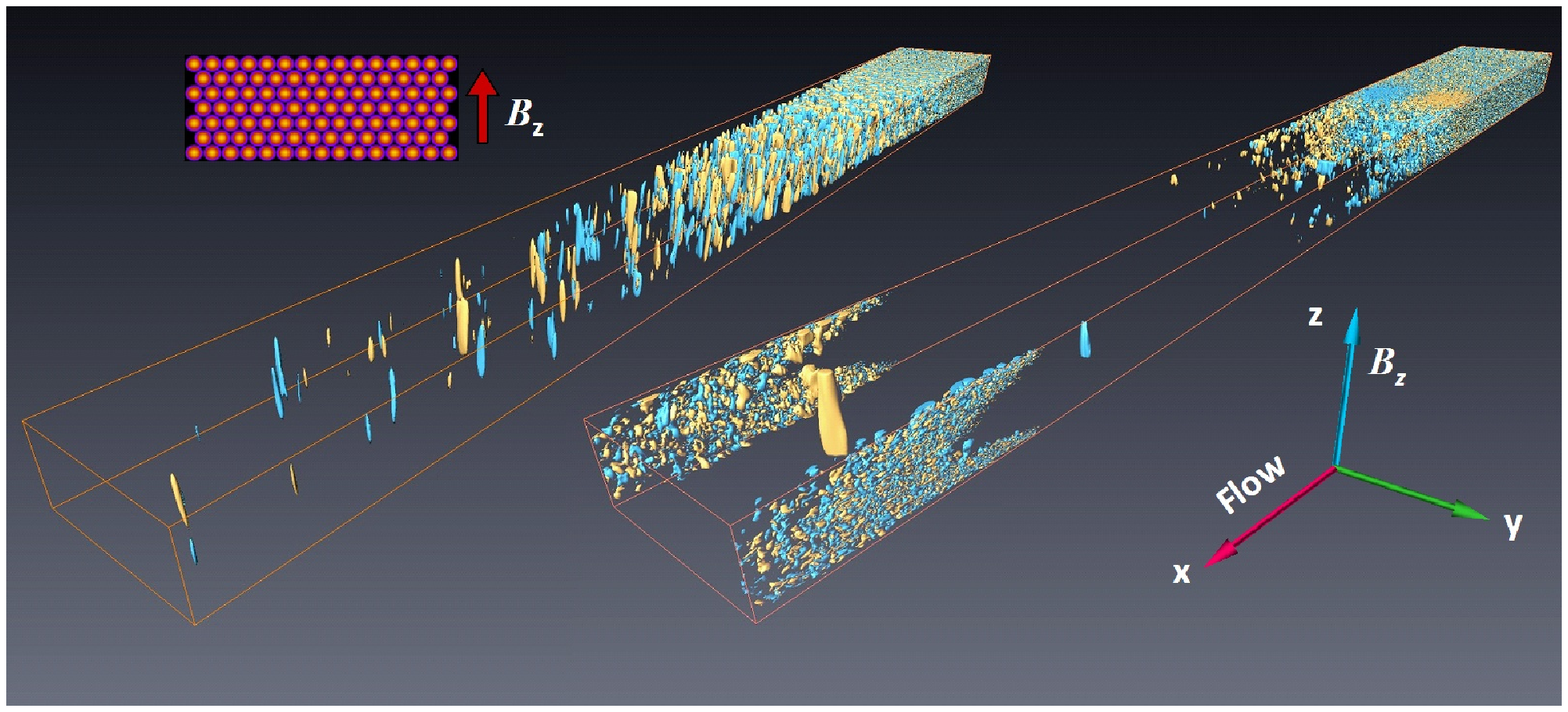}
}

\parbox{1.1\linewidth}{\hskip70mm Run 7}\parbox{0.0\linewidth}{\hskip-31mm Run 8}\vskip1mm
\centerline{
\includegraphics[width=0.99\textwidth,bb=20 45 590 270,clip=]{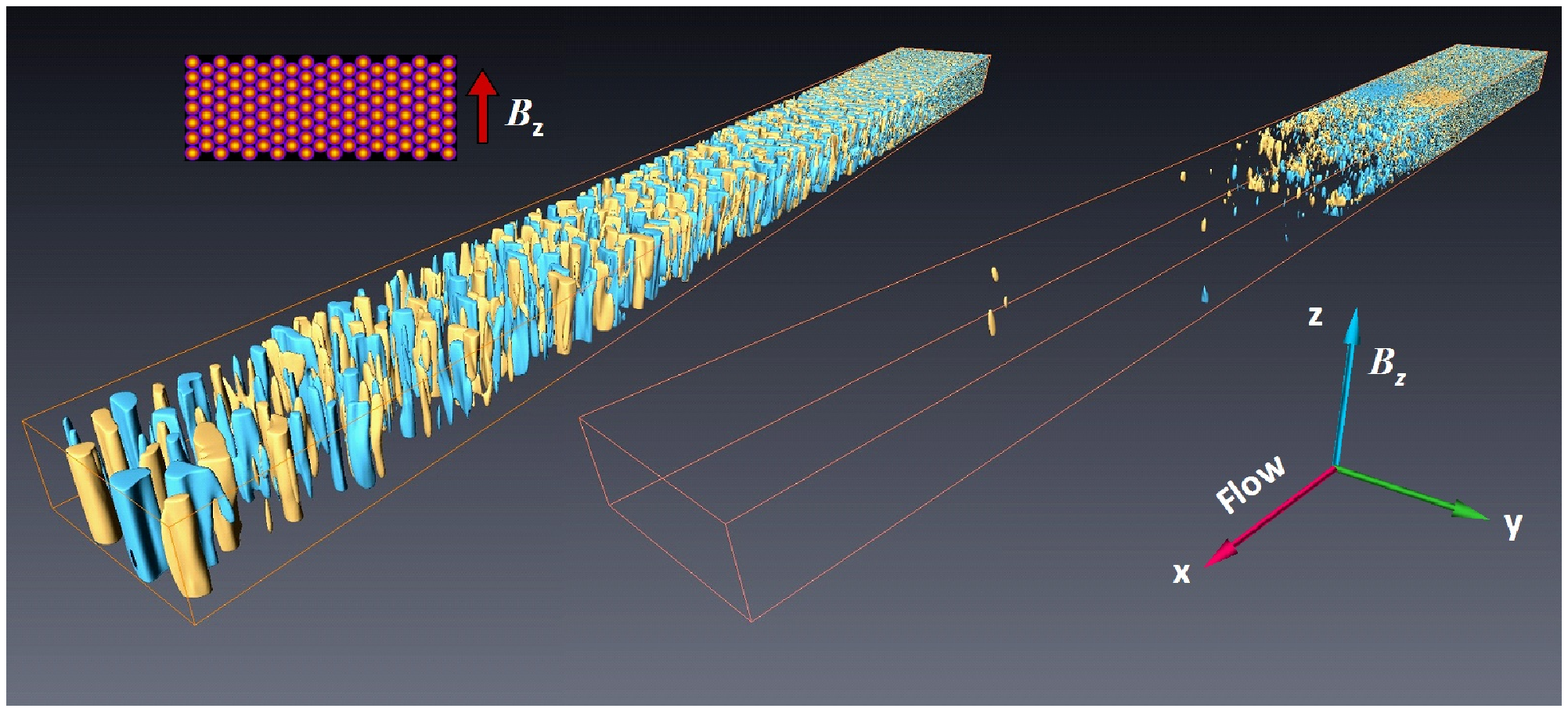}
}
\vskip-3mm
\caption{Isosurfaces of the  velocity component $v$ (transverse and perpendicular to
the main component of the magnetic field $B_{z}$) for the simulations 5, 6 (top)
and 7,8 (bottom).
Two iso-levels of the same magnitude and opposite signs (yellow -- positive, blue -- negative)
are visualized. The insert on the left shows the honeycomb patterns of Type A and B,
and the main component of the magnetic field $B_{z}$.}
\label{fig:vy3d_A}
\end{figure}

To explain these results, we will consider the spatial structure of the flows visualized in figures \ref{fig:prof_Bz}--\ref{fig:vy3d_A}.
As in section \ref{sec:results_a}, profiles of the streamwise velocity (figure \ref{fig:prof_Bz})
and isosurfaces of the transverse velocity component 
perpendicular to the magnetic field (figure \ref{fig:vy3d_A}) are shown.

We start with the simulations 6 and 8,
in which the magnet poles are shifted downstream of the honeycomb exit
(the case 2 configuration, see figures \ref{fig:experiment}b and \ref{fig:geom}a). One can see that, similarly to the simulation 4,
three-dimensional turbulence forms before the fluid enters the zone of strong magnetic field. Subsequent 
effective magnetic
damping 
results in the low amplitude of remaining velocity fluctuations reported in 
 table \ref{table:signals}.

The flows of the simulations 6 and 8 also have prominent M-shaped profiles of streamwise velocity
(see figure \ref{fig:prof_Bz}). Such a profile is expected when the flow in a duct with electrically insulating
walls enters the zone of strong transverse magnetic field \citep[see \eg][]{Branover:1978,Andreev:2006}.
The profile can also be noticed in the run 4 (see figure \ref{fig:prof}), but it is more pronounced in
the runs 6 and 8 due to the larger distance between the sidewalls (the walls parallel to the magnetic field).

The two just discussed flow features are equally observed in the simulations 6 and 8.
The only difference between the two flows is that we see significant velocity fluctuations near the sidewalls in the 
far downstream portion of the duct in the flow 6 but not in flow 8 (see figures \ref{fig:prof_Bz} and \ref{fig:vy3d_A}). 
The physical nature of this phenomenon has been verified in additional simulations. We attribute its existence 
to the strong shear layer associated with the planar side-wall jets forming in the M-shaped profile. 
Such layers are known to be be very susceptible to instabilities \citep[see e.g.][]{Kobayashi:2012}.
Similar phenomenon is also known in another configuration with planar side-wall jets, as Hunt's flow \citep[]{BraidenEPL:2016}.
The fact that side-wall turbulence appears in the run 6, but not in the run 8, is the effect of the honeycomb arrangement. 
Stronger flow instability is triggered in the run 6, since the perturbations introduced into the side-wall
layers by the honeycomb of type A are less aligned with the magnetic field and, therefore, can destabilize earlier.

In the simulations 5 and 7, the honeycomb exit is located within the zone of strong transverse magnetic field 
(the case 1 configuration, see figures \ref{fig:experiment}b and \ref{fig:geom}a). Similarly to the flow 3,
the simulations show development of quasi-two-dimensional structures that are poorly suppressed by the magnetic field
and have the from of large-scale vortices aligned with the field. Interestingly, the strength of the structures and
the amplitude of the associated velocity fluctuations is about the same in the runs 7 and 3  (see table \ref{table:signals}). 
The process of formation of the quasi-two-dimensional vortices is practically unaffected by the orientation
of the magnetic field. 

On the contrary, the effect of the initial flow anisotropy introduced by the honeycomb is quite strong. The vortices are
noticeably weaker and the fluctuation amplitude is about two times smaller in the run 5 (when the honeycomb produces 
structures elongated across the magnetic field) than in the runs 3 and 7 (when the elongation is along the field).

\section{Discussion and concluding remarks}
\label{sec:concl}

We performed numerical simulations inspired by the experiment of \citet{Sukoriansky:1986}. The main goal
was to understand the mechanisms leading to the anomalous high-amplitude velocity fluctuations detected
in the experiment when a strong magnetic field covered the entire test section including the honeycomb.
This goal has been largely achieved. The simulation results are in good qualitative agreement with
the experimental data. The presence or absence of anomalously strong fluctuations is found, respectively,
at the same flow parameters as in the experiment (cf. the experimental data in figure \ref{fig:experiment}b
and computed data in table \ref{table:signals}).

The computed spatial structure and statistical properties of the flow provide the explanation of the experimental observations. 
The jets forming at the honeycomb exit are unstable and serve as a  source of small-scale turbulence. When the magnetic field
is weak (runs 1 and 2), the kinetic energy injected into the flow is transferred to small length scales in the conventional
process of development of three-dimensional turbulence. The turbulence then decays under the combined action of viscous and
Joule dissipation.

Similar formation of three-dimensional turbulence occurs in the flows 4, 6 and 8, in which the magnetic 
field is strong but begins at a distance from the honeycomb exit. When the fluid enters the strong magnetic
field zone, the turbulence experiences strong magnetic suppression. Its subsequent evolution is characterized
by low amplitude of velocity fluctuations (see figures \ref{fig:signals}, \ref{fig:prof}, \ref{fig:vz3d}, 
\ref{fig:prof_Bz}, \ref{fig:vy3d_A} and table \ref{table:signals}) and development of weak quasi-two-dimensional
structures (see figure  \ref{fig:len_scale}).

High-amplitude velocity fluctuations develop in the runs 3, 5 and 7 when the strong magnetic field 
imposed at the exit from the honeycomb leads to rapid development of strongly anisotropic flow structures. 
This degenerates the mechanism of vortex stretching and suppresses the energy cascade to small length scales
thus preventing formation of conventional three-dimensional turbulence. The dominant flow structures evolve
into quasi-two-dimensional vortices, which are aligned with the magnetic field and, therefore, only
weakly suppressed and retain their strength and structure till the end of the computational domain,
i.e. at the streamwise distance of at least 25 shorter duct widths. It appears highly plausible that
the anomalously strong velocity fluctuations recorded in the experiment are caused by such vortices.

The difference in the flow evolution between the cases with weak and strong magnetic fields can be related to
the differences in the values of the magnetic interaction parameter (the Stuart number) $\N\equiv \Ha^2/\Rey$.
This parameter estimates the typical ratio between the Lorentz and inertial forces and, therefore,
is often used as a measure of expected transformation of turbulence by an imposed magnetic field
\citep[see \eg][]{Zikanov:1998,Vorobev:2005,Krasnov:2008a,Burattini:2010,Krasnov:2012}. The values
of $\N$ about and higher than 1 are typically required for strong transformation (there are inevitable
variations of this rule due to  various definitions of the length and velocity scales, various types
of the flow, and the variation of the transformation effect with the typical length scale). In our study,
$N=0.1088$ in the runs 1, 2 and $N=1.368$ in the runs 3-8. The fact that the suppression of three-dimensional
turbulence and dramatic changes of the flow structure are found in the simulations with strong magnetic field
but not with weak one is, therefore, fully consistent with the known trend.

We have explored the effect of the geometric features of the system on the flow's behaviour at strong
magnetic field. It has been found that the role of the orientation of the magnetic field, which can
also be interpreted as the role of the wall-to-wall distances across and along the field, is minimal.
This is demonstrated by the lack of noticeable differences between the flows in the runs 3 and 4
on the one hand and runs 7, 8 on the other hand. 

On the contrary, the initial anisotropy introduced by the honeycomb has strong effect on the flow with
the quasi-two-dimensional vortices. As demonstrated by the simulations 3, 5 and 7, the amplitude 
of the vortices is substantially reduced when the flow structures formed at the exit of the honeycomb
are elongated across rather than along the magnetic field. 

We would like to stress that the flow evolution observed in the runs 3, 5, and 7 does not include development 
of an inverse energy cascade. For inverse cascade to exist, the quasi-two-dimensional turbulence has to be
continuously forced. In our case the turbulent energy is injected locally near the honeycomb by the instability
of the jets leaving it. Part of this energy is dissipated by Joule friction, but the rest feeds quasi-two-dimensional
vortices. Downstream, the flow is unforced and is a subject to anisotropic Joule dissipation and wall friction.
Without constant supply of energy, the inverse cascade (in a strict sense of this term) does not develop,
but the vortices grow in size due to quasi-two-dimensional dynamics.
 
As we have already mentioned, the results of the simulations are in good qualitative agreement with the experimental
data of \citet{Sukoriansky:1986}. The high-amplitude fluctuations appear at the same values of $\Ha$. Assuming that the
simulation 7 is the closest analogue of the experiment, we notice that the ratios between the fluctuation amplitudes 
in the case 1 and case 2 configurations of the magnetic field are of the same order of magnitude: about 5 in the experiment
and about 2.5 in the simulations (see table \ref{table:signals}).

However, the turbulence intensity in the computed flows is about five times lower than measured in the experiment.
This is true for both low and high values of $\Ha$ and for different orientations and spatial structures of
the magnetic field. Several possible explanations are related to  both the numerical and experimental procedures.
We cannot reliably discuss the possible role of the experimental procedure due to the substantial time that
has passed since the experiment was completed. Likely numerical causes are the insufficient resolution of
the shear layers in the jets exiting the honeycomb and the assumption of laminar, with weak random noise,
nature of the jets. It is well known \citep[see \eg][]{Kim:2009} that, in numerical simulations, the instability
and mixing of submerged jets are strongly affected by the resolution and the inlet conditions. This may potentially
lead to lower energy injection from the jets  into the small-scale turbulent fluctuations. We should also mention
that in the experiment the flow between the tubes of the honeycomb is not zero, which may result in additional
shear and stronger mixing. This effect is ignored in the numerical model.

From the viewpoint of the turbulent decay theory, our work provides a good example
of non-universality of decay of MHD turbulence. The curves in figures \ref{fig:decay} and \ref{fig:decay2}
show complex behaviour of the fluctuation energy. The decay rate varies with the stage of the process
and among the velocity components. The values of the two independent non-dimensional parameters
(for example, $\N$ and $\Rey$) do not determine the decay scenario in a unique way. The process
is strongly affected by the development, or lack thereof, of quasi-two-dimensional structures.
The appearance and nature of such structures is, in turn, determined not just by the strength of
the magnetic field, but also  by the features of the flow evolution, most importantly, by the state
of the flow at the moment the magnetic field is introduced.

\begin{acknowledgments}
The work is financially supported by the DFG grants KR $4445/2-1$ and SCHU $1410/29-1$, the Helmholtz Alliance
``Liquid metal technologies'' (Limtech) and the grants CBET $1435269$ and CBET $1803730$ from the US NSF.
Computations were performed on the parallel supercomputers Jureca of the Forschungszentrum J\"ulich (NIC)
and SuperMUC of the Leibniz Rechenzentrum (LRZ), flow visualization was done at the computing center of
TU Ilmenau.
\end{acknowledgments}

\appendix
\section{}\label{appA}

\begin{figure}

\vskip3mm
\parbox{0.5\linewidth}{(a)}\parbox{0.3\linewidth}{(b)}\vskip-5mm
\centerline{
\includegraphics[width=0.45\textwidth,clip=]{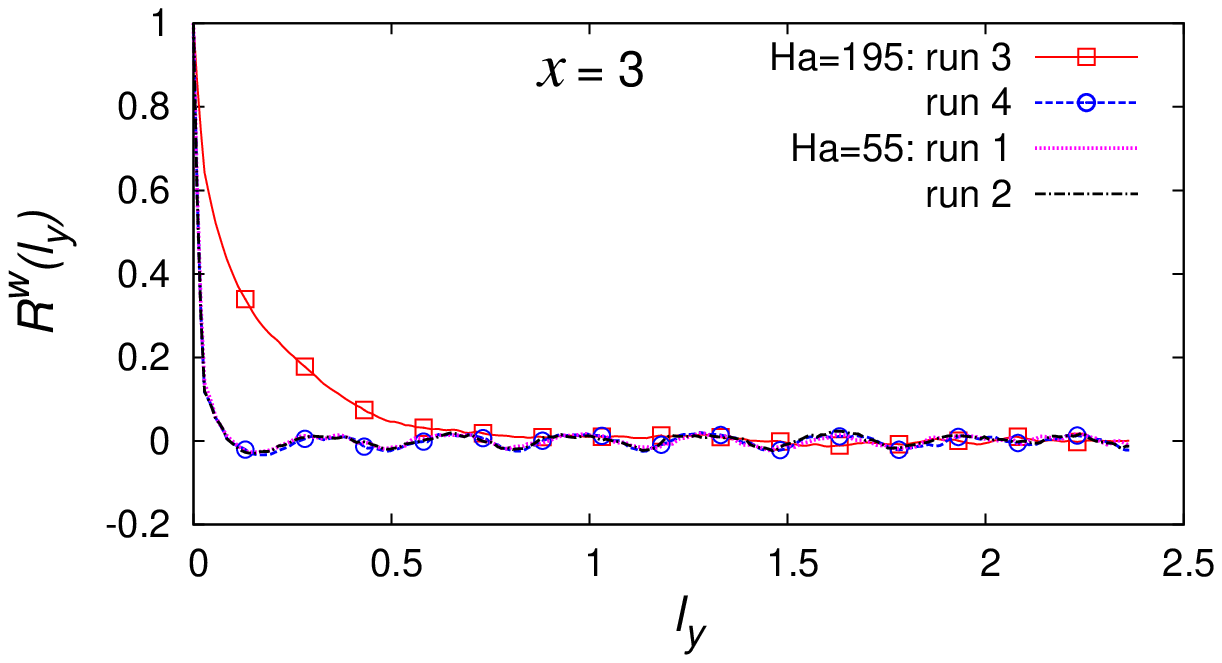}
\includegraphics[width=0.45\textwidth,clip=]{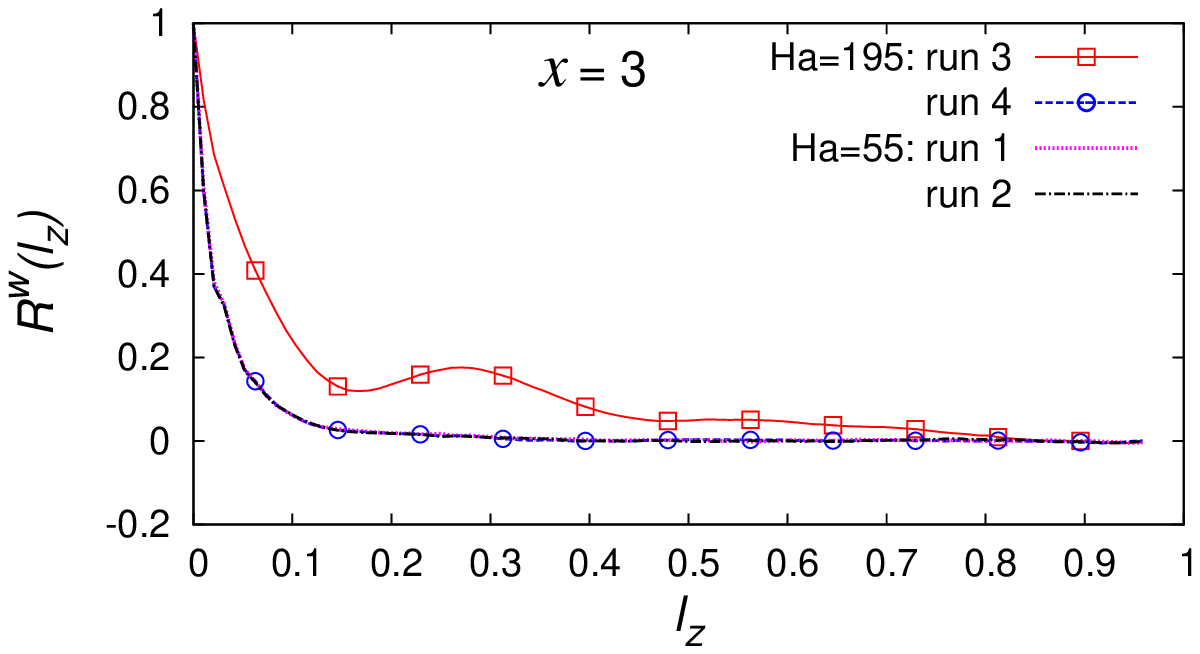}
}

\vskip-6mm
\parbox{0.5\linewidth}{(c)}\parbox{0.3\linewidth}{(d)}\vskip-5mm
\centerline{
\includegraphics[width=0.45\textwidth,clip=]{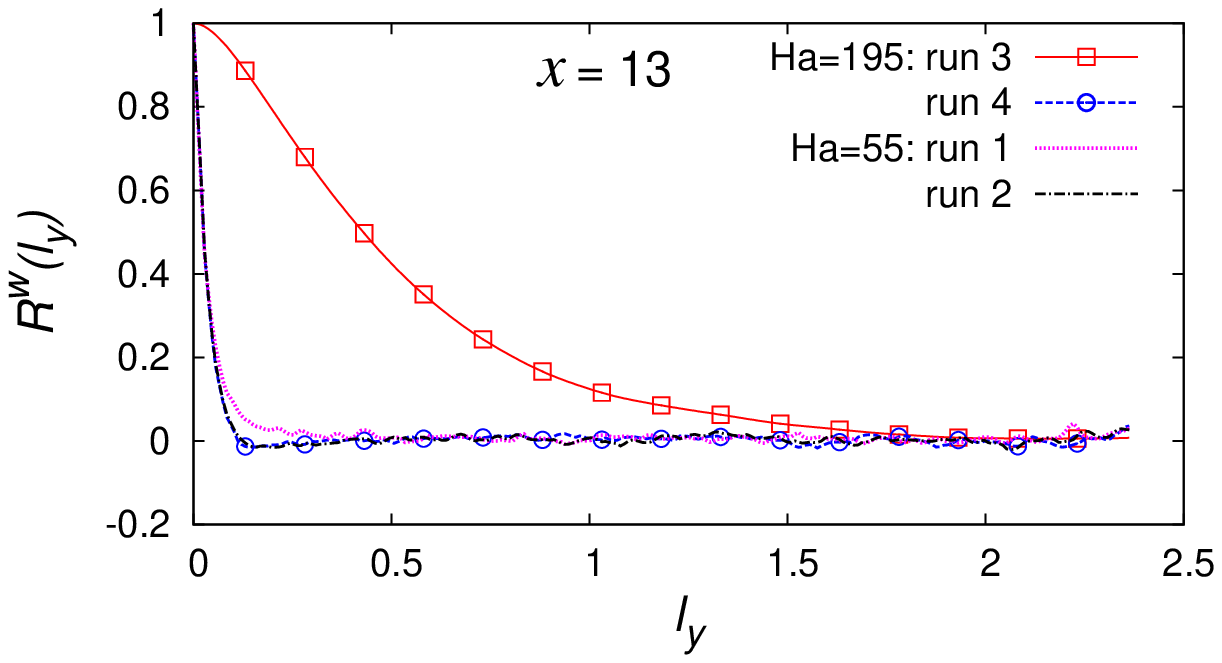}
\includegraphics[width=0.45\textwidth,clip=]{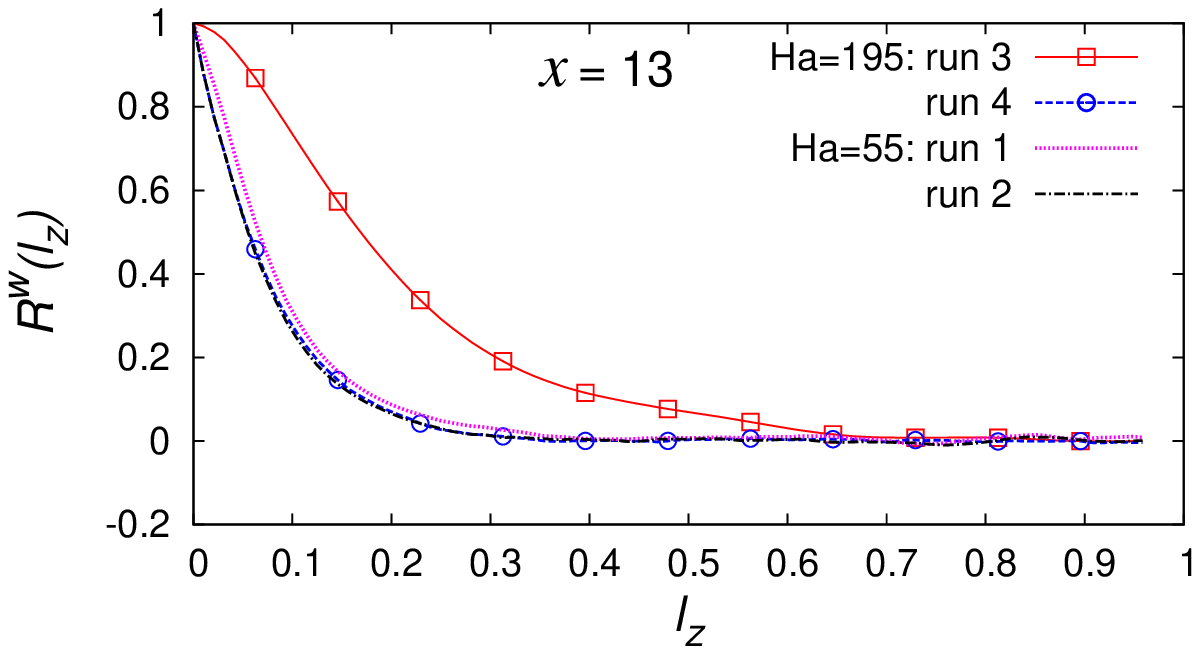}
}

\vskip-6mm
\parbox{0.5\linewidth}{(e)}\parbox{0.3\linewidth}{(f)}\vskip-5mm
\centerline{
\includegraphics[width=0.45\textwidth,clip=]{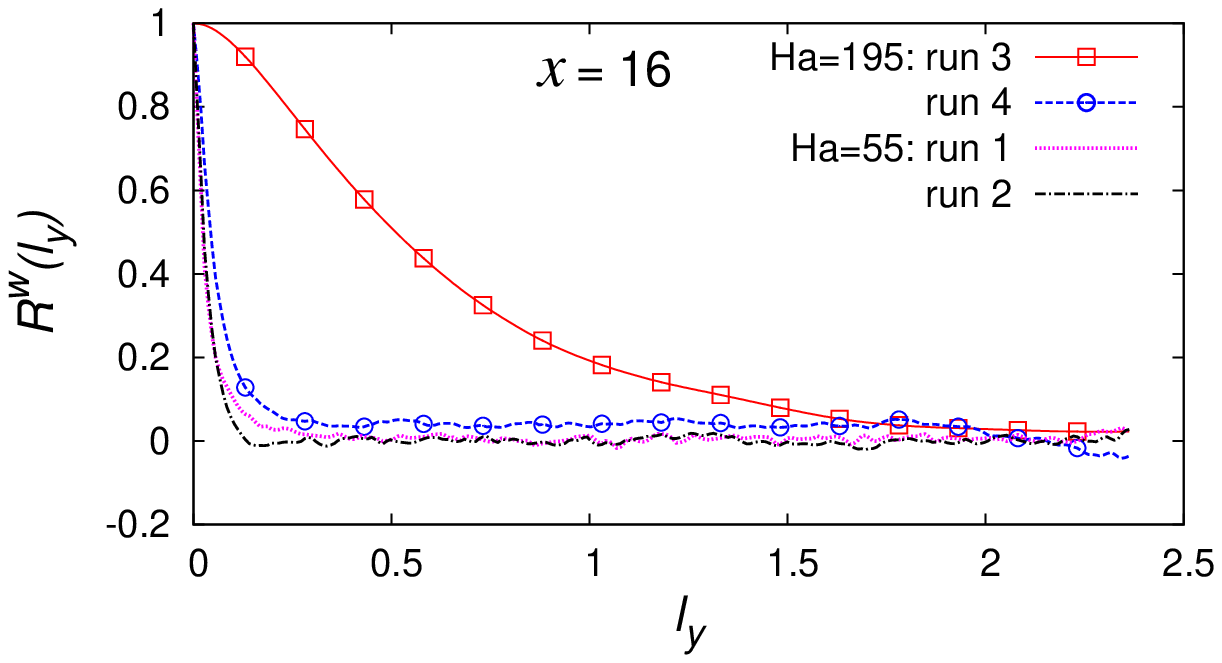}
\includegraphics[width=0.45\textwidth,clip=]{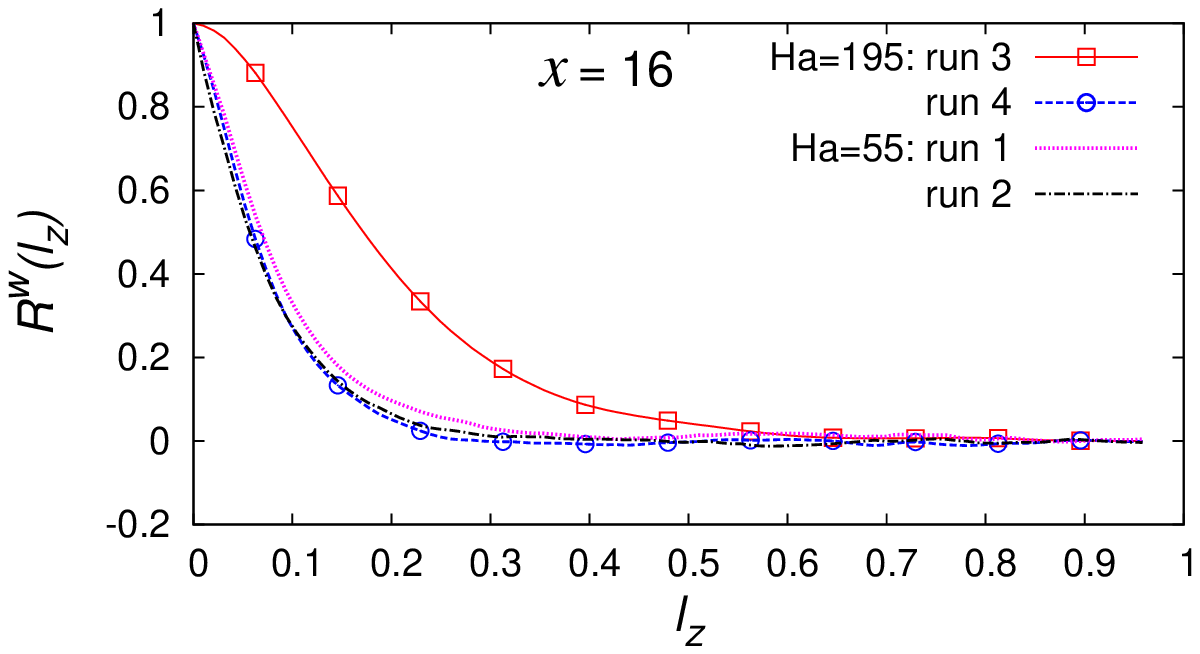}
}

\vskip-6mm
\parbox{0.5\linewidth}{(g)}\parbox{0.3\linewidth}{(h)}\vskip-5mm
\centerline{
\includegraphics[width=0.45\textwidth,clip=]{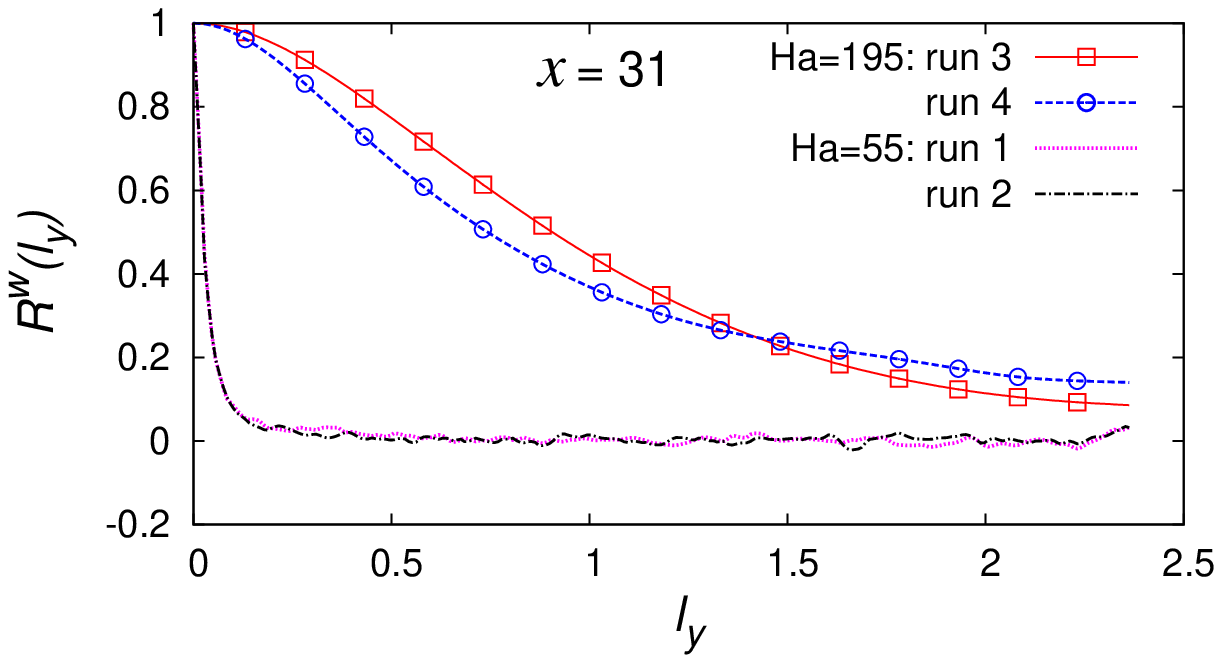}
\includegraphics[width=0.45\textwidth,clip=]{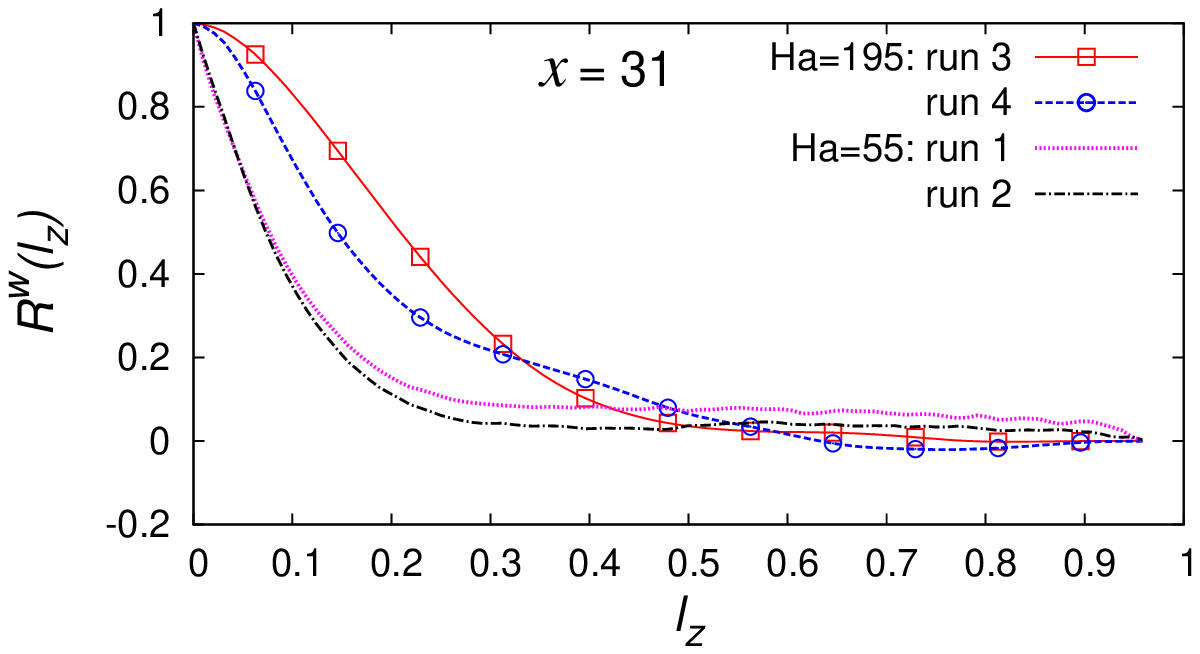}
}

\vskip-6mm
\parbox{0.5\linewidth}{(i)}\parbox{0.3\linewidth}{(j)}\vskip-5mm
\centerline{
\includegraphics[width=0.45\textwidth,clip=]{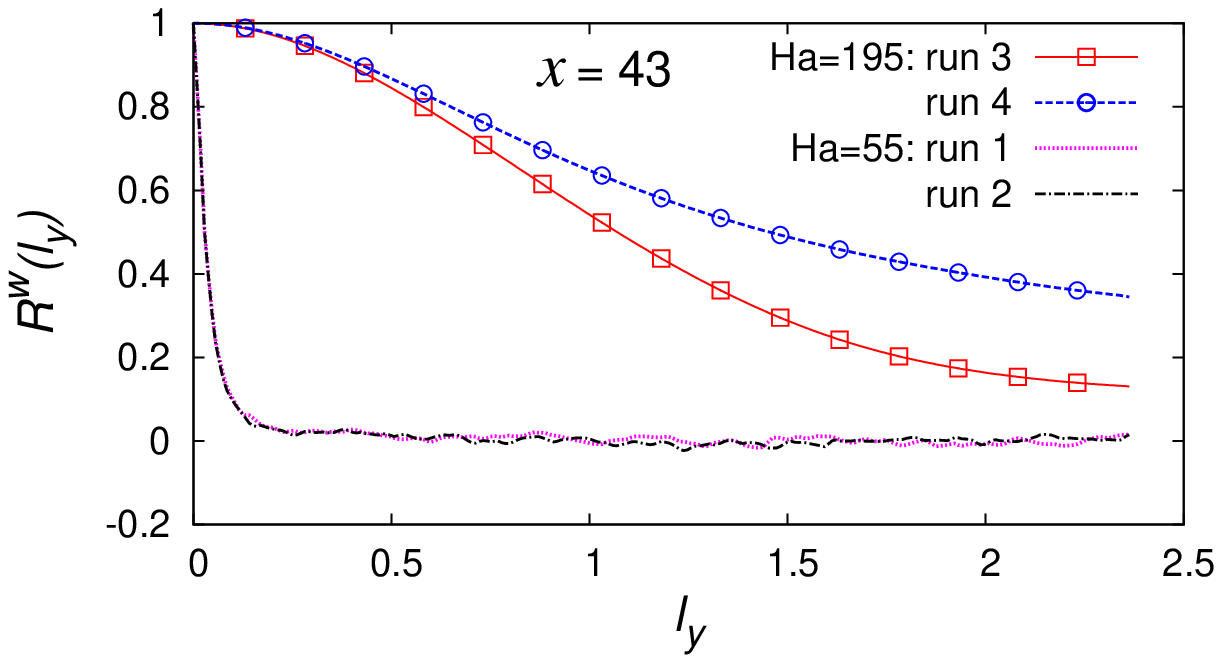}
\includegraphics[width=0.45\textwidth,clip=]{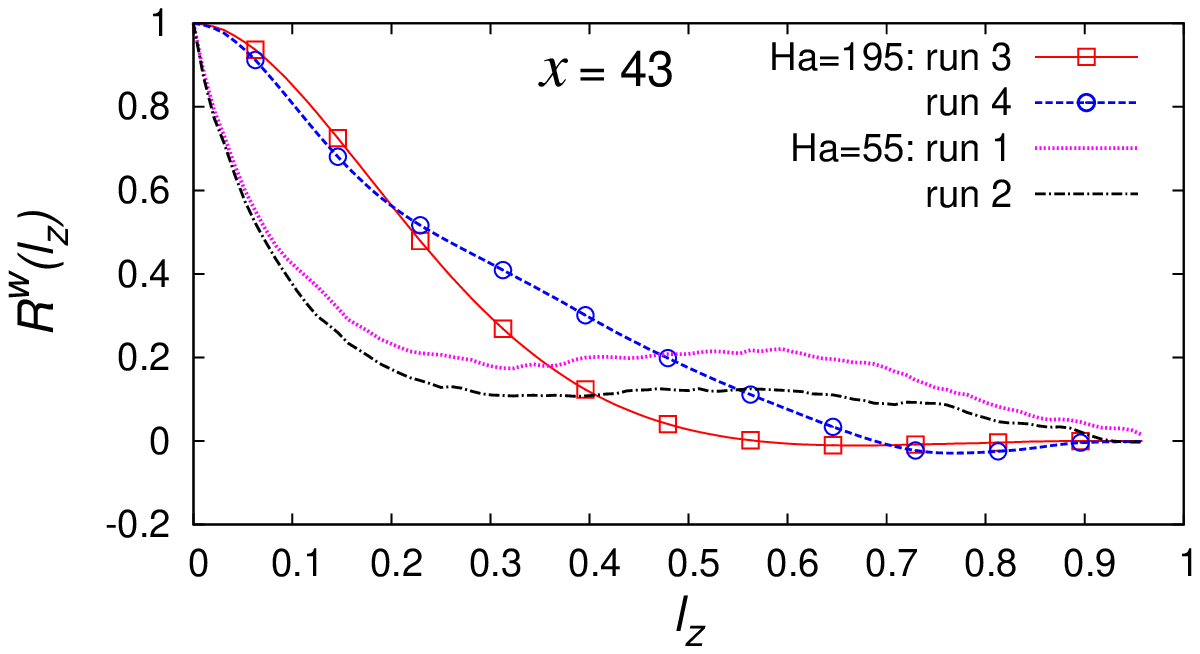}
}

\vskip-4mm
\caption{Two-points correlations in the cross-sections at $x=3, 13, 16, 31$ and $43$ for
the velocity component $w$ in runs 1-4. Left: correlation coefficients $R^w(l_y)$ versus 
distance $l_y$. Right: correlation coefficients $R^w(l_z)$ versus distance $l_z$.}
\label{fig:corr_w}
\end{figure}

\begin{figure}

\vskip3mm
\parbox{0.5\linewidth}{(a)}\parbox{0.3\linewidth}{(b)}\vskip-5mm
\centerline{
\includegraphics[width=0.45\textwidth,clip=]{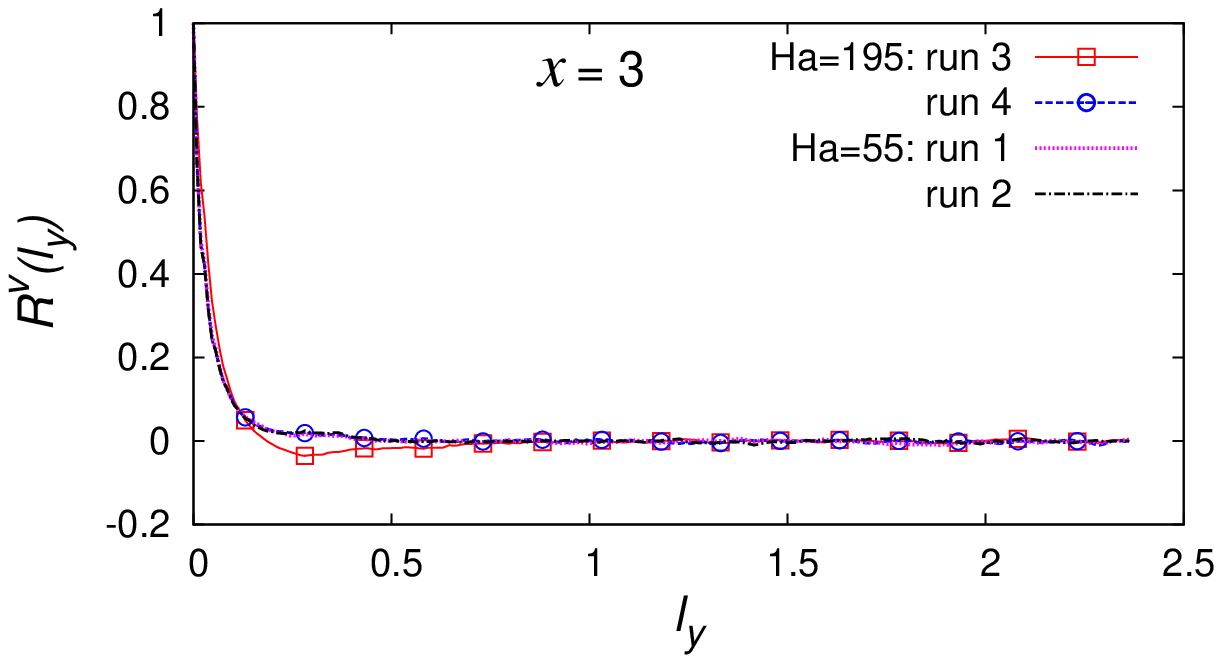}
\includegraphics[width=0.45\textwidth,clip=]{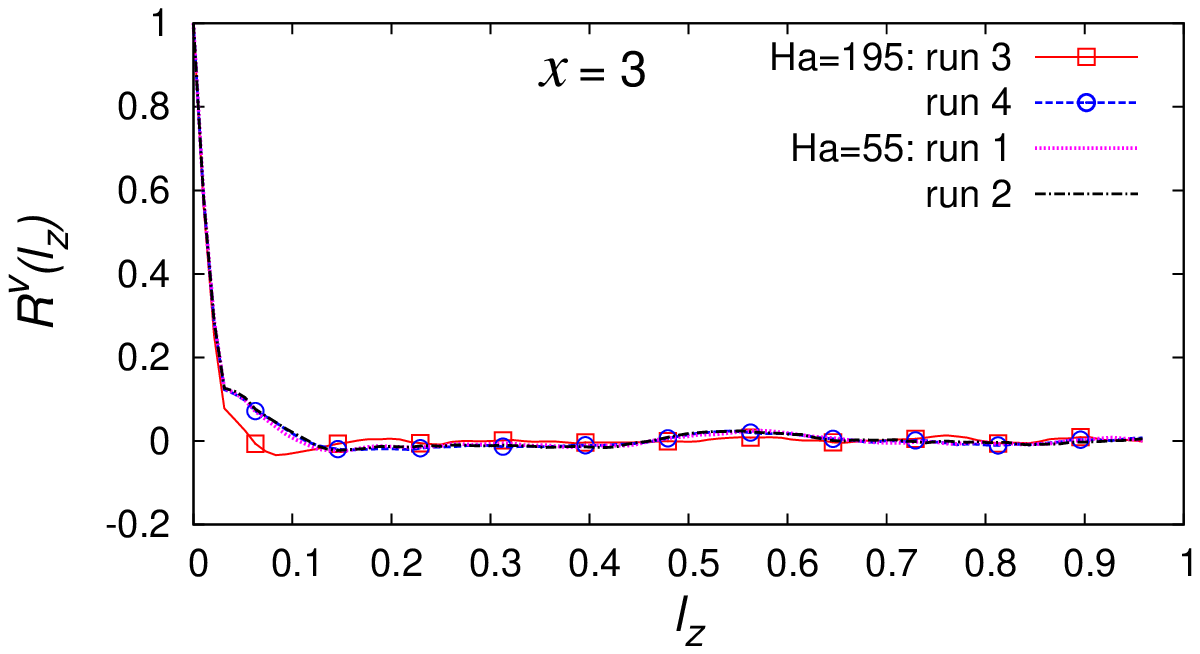}
}

\vskip-6mm
\parbox{0.5\linewidth}{(c)}\parbox{0.3\linewidth}{(d)}\vskip-5mm
\centerline{
\includegraphics[width=0.45\textwidth,clip=]{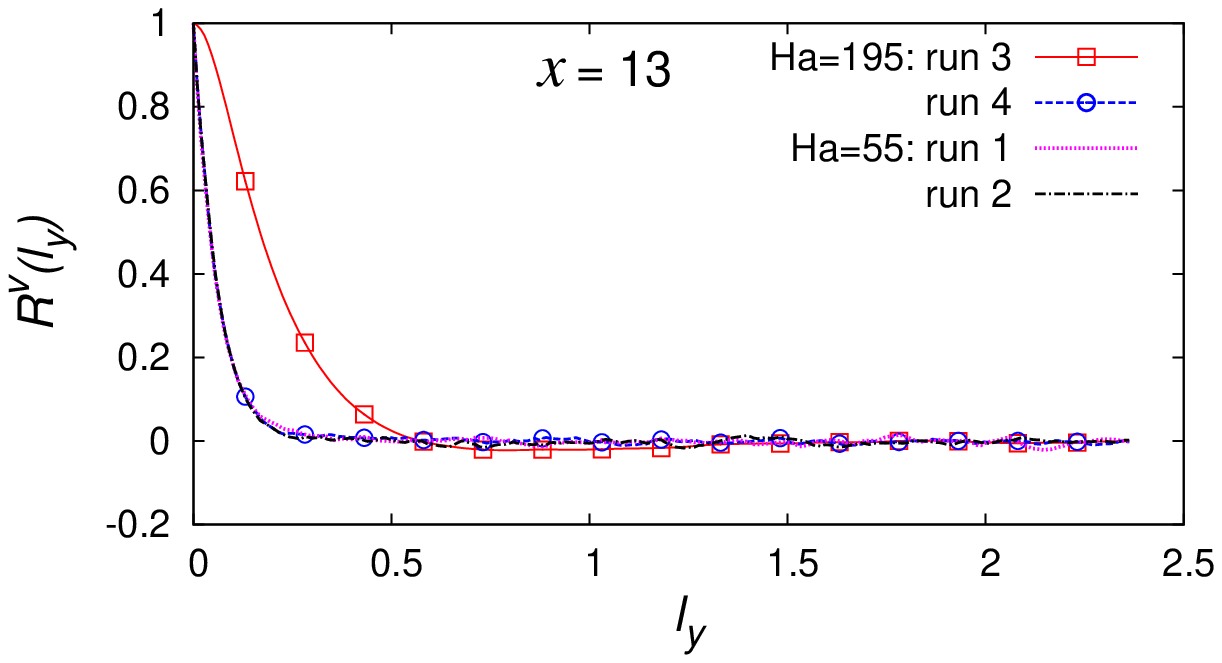}
\includegraphics[width=0.45\textwidth,clip=]{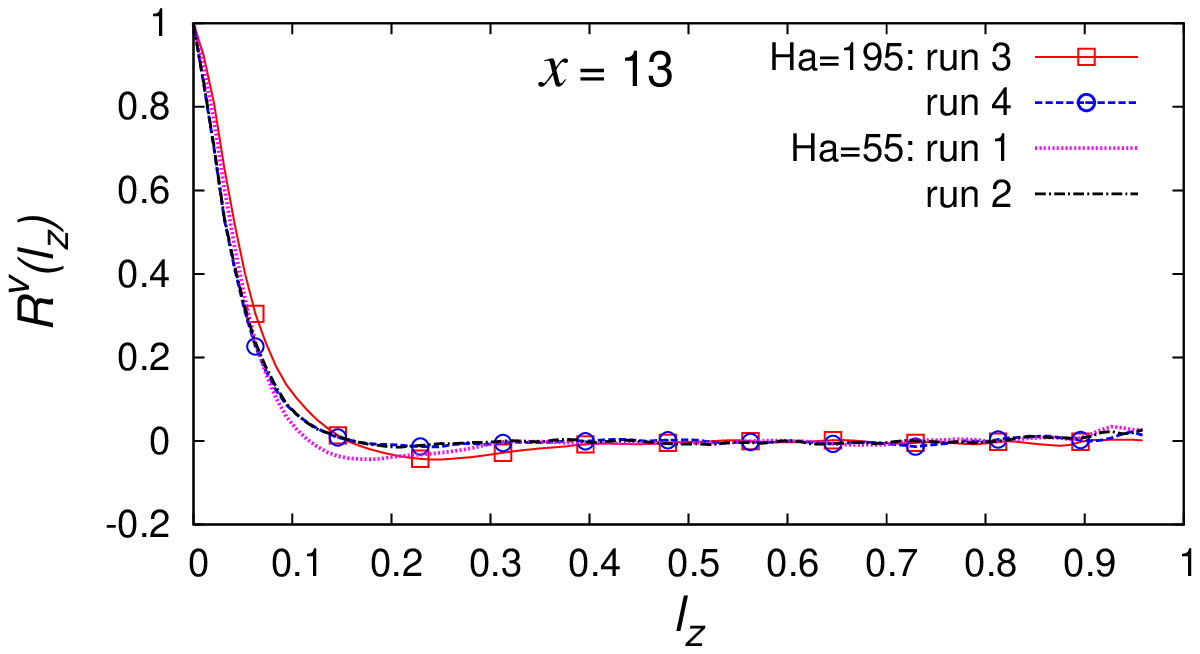}
}

\vskip-6mm
\parbox{0.5\linewidth}{(e)}\parbox{0.3\linewidth}{(f)}\vskip-5mm
\centerline{
\includegraphics[width=0.45\textwidth,clip=]{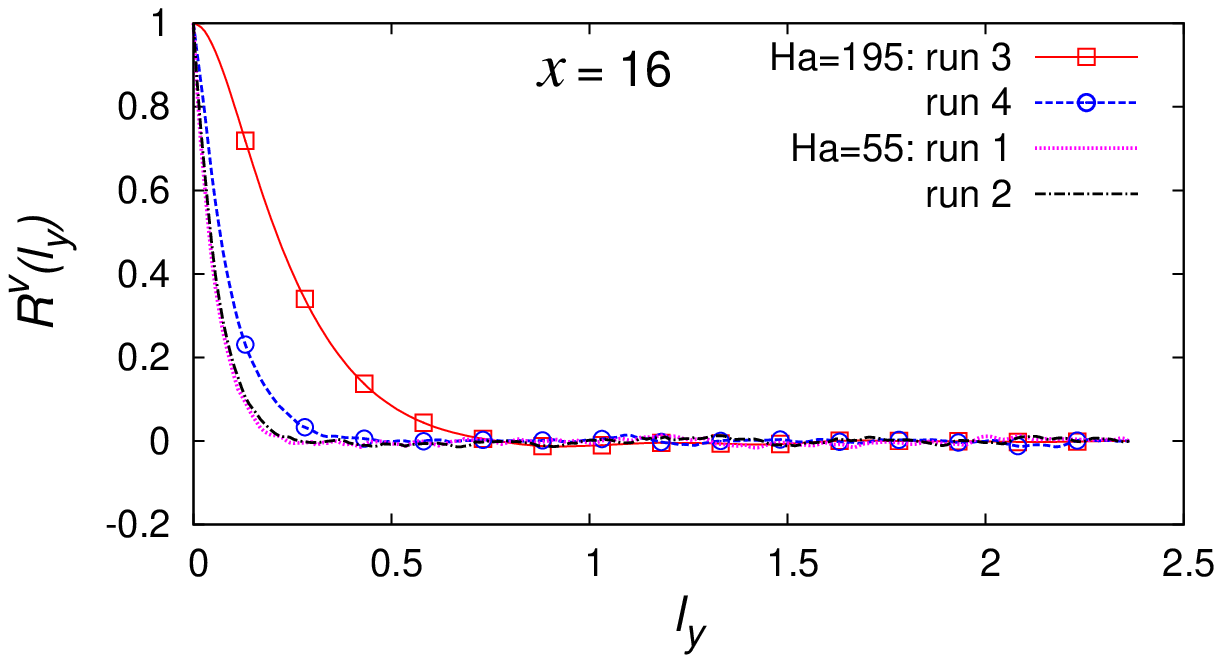}
\includegraphics[width=0.45\textwidth,clip=]{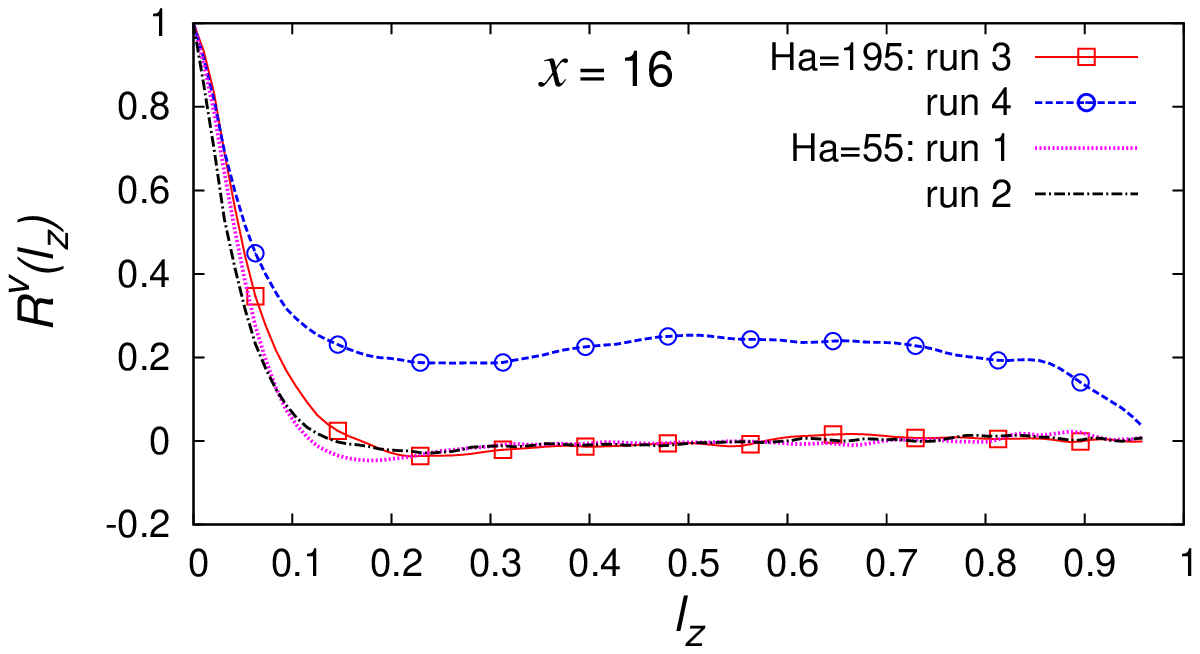}
}

\vskip-6mm
\parbox{0.5\linewidth}{(g)}\parbox{0.3\linewidth}{(h)}\vskip-5mm
\centerline{
\includegraphics[width=0.45\textwidth,clip=]{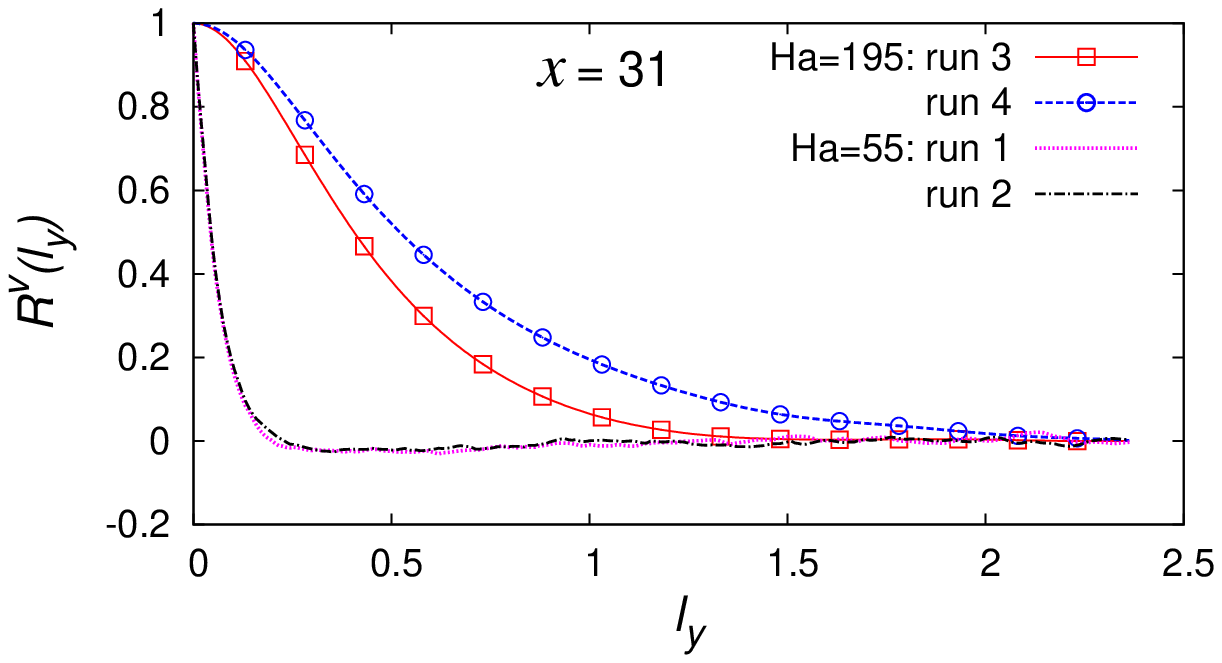}
\includegraphics[width=0.45\textwidth,clip=]{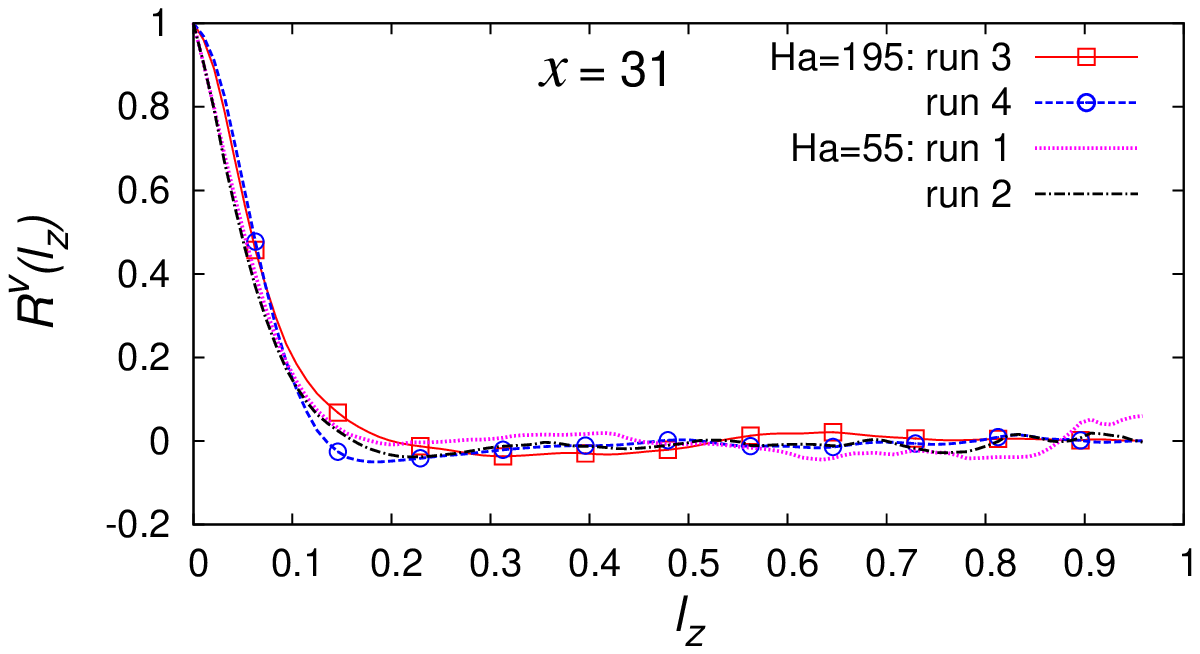}
}

\vskip-6mm
\parbox{0.5\linewidth}{(i)}\parbox{0.3\linewidth}{(j)}\vskip-5mm
\centerline{
\includegraphics[width=0.45\textwidth,clip=]{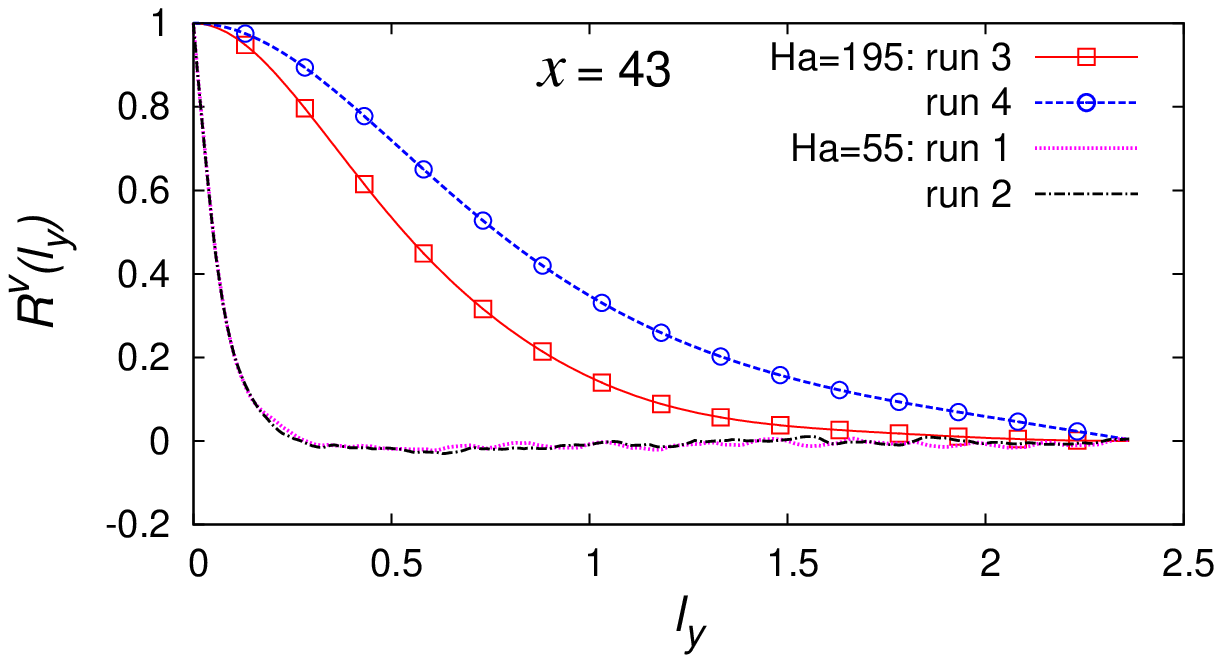}
\includegraphics[width=0.45\textwidth,clip=]{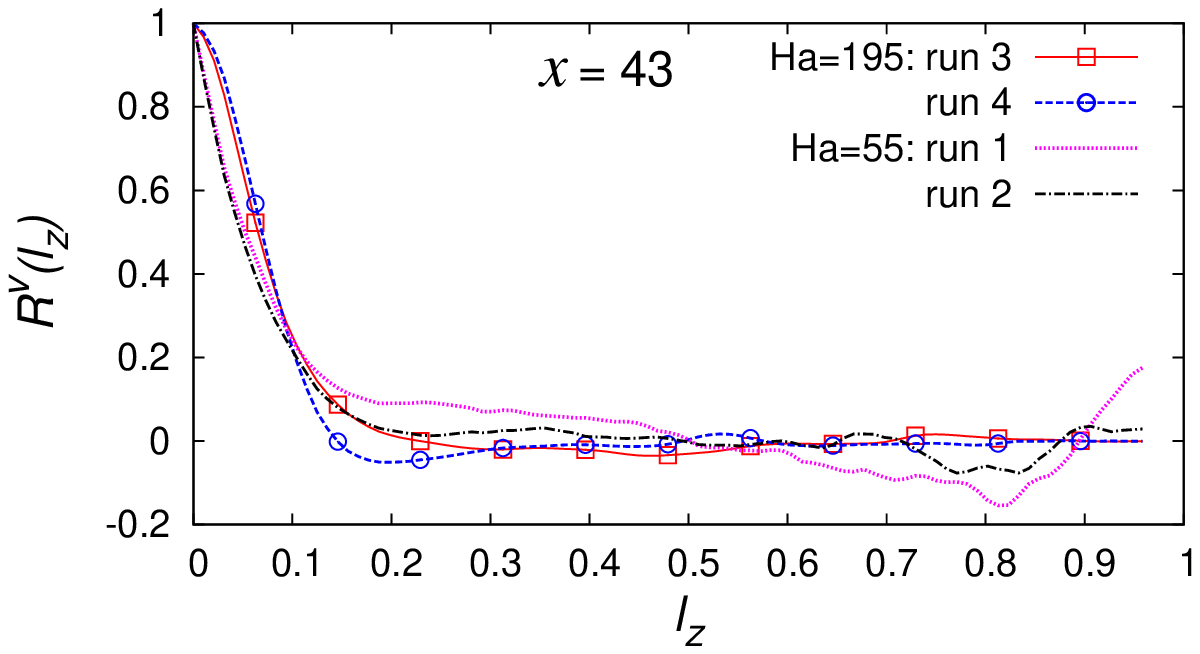}
}

\vskip-4mm
\caption{Two-points correlations in the cross-sections at $x=3, 13, 16, 31$ and $43$ for
the velocity component $v$ in runs 1-4. Left: correlation coefficients $R^v(l_y)$ versus 
distance $l_y$. Right: correlation coefficients $R^v(l_z)$ versus distance $l_z$.}
\label{fig:corr_v}
\end{figure}

\begin{figure}

\vskip2mm
\parbox{0.22\linewidth}{$x=13$ \hskip15mm $u$}\parbox{0.78\linewidth}{\hskip38mm $v$ \hskip38mm $w$}
\centerline{
\includegraphics[width=0.3\textwidth,bb=50 20 550 265,clip=]{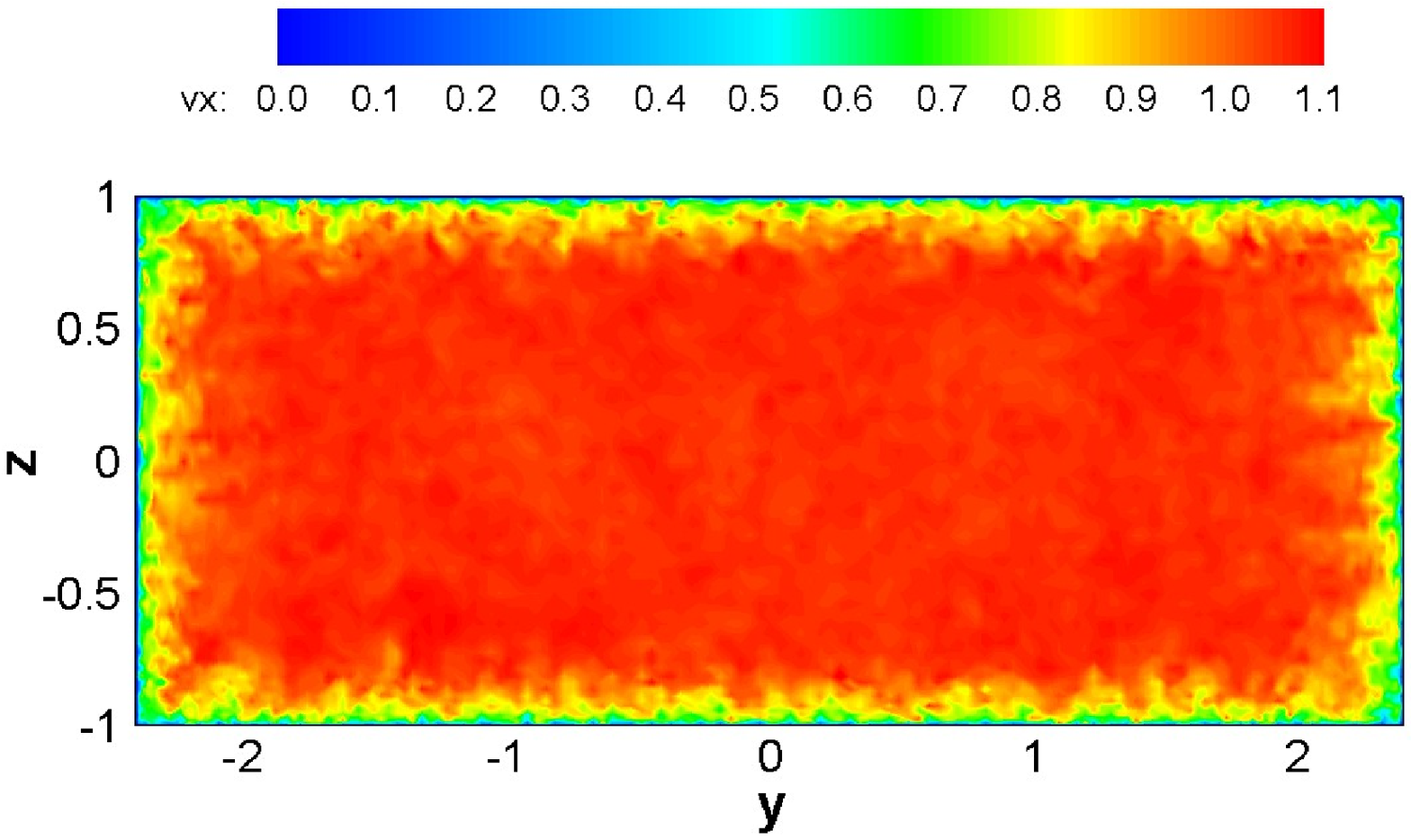}
\includegraphics[width=0.3\textwidth,bb=50 20 550 265,clip=]{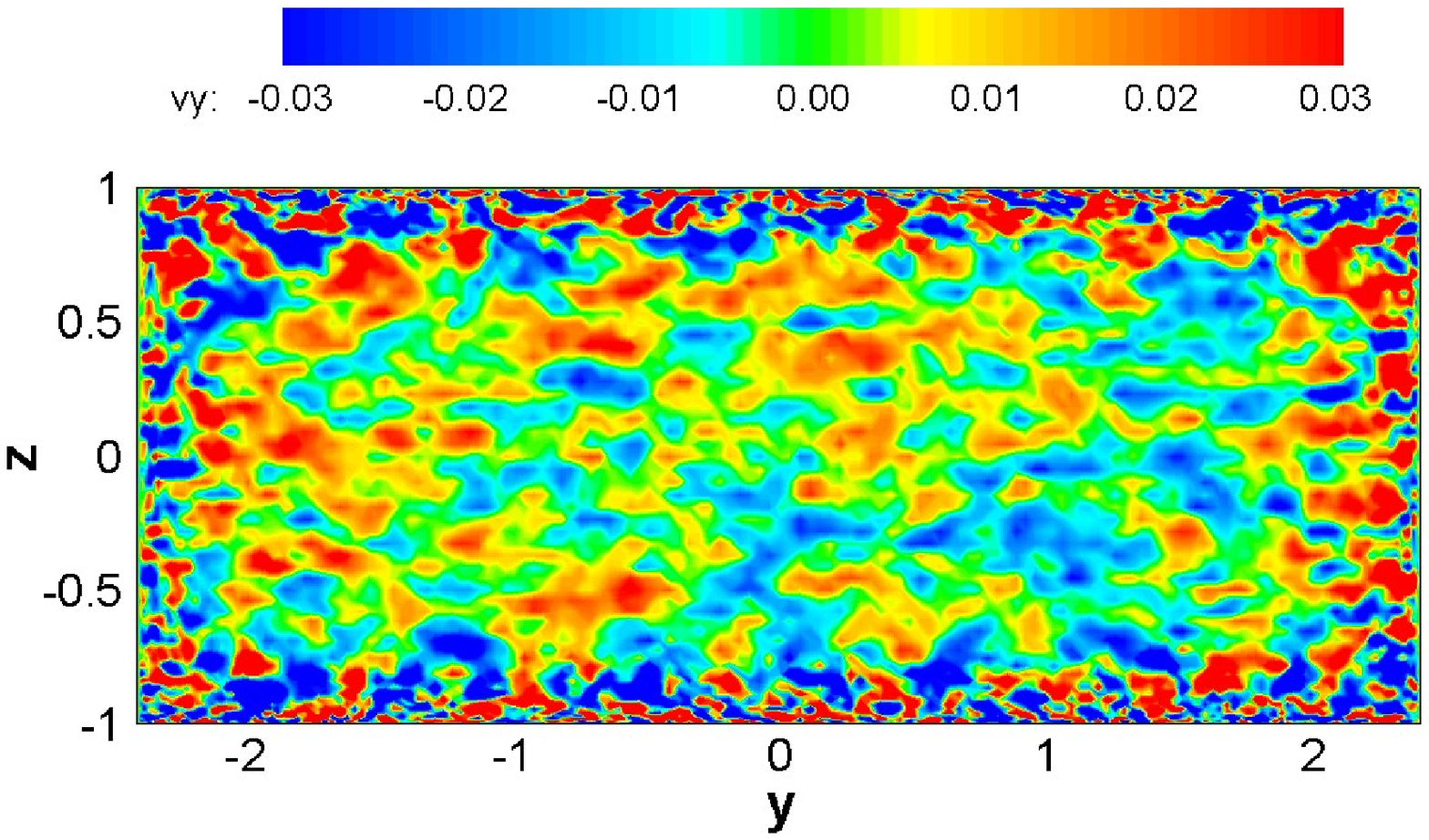}
\includegraphics[width=0.3\textwidth,bb=50 20 550 265,clip=]{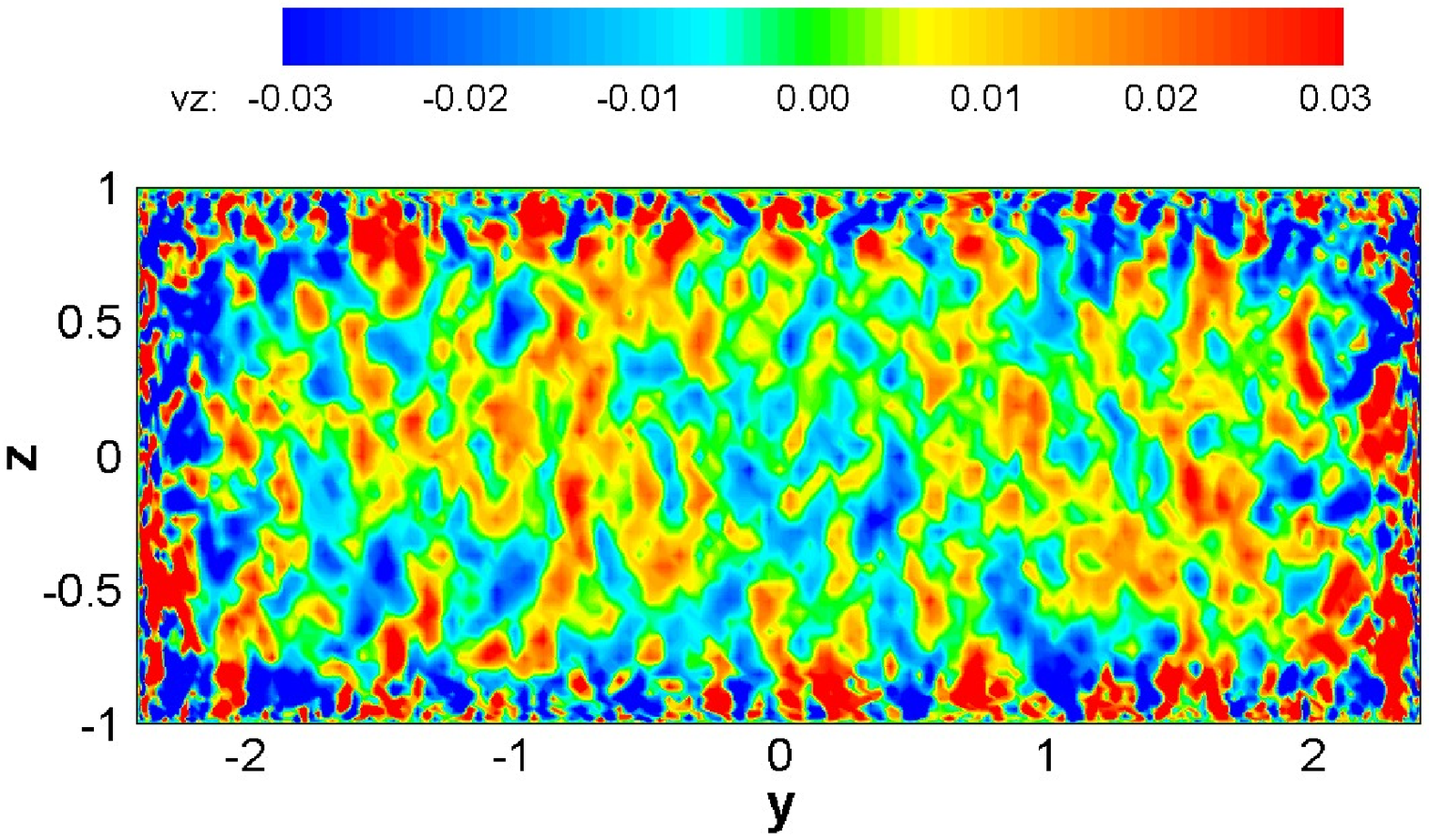}
}
\vskip-5mm
\parbox{0.22\linewidth}{$x=14.5$ \hskip13mm $u$}\parbox{0.78\linewidth}{\hskip38mm $v$ \hskip38mm $w$}
\centerline{
\includegraphics[width=0.3\textwidth,bb=50 20 550 265,clip=]{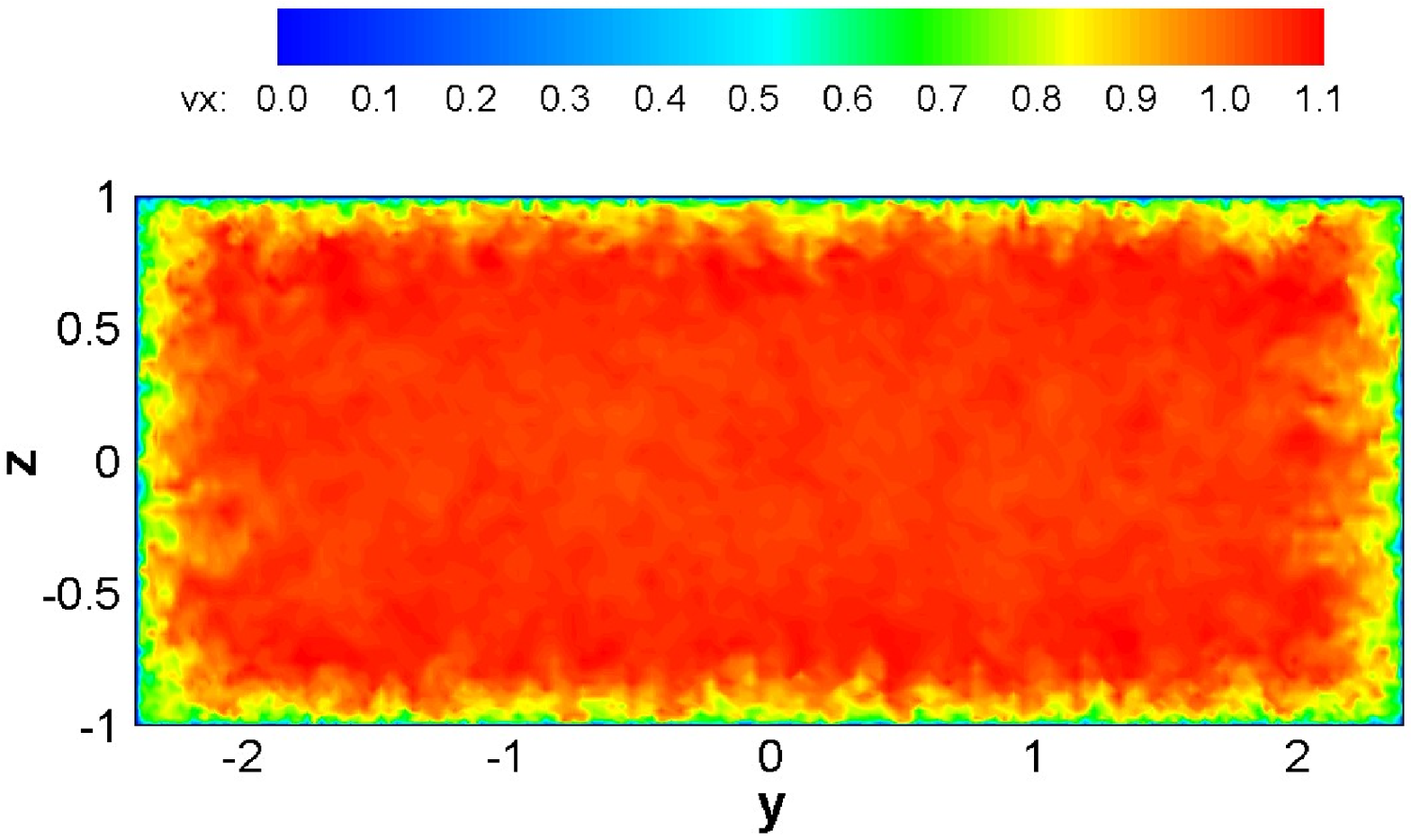}
\includegraphics[width=0.3\textwidth,bb=50 20 550 265,clip=]{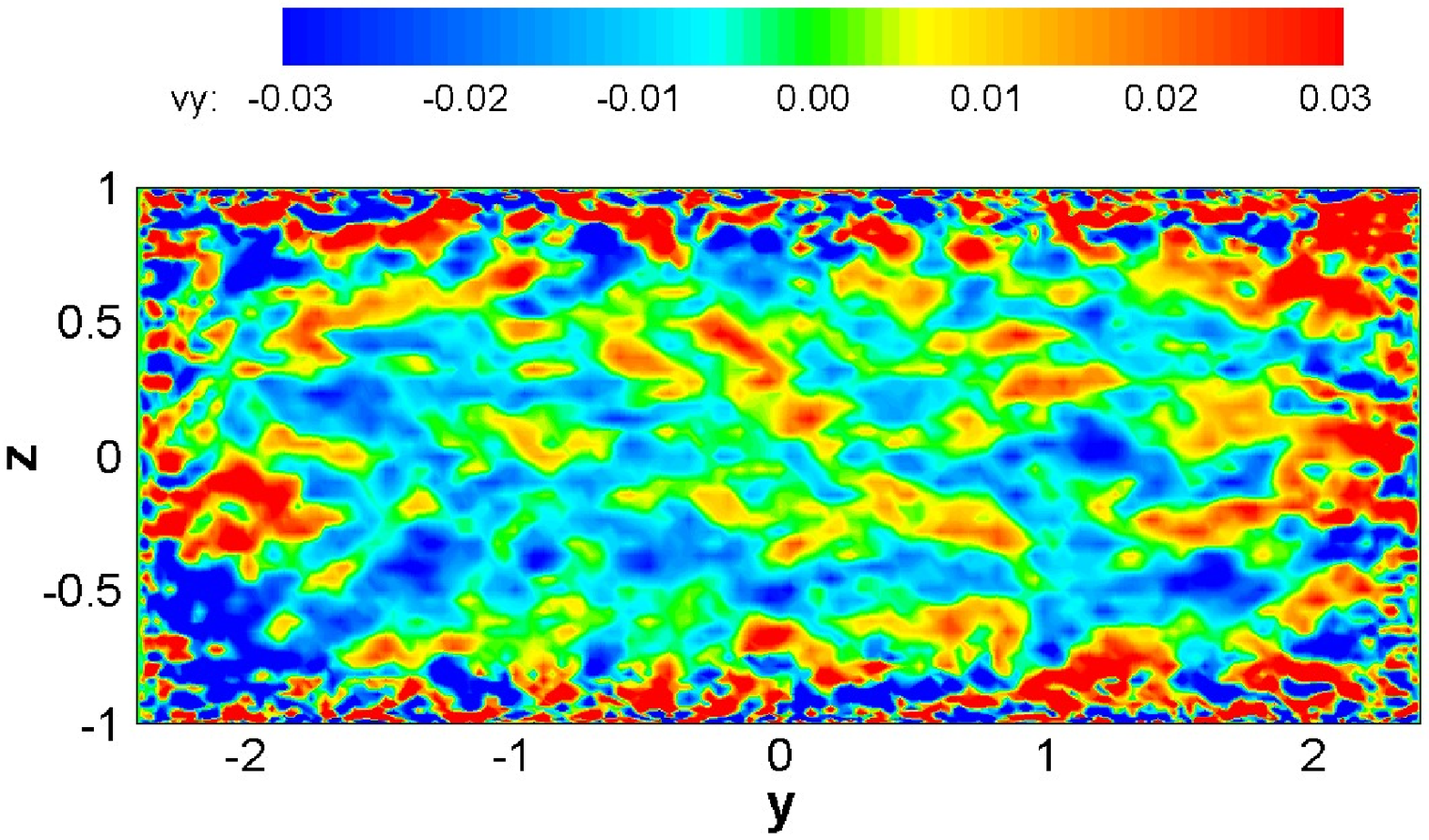}
\includegraphics[width=0.3\textwidth,bb=50 20 550 265,clip=]{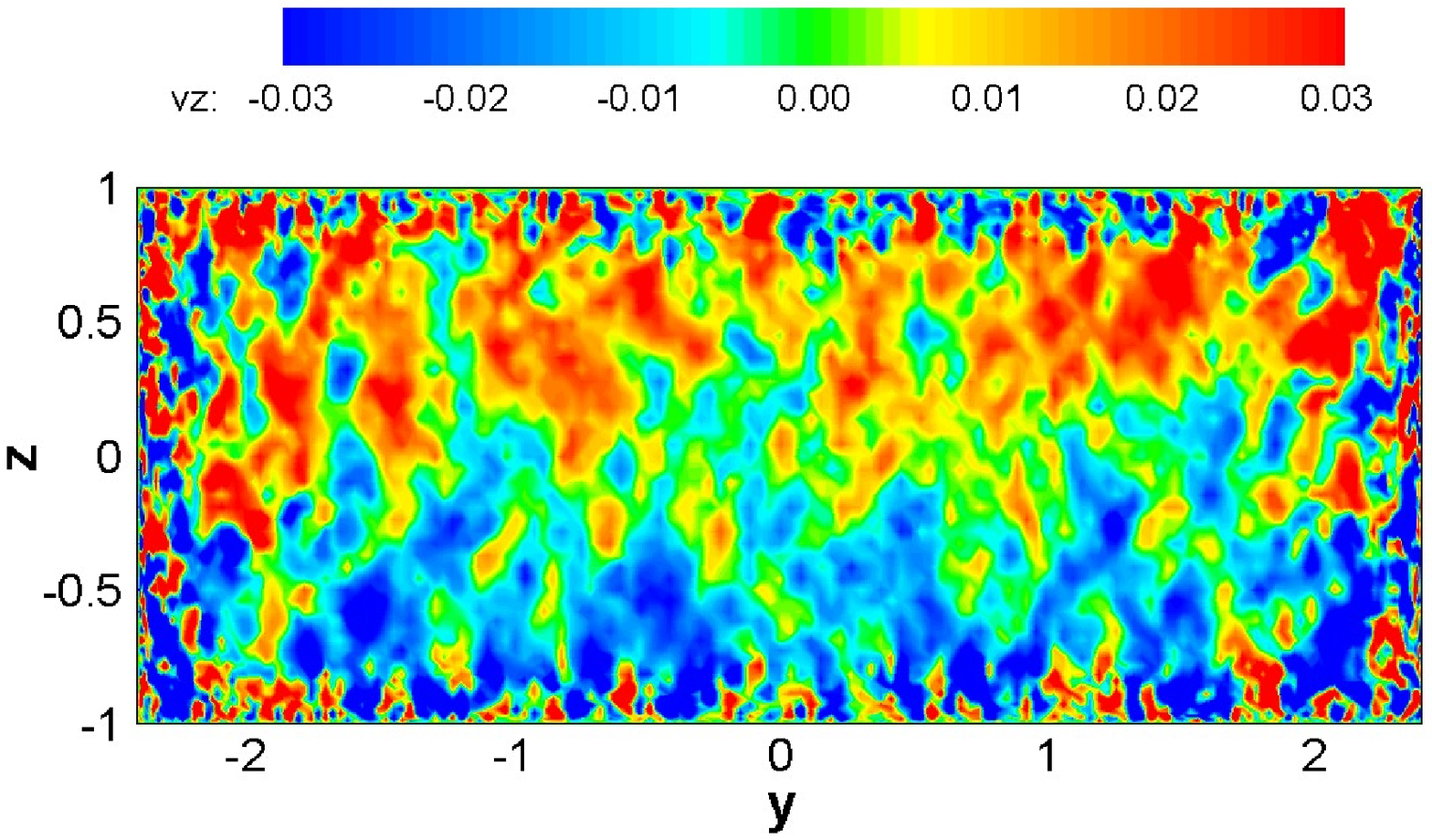}
}
\vskip-5mm
\parbox{0.22\linewidth}{$x=16$ \hskip15mm $u$}\parbox{0.78\linewidth}{\hskip38mm $v$ \hskip38mm $w$}
\centerline{
\includegraphics[width=0.3\textwidth,bb=50 20 550 265,clip=]{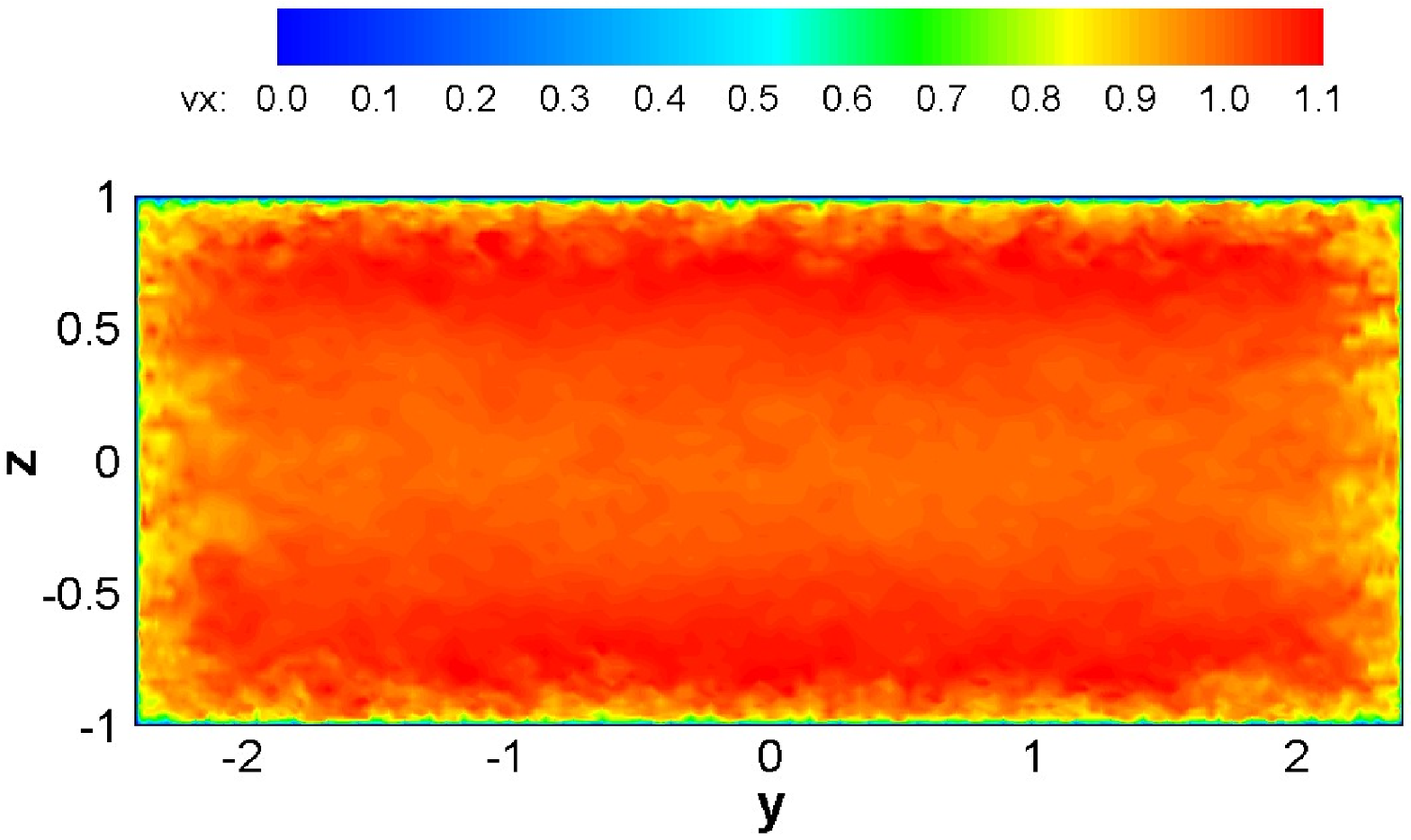}
\includegraphics[width=0.3\textwidth,bb=50 20 550 265,clip=]{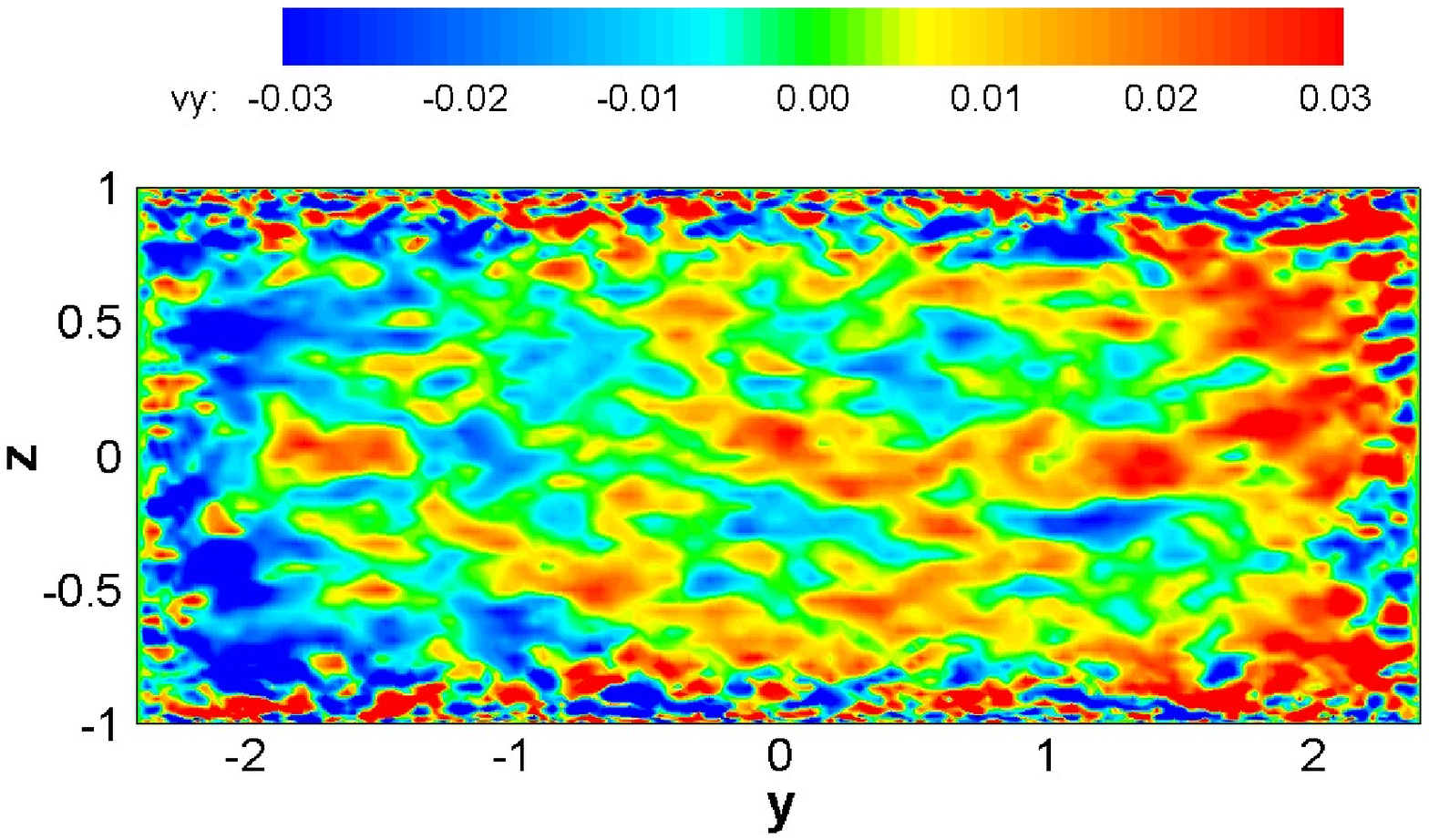}
\includegraphics[width=0.3\textwidth,bb=50 20 550 265,clip=]{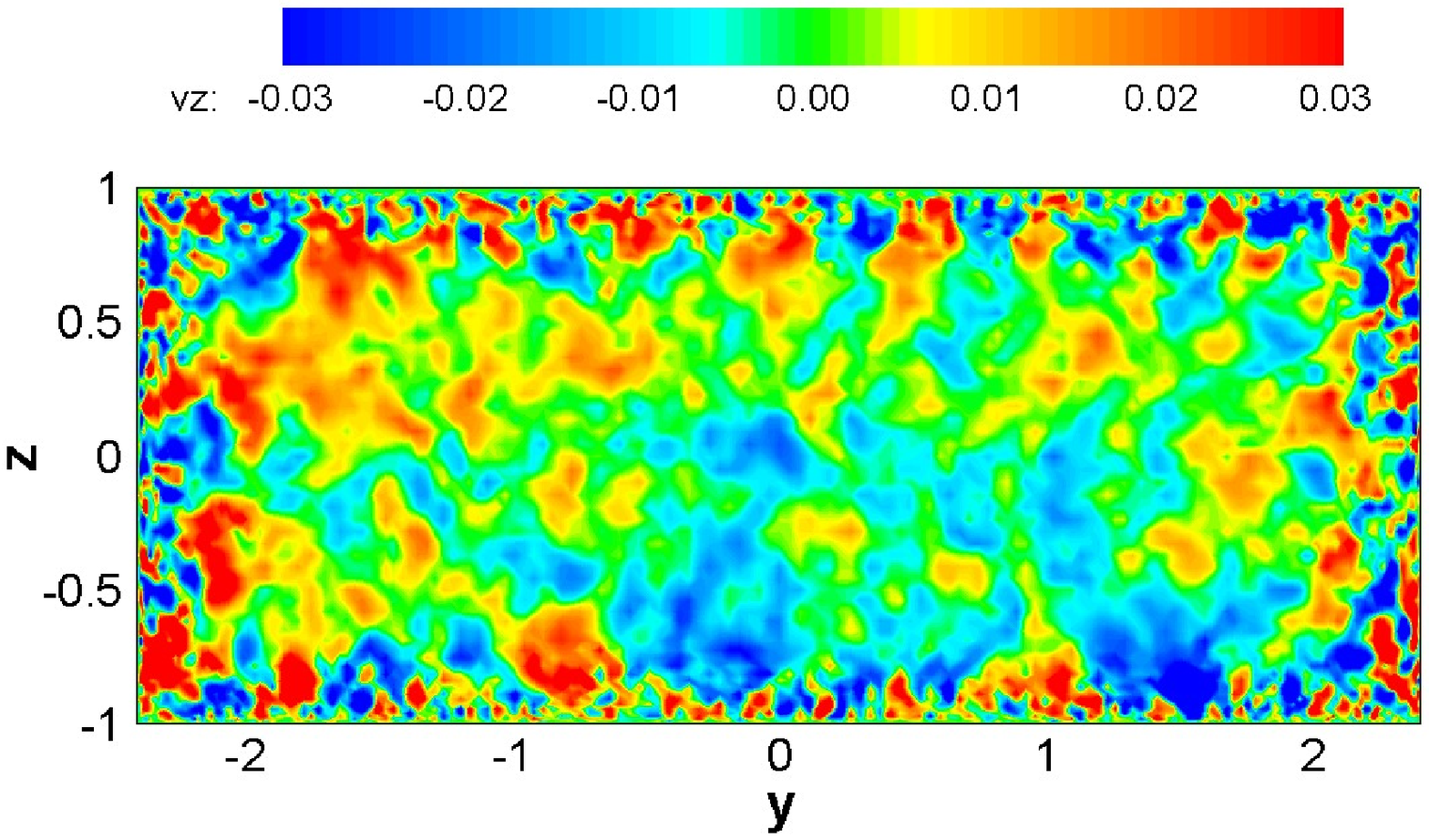}
}
\vskip-2mm

\hskip8.5mm
\includegraphics[width=0.25\textwidth]{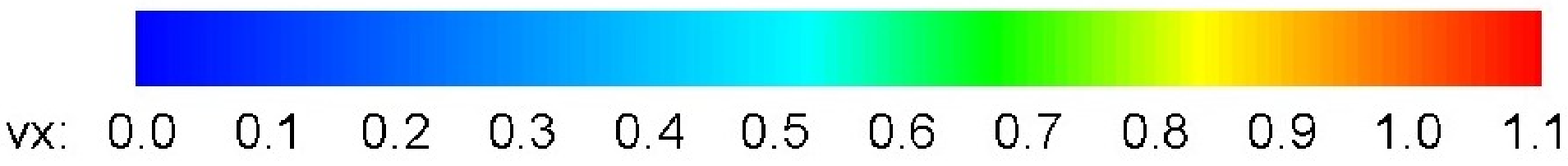}
\hskip27.5mm
\includegraphics[width=0.25\textwidth]{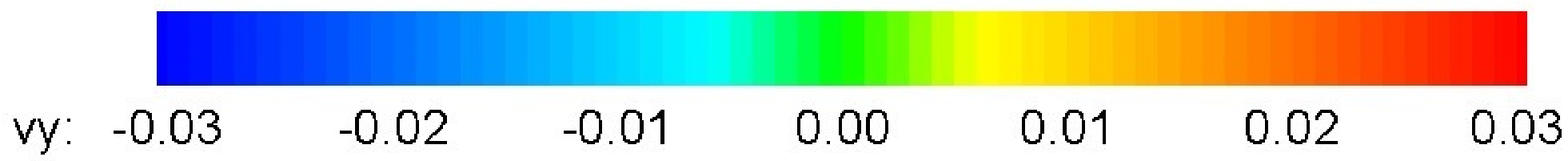}

\vskip-4mm
\caption{Transformation of the flow in the run 4 at the entrance into the strong magnetic field zone.
Instantaneous distributions of the three velocity components at $x=13$, $14.5$ and $16$ are shown.}
\label{fig:slice_YZ}
\end{figure}

The two-point correlation functions obtained for the transverse velocity components $w$ and $v$ in
the simulations 1-4 are shown in figures \ref{fig:corr_w} and \ref{fig:corr_v}. Long-range correlations
remain weak in the flows 1 and 2 during the entire decay and that there is practically no difference
between the two curves. This confirms the essentially three-dimensional small-scale structure of
turbulence in these flows.

The development of of strong correlations along the magnetic field in
the flows 3 and 4 is consistent with the results of earlier simulations and theoretical
models \citep[see \eg][]{Moffatt:1967,Davidson:1997,Zikanov:1998,Krasnov:2008a,Burattini:2010}
of transformation of turbulent flow under the impact of a strong magnetic field. While the growth
of the typical  scale of the turbulent structures in the field direction is always stronger and
the only one caused directly by the Joule dissipation, the growth of the typical transverse size
is caused by the  enlargement of the quasi-two-dimensional vortices.

The results obtained for the correlation coefficient $R^v$ in flow 4 at $x=16$ (see figure \ref{fig:corr_v}f)
may appear surprising. The  flow has nearly constant significant correlations ($R^v\approx 0.2$) over
almost the entire duct width. This is not observed for any other computed correlation coefficient
in any other cross-section. The reason for this behavior is illustrated in figure \ref{fig:slice_YZ}.
From $x=13$ to $x=16$, the streamwise velocity $u$ changes its profile in the way typical for a duct flow
entering a strong magnetic field \citep[see e.g.][for a discussion of the flow transformation]{Andreev:2006}.
Along the $y$-axis parallel to the magnetic field, the Hartmann profile with nearly uniform velocity in
the core and thin Hartmann boundary layers develops. Along the $z$-axis, the profiles acquires the typical
M-shape. The redistribution of the streamwise velocity is accompanied by a non-zero mean flow toward
the walls at $y=\pm 1$ (clearly visible in the distribution of $v$ at $x=16$) and in the $z$-direction
(visible in the distribution of $w$ at $x=14.5$, i.e. slightly upstream of the beginning of full-strength
magnetic field, in agreement with the scenario of formation of the M-shaped profile). The elevated
correlation coefficient $R^v$ in flow 4 at $x=16$ is caused by the flow in the $y$-direction. 

\vskip-3mm
\bibliographystyle{jfm}

\end{document}